\documentclass[modern]{aastex631}
\usepackage{amsmath,amssymb}
\usepackage{pifont}
\usepackage{xcolor}
\usepackage{animate}
\usepackage{mdframed}
\usepackage{pythonhighlight}
\usepackage{hyperref}
\usepackage{graphicx} % Required for inserting images
\usepackage{multirow}
\usepackage{algorithm}
\usepackage{algpseudocode}
\usepackage{booktabs}
\usepackage{tabularx}
\usepackage{comment}

\allowdisplaybreaks

\begin{document}

\title{Efficient PSF Modeling with ShOpt.jl: \\A PSF Benchmarking Study with JWST NIRCam Imaging}

\author[0000-0002-8494-3123]{Edward M.\ Berman}
\affiliation{Northeastern University, 100 Forsyth St. Boston, MA 02115, USA}

\author[0000-0002-9883-7460]{Jacqueline E.\ McCleary}
\affiliation{Northeastern University, 100 Forsyth St. Boston, MA 02115, USA}

%%%%%%%%%%%%%%%%%%%%%%%%%%%%%%%%%%%%%%%%%%%%%%%%%%%%%%%%%%%%%%%%%%%%%%%%
% Heavily relied upon insights and conversations. Order TBD
%%%%%%%%%%%%%%%%%%%%%%%%%%%%%%%%%%%%%%%%%%%%%%%%%%%%%%%%%%%%%%%%%%%%%%%%

\author[0000-0002-6610-2048]{Anton M.\ Koekemoer}
\affiliation{Space Telescope Science Institute, 3700 San Martin Dr., Baltimore, MD 21218, USA} 

\author[0000-0002-3560-8599]{Maximilien Franco}
\affiliation{The University of Texas at Austin, 2515 Speedway Blvd Stop C1400, Austin, TX 78712, USA}

\author[0000-0003-4761-2197]{Nicole E.\ Drakos}
\affiliation{Department of Physics and Astronomy, University of Hawaii, Hilo, 200 W Kawili St, Hilo, HI 96720, USA}

\author[0000-0001-9773-7479]{Daizhong Liu}
\affiliation{Max-Planck-Institut f\"ur Extraterrestrische Physik (MPE), Giessenbachstr. 1, D-85748 Garching, Germany}

\author[0000-0002-8987-7401]{James W.\ Nightingale}
\affiliation{Department of Physics, Institute for Computational Cosmology, Durham University, South Road, Durham DH1 3LE, UK}

\author[0000-0002-7087-0701]{Marko Shuntov}
\affiliation{Institut d'Astrophysique de Paris, CNRS, Sorbonne Universit\'e, 98bis Boulevard Arago, 75014, Paris, France}

\author[0000-0001-8450-7885]{Diana Scognamiglio}
\affiliation{Jet Propulsion Laboratory, California Institute of Technology, 4800, Oak Grove Drive - Pasadena, CA 91109, USA}

\author[0000-0002-6085-3780]{Richard Massey}
\affiliation{Department of Physics, Centre for Extragalactic Astronomy, Durham University, South Road, Durham DH1 3LE, UK}

\author[0000-0003-3266-2001]{Guillaume Mahler}
\affiliation{Department of Physics, Institute for Computational Cosmology, Durham University, South Road, Durham DH1 3LE, UK}
\affiliation{Department of Physics, Centre for Extragalactic Astronomy, Durham University, South Road, Durham DH1 3LE, UK}

\author[0000-0002-9489-7765]{Henry Joy McCracken}
\affiliation{Institut d’Astrophysique de Paris, UMR 7095, CNRS, and Sorbonne Universit{\'e}, 98 bis boulevard Arago, F-75014 Paris, France}

\author[0000-0002-4271-0364]{Brant E.\ Robertson}
\affiliation{Department of Astronomy and Astrophysics, University of California, Santa Cruz, 1156 High Street, Santa Cruz, CA 95064, USA}

\author[0000-0002-9382-9832]{Andreas L.\ Faisst}
\affiliation{Caltech/IPAC, 1200 E. California Blvd., Pasadena, CA 91125, USA}

\author[0000-0002-0930-6466]{Caitlin M.\ Casey}
\affiliation{The University of Texas at Austin, 2515 Speedway Blvd Stop C1400, Austin, TX 78712, USA}
\affiliation{Cosmic Dawn Center (DAWN), Denmark}

\author[0000-0001-9187-3605]{Jeyhan S.\ Kartaltepe}
\affiliation{Laboratory for Multiwavelength Astrophysics, School of Physics and Astronomy, Rochester Institute of Technology, 84 Lomb Memorial Drive, Rochester, NY 14623, USA}

\collaboration{20}{COSMOS-Web: The JWST Cosmic Origins Survey}

\correspondingauthor{Edward Berman}
\email{berman.ed@northeastern.edu}

%\author{F.X Timmes}
%\affiliation{Arizona State University}
%\affiliation{AAS Journals Associate Editor-in-Chief}

%\author{Amy Hendrickson}
%\altaffiliation{AASTeX v6+ programmer}
%\affiliation{TeXnology Inc.}

%\author{Julie Steffen}
%\affiliation{AAS Director of Publishing}
%\affiliation{American Astronomical Society \\
%1667 K Street NW, Suite 800 \\
%Washington, DC 20006, USA}

%% Note that the \and command from previous versions of AASTeX is now
%% depreciated in this version as it is no longer necessary. AASTeX 
%% automatically takes care of all commas and "and"s between authors names.

%% AASTeX 6.31 has the new \collaboration and \nocollaboration commands to
%% provide the collaboration status of a group of authors. These commands 
%% can be used either before or after the list of corresponding authors. The
%% argument for \collaboration is the collaboration identifier. Authors are
%% encouraged to surround collaboration identifiers with ()s. The 
%% \nocollaboration command takes no argument and exists to indicate that
%% the nearby authors are not part of surrounding collaborations.

%% Mark off the abstract in the ``abstract'' environment. 
\begin{abstract}

With their high angular resolutions of 30--100 mas, large fields of view, and complex optical systems, imagers on next-generation optical/near-infrared space observatories, such as the Near-Infrared Camera (NIRCam) on the James Webb Space Telescope (JWST), present both new opportunities for science and also new challenges for empirical point spread function (PSF) characterization. In this context, we introduce \texttt{ShOpt}, a new PSF fitting tool developed in Julia and designed to bridge the advanced features of PIFF (PSFs in the Full Field of View) with the computational efficiency of PSFEx (PSF Extractor). Along with \texttt{ShOpt}, we propose a suite of non-parametric statistics suitable for evaluating PSF fit quality in space-based imaging. Our study benchmarks \texttt{ShOpt} against the established PSF fitters PSFEx and PIFF using real and simulated COSMOS-Web Survey imaging. We assess their respective PSF model fidelity with our proposed diagnostic statistics and investigate their computational efficiencies, focusing on their processing speed relative to the complexity and size of the PSF models. We find that \texttt{ShOpt} can already achieve PSF model fidelity comparable to PSFEx and PIFF while maintaining competitive processing speeds, constructing PSF models for large NIRCam mosaics within minutes.

\end{abstract}

%% Keywords should appear after the \end{abstract} command. 
%% The AAS Journals now uses Unified Astronomy Thesaurus concepts:
%% https://astrothesaurus.org
%% You will be asked to selected these concepts during the submission process
%% but this old "keyword" functionality is maintained in case authors want
%% to include these concepts in their preprints.

%% From the front matter, we move on to the body of the paper.
%% Sections are demarcated by \section and \subsection, respectively.
%% Observe the use of the LaTeX \label
%% command after the \subsection to give a symbolic KEY to the
%% subsection for cross-referencing in a \ref command.
%% You can use LaTeX's \ref and \label commands to keep track of
%% cross-references to sections, equations, tables, and figures.
%% That way, if you change the order of any elements, LaTeX will
%% automatically renumber them.
%%
%% We recommend that authors also use the natbib \citep
%% and \citet commands to identify citations.  The citations are
%% tied to the reference list via symbolic KEYs. The KEY corresponds
%% to the KEY in the \bibitem in the reference list below. 

\section{Introduction} \label{sec:intro}

The inherent limitations of optical systems introduce artifacts into telescope imaging. Effects like diffraction, optical aberrations, atmospheric turbulence (if applicable), and telescope jitter are summarized in the telescope’s point spread function (PSF): the response of the optical system to an idealized point of light. Good PSF modeling during image processing reduces the impact of things like atmospheric turbulence and optical aberrations during scientific analysis. Failure to model the PSF correctly can lead to inaccurately measured positions, sizes, and shapes of small targets like galaxies.

The central importance of PSF characterization to astrophysics means that there is a wealth of PSF fitters available. These generally fall into two classes: forward-modeling approaches, which use physical optics propagation based on models of optical elements, and empirical approaches, which use observed stars as fixed points to model and interpolate the PSF across the rest of the image. In both cases, the PSF model may be validated by comparing a set of reserved stars to the PSF model's prediction.

 Empirical characterization tools like PSFEx \citep{2011ASPC} and PIFF \citep{Jarvis_2020} \footnote{\url{https://github.com/astromatic/psfex} and \url{https://github.com/rmjarvis/Piff}} are widely popular in astrophysics. However, the quality of PIFF and PSFEx models tends to be quite sensitive to the values of hyperparameters used to run the software, with optimal parameter selection sometimes relying on brute-force guess-and-check runs. PIFF, using the modelling and interpolation scheme used for the Dark Energy Survey Year 3 observations, is also notably inefficient for large, well-sampled images, taking hours in the worst cases.

%The interplay between the precision of predictive PSF models and the real-world adaptability of empirical PSF models offers a well-rounded approach to understanding PSF behavior, as each validates and refines the other. 

%Recently, researchers at STScI developed forward models of JWST PSFs by considering the optics of the telescope.  The main drawback of using these models is that they fail to account for effects like scattered light from the optical tube and “unknown unknowns” that might emerge.

%However, the quality of PIFF and PSFEx models tends to be quite sensitive to the values of hyperparameters used to run the software, with optimal parameter selection relying on brute-force guess-and-check runs. PIFF is also notably inefficient for large, well-sampled images, taking hours in the worst cases. 
%While there are many existing empirical PSF fitters, they were built either as general purpose tools or for very dramatically different data sets, see \cite{2011ASPC, Jarvis_2020}. 

% Weirdly introduced...
Because space telescopes are unimpeded by the atmosphere, their PSFs are also often characterized with forward-modeling approaches like Tiny Tim \citep{krist201120} and WebbPSF \citep{2012SPIEWebb,10.1117/12.2056689, ji2023jades}. WebbPSF models are continually updated based on telescope telemetry, ensuring high accuracy in all bandpasses and for all instruments regardless of image noise. While robust, forward modeling is not infallible and may occasionally miss short-timescale variations and other ``unknown unknowns'' that can be captured with empirical PSF models, albeit at the cost of much higher noise.

The James Webb Space Telescope (JWST) represents a giant leap forward in our ability to explore the cosmos. Equipped with groundbreaking instruments like the Near Infrared Camera (NIRCam) and the Mid-Infrared Imager (MIRI), the telescope is poised to unlock unprecedented insights into the early universe \citep{robertson2022galaxy}. At the same time, these advances usher in a new set of complexities for PSF characterization:
\begin{enumerate}
    \item Modeling the PSFs of space observatories like JWST is non-trivial since they generally operate near their diffraction limits and exhibit intricate optical patterns that defy the analytic approximations acceptable for PSFs of ground-based imaging. In particular, the PSFs exhibit steep spatial gradients, making them highly sensitive to the hyperparameters of commonly used analytic profiles, where slight variations can lead to substantial inaccuracies. The assessment of PSF models for space-like imaging presents an additional challenge, as most metrics of fit quality, e.g., full width at half maximum or second moments of intensity, treat the PSF as an elliptical Gaussian -- a poor approximation for the ``spiky'' NIRCam and MIRI PSFs. While there are some useful moment based metrics for assessing fit quality \citep[e.g.][]{zhang2023impact}, there are no universally adopted non-parametric diagnostics for assessing pixel-level model biases.
    %This calls for a new suite of diagnostics. 
    \item The NIRCam detector pixel scales are 0.031\arcsec/pixel and 0.063\arcsec/pixel for the short and long wavelength channels, respectively \citep{2003SPIE, 2005SPIE, 2012SPIE}. At these fine scales, fully capturing the intensity profile of the NIRCam PSF requires a much larger number of pixels than is needed for typical surveys, whose detectors have pixel scales 3--10 times larger than NIRCam \citep{Jarvis_2020, Fu_2022, York_2000, mccleary2023lensing}. As a consequence, the number of pixels need to model the full size of the PSF is much greater.
    %the wings of the PSF at this detector pixel scale, our vignettes must be $131 \times 131$ to $261 \times 261$ pixels across. Because of NIRCam's small pixel scale, fully capturing the  are needed The vignette sizes we're considering are much larger than those in prior surveys  
    This new regime prompts a necessary evaluation of how current PSF fitters perform at this new scale. That is, can the PSF fitters still capture the full dynamic range of the distortion and do so in a reasonable time.  
\end{enumerate}

To meet these challenges, we introduce \texttt{ShOpt}\footnote{The name ShOpt is a contraction of ``shear optimization.'' That is, we are finding the best-matching PSF by formulating optimization problems over the space of all possible shears. The name was inspired by the \texttt{manopt} library, a contraction of manifold optimization. It is an apt comparison given the manifold learning we describe in Section \ref{subsec:analytic}.}, 
a new PSF modeling tool that strives to retain the best of existing PSF modeling software while advancing their mathematical formulation and increasing their computational efficiency. Written in the high-level Julia language using a functional programming style\footnote{By functional, we mean the functional programming paradigm. While Julia provides support for object-oriented design patterns with \texttt{structs}, Julia code is generally written with design patterns that put the emphasis on reusable functions.}, \texttt{ShOpt} offers both accessibility and speed, positioning it as a valuable tool for the astrophysical community. 
\texttt{ShOpt} introduces manifold-based algorithms for enhanced efficiency in analytic profile fitting. \texttt{ShOpt} also employs three distinct techniques for pixel-basis fitting: principal component analysis (PCA), an autoencoder, and kernel smoothing.

Along with \texttt{ShOpt}, we introduce a suite of non-parametric PSF fit statistics appropriate for space-based imaging, namely reduced $\chi^2$, mean relative error, and mean absolute error. Our proposed PSF characterization statistics are inspired by strong gravitational lensing analyses and move beyond the conventional metrics based on elliptical Gaussians. In this study, we evaluate the performance of \texttt{ShOpt}, PSFEx, and PIFF using data from the COSMOS-Web survey \citep{casey2023cosmosweb}, using our proposed PSF fit statistics to gauge their relative performance. In addition, we time the PSF fitters to measure the computational efficiency of each tool.

To summarize, our contributions are the following:
\begin{itemize}
    \item We introduce \texttt{ShOpt}, and with it, a number of methods for efficient empirical PSF characterization. 
    \item We benchmark the model fidelity of different PSF fitters using non-parametric approaches. We use real and simulated catalogs from the COSMOS-Web Survey and measure $\chi^2$, mean relative error, and mean absolute error. We supplement these statistics with conventional second moment HSM fits \citep{mandelbaum2005systematic, hirata2003shear}.
    \item We benchmark the computational efficiency of different PSF fitters.
\end{itemize}

The remainder of this paper is structured as follows. In Section \ref{sec:style}, we establish the workflow of \texttt{ShOpt} and the notation used in this paper. In Section \ref{sec:psfHat}, we develop our methods for Gaussian and pixel-basis PSF fits, and in Section \ref{sec:FOV} our methods for fitting their variation across the field of view. In Section \ref{sec:shDH}, we describe our benchmarking methods and data sets. Our algorithmic choices for \texttt{ShOpt} are justified using big-$\mathcal{O}$ time complexity analysis in Section \ref{sec:runtime}, specifically detailing \texttt{ShOpt}'s speed variation with different input parameters. In Section \ref{sec:datacleaning}, we describe our benchmarking methods and data sets. We present our results in Section \ref{sec:results}, and discussion and conclusions in Section \ref{sec:conclusions}.

\section{ShOpt Notation, Workflow, and Overview} \label{sec:style}

\subsection{Notation and Preliminaries}
We represent star locations in pixel coordinates as $(x,y)$ and in celestial coordinates (expressed in degrees) as $(u,v)$. In terms of a nominal position source, positive $u$ is west, and positive $v$ is to the north. ``Vignette'' refers to small, localized images, centered around individual stars or celestial objects, that are extracted from larger astronomical images.

We use the following notation for working with manifolds. For two sets $A$, $B$,
\begin{eqnarray}
    B_2 &\equiv& \{ \left[x,y\right] : x^2 + y^2 < 1 \} \subset \mathbb{R}^2 \\[0.5em]
    \mathbb{R}_+ &\equiv& \{ x : x > 0 \} \subset \mathbb{R}\\[0.5em]
    A \times B &\equiv& \{(a, b): a \in A, b \in B \}
\end{eqnarray}

Throughout this paper, we describe the shape of a smoothly varying (analytic) PSF profile in terms of the variables $\left[s, g_1, g_2\right]$, where $(g_1, g_2)$ are the polarization states of an elliptical source with reduced shear $\mathbf{g} = g_1 + ig_2 = ge^{i2\theta}$, where $\theta$ is the angle of the major axis from [West/some fiducial orientation] and
\begin{equation}
    g \equiv \frac{1 - q}{1 + q},
\end{equation}
for an ellipse with (major/minor) axis ratio $q$, such that $0\leq g<1$ \citep{Bernstein_2002}. While shear is typically treated as a number in the complex plane $\mathbb{C}$, we make an equivalent characterization that shear is a vector $\left[g_1, g_2\right] \in \mathbb{R}^2$.  This vector representation corresponds to the real and imaginary parts of the complex shear. The free parameter $s$ represents the size of the ellipse (the geometric mean of the major and minor axis lengths). We introduce free parameters $\left[ \sigma, e_1, e_2\right]$ that reparameterize $\left[s, g_1, g_2\right]$, noting that we adopt a slightly different relationship between ellipticity $e$ and shear $g$ than may be seen elsewhere in the literature, dropping of a factor of $2$ in Equation~\ref{eq:g(e)} for the purposes of an easier calculation of the inverse map (details in Section \ref{sec:psfHat}).

Additionally, we are often concerned with a more realistic pixel basis, where each pixel in an $n \times n$ star image is a basis element. 

For our non-parametric summary statistics defined in Equations \ref{eq:Reducedchisq}--\ref{eq:MAE}, we use $v_{i,j,k}$ to represent the $i \times j$ pixels in vignette $k$, and $p_{i,j,k}$ to represent the $i \times j$ pixels in PSF model $k$. $\sigma_{i,j,k}$ represents the uncertainty in the model at pixel $(i,j)$ and vignette $k$. The total number of vignettes is $K$. The number of pixels in a vignette is denoted $N_\mathrm{pix}$.
\begin{align}
    \overline{\chi^2}_\nu = \frac{1}{K}\sum_k \frac{1}{d.o.f.}\sum _{(i,j)}^{N_\mathrm{pix}}  \frac{\left(v_{i,j,k} - p_{i,j,k}\right)^2}{\left(\sigma_{i,j,k}\right)^2} \label{eq:Reducedchisq} \\[0.5em]
    \text{median }\chi^2_\nu = \text{median}~_K \left[ \frac{1}{d.o.f.}\sum _{(i,j)}^{N_\mathrm{pix}}  \frac{\left(v_{i,j,k} - p_{i,j,k}\right)^2}{\left(\sigma_{i,j,k}\right)^2} \right]  \label{eq:MedianReducedchisq}\\[0.5em]
    \text{Mean Relative Error} = \frac{1}{N_\mathrm{pix}}\sum _{(i,j)}^{N_\mathrm{pix}} \frac{1}{K}\sum_k \frac{v_{i,j,k} - p_{i,j,k}}{v_{i,j,k}} \label{eq:MRE} \\[0.5em]
    \text{Mean Absolute Error} = \frac{1}{N_\mathrm{pix}}\sum _{(i,j)}^{N_\mathrm{pix}} \frac{1}{K}\sum_k \frac{\mid v_{i,j,k} - p_{i,j,k}\mid}{\mid v_{i,j,k} \mid } \label{eq:MAE}
\end{align}

We shall henceforth refer to mean relative error as MRE and mean absolute error as MAE. To enable flux-based comparisons between PSF vignettes, which most fitters normalize to unity, and star vignettes, we add the appropriate star flux to the PSF vignettes, then add Gaussian noise with mean and variance taken from the sky background in the vicinity of the star.

\subsection{Code Overview}

\texttt{ShOpt} takes inspiration from robotics algorithms, such as SE-Sync \citep{doi:10.1177/0278364918784361}, that run on manifold-valued data.  The manifold properties of shears are described in \cite{Bernstein_2002}; we expand on their work to provide more robust multivariate analytic fits to PSF intensity profiles. We specifically use multivariate Gaussians, as they are cheap to compute, but for a more rigorous treatment of optimization methods on manifold-valued data, see \cite{AbsMahSep2008} and \cite{boumal2023intromanifolds}. \texttt{ShOpt} also provides three modes for fitting PSFs in the pixel basis: \texttt{PCA} mode, \texttt{autoencoder} mode, and \texttt{smoothing} mode, detailed in Section \ref{sec:psfHat}. \texttt{PCA} mode approximates the original image by summing the first $n$ principal components, where $n$ is supplied by the user. We also introduce \texttt{autoencoder} mode, which uses a neural network with an autoencoder architecture to learn the PSF. A square image of side length $n$ can be thought of as vectors in $\mathbb{R}^{n^2}$, where each pixel is a basis element. The architecture is built so that the image is encoded into a vector space represented by a basis with dim$(V) < n^2$ before being decoded back into the dimension of our original image. The nonlinearity of the network ensures that the key features are learned instead of some linear combination of the sky background. Both \texttt{PCA} and \texttt{autoencoder} modes provide the end user with tunable parameters that allow for accurate reconstruction of the model vignettes without overfitting to noise. \texttt{Smoothing} mode applies a Lanczos kernel to the input vignettes, and uses the smoothed output for the pixel basis fit.

\vspace{1ex}
\begin{mdframed}[
    frametitle={Why Julia?},
    outerlinewidth=0.6pt,
    innertopmargin=6pt,
    innerbottommargin=6pt,
    roundcorner=4pt]
 \texttt{ShOpt} is written in Julia. The Julia programming language is a high-level and functional language like Python, which makes it accessible to a community of open-source developers. At the same time, Julia is equipped with a just-in-time compiler, which helps Julia code execute quickly by recycling compiled code. This offers a speed advantage over Python, which is first transpiled to C before being translated into machine code. Julia also offers some of the most sophisticated tools for problems at the intersection of numerical linear algebra and optimization, such as the \texttt{ForwardDiff.jl} and \texttt{Optim.jl} \citep{RevelsLubinPapamarkou2016, Mogensen2018} libraries. Julia also has an abundance of support for working with manifolds such as \texttt{manopt}, which may be pertinent in future releases of \texttt{ShOpt} \citep{Bergmann2022}. Finally, much of today's production code is written with the help of program synthesis tools such as GitHub Copilot. It is clear that these tools will soon make Julia more accessible than ever before to astrophysicists and other potential future \texttt{ShOpt} contributors. Work is being done to strengthen these tools for programming languages with much less training data available, such as \texttt{Julia}. \cite{cassano2023knowledge} successfully demonstrated how to transfer knowledge from programming languages with lots of publicly available training data (e.g., Python) to programming languages with much less training data available (e.g., Racket, OCaml, Lua, R, and most importantly, Julia). 
\end{mdframed}
\vspace{1ex}

The source code for \texttt{ShOpt} can be found on GitHub\footnote{\href{https://github.com/EdwardBerman/shopt}{https://github.com/EdwardBerman/shopt}
    and  \href{https://edwardberman.github.io/shopt/}{https://edwardberman.github.io/shopt/}
} and is detailed in our companion paper \citep{berman2023shoptjl}.

\texttt{ShOpt} accepts as input a FITS-format catalog containing star vignettes, positions, and signal-to-noise ratios (SNR). One concession \texttt{ShOpt} makes is that it makes more sense to build on the \texttt{Astropy} infrastructure than to rebuild everything from scratch. As such, \texttt{PyCall.jl} is used to handle FITS input and output. Since this is only done once and not iterated over in any main loop, this does not significantly affect performance. 

After an initial quality check based on SNR, the star vignettes are fit with a multivariate Gaussian to remove outlier stars. While these Gaussian fits are not used to create a PSF model, they are helpful in screening out stars that are saturated, too bright, or too faint. In situations where the image noise is high, the SNR threshold must be kept low to include enough stars for fitting. After these pre-processing steps, stars are fit in the pixel basis using either the \texttt{PCA} or \texttt{autoencoder} modes. The output learned stars are then fit with a multivariate Gaussian, which serves both to reject bad pixel-basis fits and to record the rough size and shape of the PSF. Surviving stars act as fixed points for polynomial interpolation of the PSF's spatial variation across the field of view. Finally, the learned data is saved to a summary FITS file, along with diagnostic plots and logging statements. This is summarized in Algorithm \ref{alg:dataprocess}.

\begin{algorithm}[H]
\caption{ShOpt Workflow}\label{alg:dataprocess}
\begin{algorithmic}[1]
\State $starCatalog \gets \text{loadStarCatalog()}$
\State $filteredCatalog \gets \text{filterBySignalToNoise}(starCatalog)$

\For{each vignette in filteredCatalog}
    \State $gaussianFit \gets \text{fitMultivariateGaussian}(vignette)$
    \If{not $\text{isGoodFit}(gaussianFit)$}
        \State $\text{remove}(vignette)$
    \EndIf
\EndFor

\State $pixelGrid \gets \text{createPixelGrid}(filteredCatalog)$
\State $reducedData \gets \text{dimensionalityReduction}(pixelGrid)$

\For{each grid in reducedData}
    \State $gaussianFit \gets \text{fitMultivariateGaussian}(grid)$
    \If{not $\text{isGoodFit}(gaussianFit)$}
        \State $\text{remove}(grid)$
    \EndIf
\EndFor

\For{num iterations}

\For{each point in fieldOfView}
    \State $interpolatedStar \gets \text{interpolate}(point, reducedData)$
    \If{$\text{isOutlier}(interpolatedStar)$}
        \State $\text{remove}(Star)$
    \EndIf
\EndFor
\EndFor
\State $\text{plotAndSaveData}()$

\end{algorithmic}
\end{algorithm}

\section{Parameter Estimation for PSF Modeling} \label{sec:psfHat}
\subsection{Analytic Profile Fitting} \label{subsec:analytic}

\texttt{ShOpt}'s analytic model has two components, a shear transformation and a radial function. We elect to fit an elliptical Gaussian $f_{Gaussian}(r)$ for our radial function:
\begin{equation}
    f_{Gaussian}(r) = Ae^{-r^2/2}
\end{equation}
where A makes the image sum to unity. There are other radial profiles that we could choose from, however, any radial profile parameterized by $\left[s, g_1, g_2\right]$ contains some azimuthal symmetry that makes the ``wings'' of the PSF difficult to model. Thus, it is not worth the computational cost to fix a radial profile that is any more elaborate than an elliptical Gaussian as there are diminishing returns for accuracy. 

For any analytic model, the shear is always the same, 
\begin{align}
    \begin{bmatrix} u'\\ v' \end{bmatrix} &= \frac{s}{\sqrt{1 - \left(g_1\right)^2 - \left(g_2\right)^2}}\begin{bmatrix}
    1 + g_1 & g_2\\
    g_2 & 1 - g_1
    \end{bmatrix}\begin{bmatrix} u\\ v \end{bmatrix} \label{eq:GC}\\
\end{align}
and  \begin{equation}
    r = \sqrt{\left(u'\right)^2 + \left(v'\right)^2}.
\end{equation} 

\begin{figure}[!htb]
    \centering
    \includegraphics[width=0.7\textwidth]{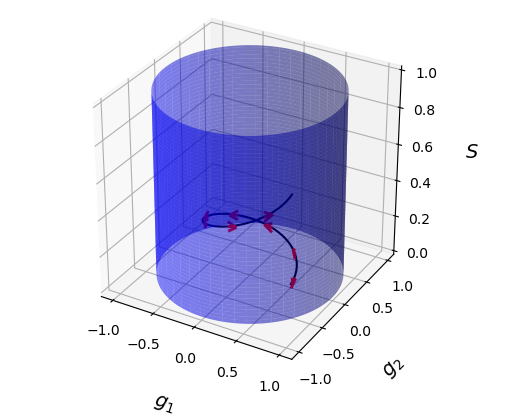}
    \caption{Parameter space for fitting an analytic profile, with sample iterations of the algorithm converging toward a learned $\left[s, g_1, g_2\right]$. Note that the cylinder extends upwards toward infinity but is bounded from below by $0$.}
    \label{fig:pspace}
\end{figure}

The shear matrix in Equation \ref{eq:GC} is known to be positive definite, and $s$ is strictly positive. Therefore, our parameters are constrained to $\mathbb{R}_+ \times B_2$. The parameter space can be visualized as a solid cylinder extending infinitely far from the origin in one direction, as seen in Figure \ref{fig:pspace}.
    
\begin{figure}[!htb]
    \centering
    \includegraphics[scale=0.45]{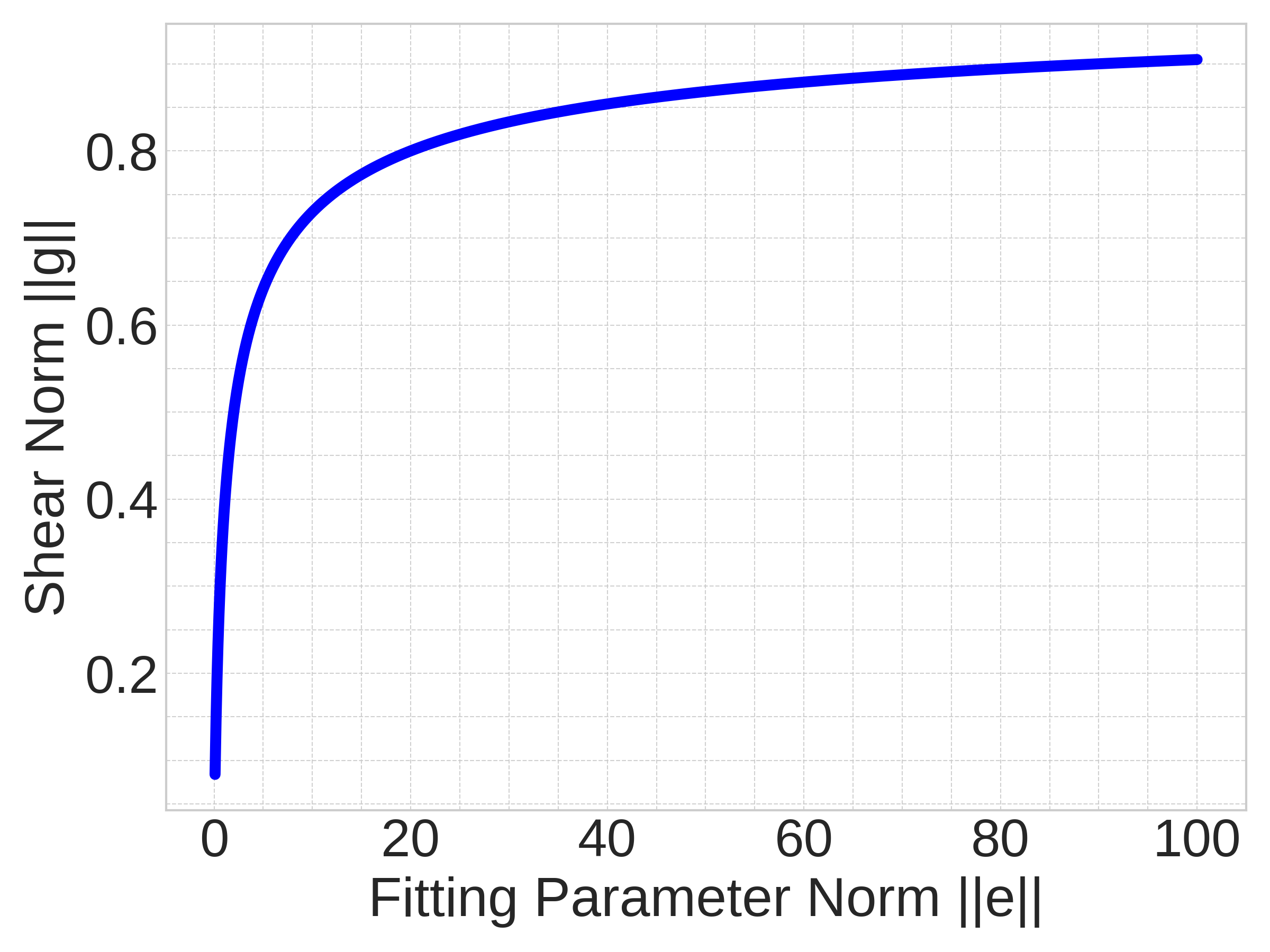}
    \caption{Reparameterization function. For any ellipticity vector $\left[e_1 , e_2\right]$ the associated shear vector $\left[g_1, g_2\right]$ has a norm in $[0,1)$.}
    \label{fig:rp}
\end{figure}

The finite domain of $g$ poses a problem for fitting. As such, we adopt a fitting parameter $e$ that allows us to map any value in $\mathbb{R}^2$ onto $g$:
\begin{equation}
        e_i = \frac{g_i}{1 - ||g||^2} .
\end{equation} 
This map\footnote{We could look at this as a map from $B_n \to \mathbb{R}^n $ but in this case we are only concerned about vectors $g = \left[ g_1, g_2\right]$ and $e = \left[ e_1, e_2\right]$} from $B_2 \to \mathbb{R}^2$ is relatively smooth, so that any automatic differentiation software can handle it during the fitting process. We can construct an inverse map via 
\begin{equation}
        ||g||^2 = \frac{2||e||^2}{1 + 2||e||^2 + \sqrt{1 +4||e||^2}}.
        \label{eq:g(e)}
\end{equation}
The inverse map allows us to constrain our update steps inside of our parameter space which leads to quicker convergence and the ability to handle noisier data more effectively. Equation \ref{eq:g(e)} also keeps the nice properties of smoothness and monotonicity desirable for activation functions, see Figure \ref{fig:rp}. Equation \ref{eq:g(e)} is not the canonical map from $\mathbb{R}^2$ to $B_2(r)$; two sigmoid functions could also have been used. We chose this particular function because it avoids the finite domain problem, is easily differentiable by the tools we use, and has a loose geometric interpretation described in \citep{bonnet1995statistical}. 

Note that this inverse only specifies what the new norm should be; the components still need to be adjusted accordingly.  We reparameterize\footnote{Technically, for $s$ to be strictly positive we would set $s := \sigma^2 + \varepsilon$, where $\varepsilon$ is some small positive perturbation. Numerically, it makes little difference: stars with very small $s$ are removed during pre-processing.} as follows: 
     \begin{equation}
        s \equiv \sigma^2 \label{eq:s}
    \end{equation}
     \begin{equation}
        g_i \equiv e_i\frac{||g||}{||e||}
    \end{equation}

The values of $\sigma, e_1, e_2 $ are obtained from each update step and $\lVert e \rVert$ is then determined from the usual $L_2$ norm. The parameter $s$ is obtained from Equation \ref{eq:s} and $\lVert g \rVert$ from Equation \ref{eq:g(e)}. Finally, $\lVert g \rVert$ is used with $\lVert e \rVert $ to calculate $g_1$ and $g_2$. 

We have that $r$ is a function of $\left[ u, v, s, g_1, g_2\right]$ and $f$ is a function of $r$ such that 
\begin{equation}
    f(r) \equiv I_p(u,v,s, g_1, g_2). \label{eq:I}
\end{equation}
Even though we solve for parameters $\left[ \sigma, e_1, e_2\right]$, the loss is still dictated by $\left[s, g_1, g_2\right]$.
\begin{equation}
  \text{cost}(s_t, g_{1_t}, g_{2_t}) = \frac{1}{2} \sum_{u_{pix}} \sum_{v_{pix}} (I_p(u , v)^* - I_p(u, v, s_t, g_{1_t}, g_{2_t}))^2  
\end{equation}
Here $*$ denotes the ground truth (i.e. star) in the star vignette and the subscript $t$ denotes iteration $t$ in an LBFGS run. Parameters are found using Julia's \texttt{Optim.jl} library \citep{Mogensen2018} and the gradient is computed using Julia's \texttt{ForwardDiff.jl} library \citep{RevelsLubinPapamarkou2016}.

\subsection{Pixel Grid Fits}
\texttt{PCA} mode, \texttt{autoencoder} mode, and \texttt{smoothing} mode all provide ``pixel-grid'' mechanisms for reconstructing the PSF in the image pixel basis. We outline each of these modes below. 

\subsubsection{PCA Mode}
\texttt{PCA} mode approximates a star image with a rank-$n$ approximation of the original image using principal component analysis, where $n$ is supplied by the user. Modeling the vignette in this way gives the user a method of determining how much they want their pixel grid model to represent the original star vignette. Choosing a lower rank approximation will yield less detail on the shape of the PSF, but will do so quickly and without much noise contamination. On the other hand, a higher rank approximation will capture more of the fine details, but at the risk of capturing some of the noise. It should be noted that the choice of $n$ has little overhead cost on the \texttt{ShOpt} computation time.

To minimize aliasing effects that might appear in low-rank approximations, the output image is convolved with a smoothing Lanczos kernel $\mathcal{L}$:
\begin{equation}
    \mathcal{L}(x, a) = 
 \begin{cases} 
0 & \text{if } |x| \geq a \\
1 & \text{if } x = 0 \\
a  \text{sinc}(\pi x) \text{sinc}\left(\frac{\pi x}{a}\right) & \text{otherwise}
\end{cases},
\label{eq:lk}
\end{equation}
where $a$ is a tunable parameter for the size of the kernel and $x$ is the distance from the center of the kernel. 

\subsubsection{Autoencoder Mode}
\texttt{Autoencoder} mode uses deep learning to reconstruct PSF vignettes pixel by pixel. We adopt an autoencoder architecture because the projection into a latent space forces the machine learning algorithm to learn the key features of the image regardless of the noise present. The operations are carried out by Julia's \texttt{Flux.jl} machine learning library \citep{innes:2018}.

In our specific architecture, the input star vignette is first flattened and then passed through the network, which has an encoder with one layer containing $128$ nodes and a second layer containing $64$ nodes. The encoder feeds into a latent space with $32$ nodes, which is then decoded back into a layer of $64$ nodes, $128$ nodes, and finally back into an image (vignette) of the same dimension as the input vignette. We enforce both the input and output vignettes to sum to unity so that the relative distributions of intensities can be compared. The loss function is a mean squared error:
\begin{equation}
    \text{cost}(x) = \frac{1}{N_\mathrm{pix}}\sum_{N_\mathrm{pix}} (x_p - \hat{x}_p)^2,
\end{equation}
where $\hat{x}$ denotes the image after it has been put through the autoencoder. 

The network consists of two activation functions. A \texttt{leakyrelu} \citep{Maas2013RectifierNI} function is used for all layers except for the last one. This choice reflects that most of the vignette pixel values are positive, and the ones that are not are usually close to zero or eliminated in data pre-processing. The final layer uses a \texttt{tanh} activation function to ensure that output values stay bounded between $(-1,1)$. The network trains until it either hits a specified number of epochs or until it hits a stopping gradient. We encourage exploration of the number of epochs and the minimum stopping gradient to find an appropriate middle ground between accuracy and time of completion. Stopping gradients between $10^{-5}$ and $10^{-6}$ are usually sufficient to get a good approximation in reasonable time and in $100$ epochs or less. This was revealed through the use of diagnostic plots that are introduced in Section \ref{sec:results}. The activation functions and number of nodes are not tunable by default; changing those is not recommended.

\subsubsection{Smoothing Mode}
In the case of well-sampled images like the NIRCam data considered in this analysis, we find that a basic smoothing is sufficient before interpolation. This is implemented in ShOpt as \texttt{Smoothing} mode, which uses only the Lanczos kernel introduced in Equation \ref{eq:lk} before  produce the pixel grid fit. While \texttt{PCA} and \texttt{Autoencoder} can yield denoised PSF models from low dimensional reconstructions, \texttt{Smoothing} mode is a simpler technique that avoids some of their limitations, albeit at the expense of noisier models.

\section{Interpolation Across the Full Field of View} \label{sec:FOV}
To fit for the spatial variation of the size and shape of the PSF, \texttt{ShOpt} first produces a rough estimate by interpolating the analytic fit parameters $\left[s, g_1, g_2 \right]$ across the field of view using a polynomial of order $n$, where $n$ is supplied by the user. For an order-$n$ polynomial $p(u,v))$, the cost function takes the form
\begin{equation}
    \mathrm{cost}_p(\left[ a_0, \hdots ,a_{(n + 1)(n + 2)//2} \right]) = \left( \left[ \sum_{(i,j), i + j  \leq n} a_{ij}u^i v^j \right]  - p(u,v)^* \right)^2\label{eq:polycost}
\end{equation}
where $*$ denotes the ground truth obtained by pixel basis fits. This gives us different polynomials in $u$ and $v$ for $\left[s, g_1, g_2\right]$. There is a tunable parameter for the stopping gradient that leaves the tradeoff between speed and accuracy of the polynomial interpolation to the end user. As elsewhere, the minimum of the cost function is found with LBFGS and the \texttt{Optim.jl} library.

Subsequently, the pixel-grid model is interpolated across the field of view. By definition, each pixel in pixel-grid models is a basis element, so the natural thing to do is to give each pixel in an $n \times n$ vignette its own polynomial. However, we found that this approach for solving for the polynomial coefficients produced systematically biased models, possibly due to the conditioning of the pixel intensity values. Thus, we opt for an alternative approach. To solve for the coefficients of the polynomial in the pixel basis fit, we may specify an matrix $m \times n$ matrix, where $m$ is the number of stars we are interpolating over and $n$ is still order of the polynomial. We denote this design matrix by $A$. If the vector $x$ represents the coefficients of the polynomials and the vector $b$ represents the intensity values, then we may use the known least squares solution to the matrix equation
\begin{equation}
    Ax = b. \label{eq:matrix}
\end{equation} 

Not all stars are good exemplars of the PSF due to things like saturation, color effects, and noise. To combat this, \texttt{ShOpt} does its polynomial interpolation over several iterations according to Algorithm \ref{alg:star_filtering}.
\begin{algorithm}[H]
\caption{Iterative Polynomial Interpolation and Star Filtering Process}
\label{alg:star_filtering}
\begin{algorithmic}[2]
\State NumIterations $\gets$ [define number of iterations]

\For{$i \gets 1$ \textbf{to} NumIterations}
    \State PSFModel $\gets$ PolynomialInterpolate(TrainingStars)
    \State RenderedStars $\gets$ RenderPSF(PSFModel)
    \State MSE $\gets$ ComputeMSE(TrainingStars, RenderedStars) \Comment{MSE: Mean Squared Error}
    \State WorstStars $\gets$ GetWorstPerformingStars(MSE, 10\% or N sigma)
    \State TrainingStars $\gets$ TrainingStars - WorstStars
\EndFor

\State FinalPSFModel $\gets$ PolynomialInterpolate(TrainingStars) 

\State \Return TrainingStars, FinalPSFModel
\end{algorithmic}
\end{algorithm}

The number of iterations to refine the polynomial interpolation is specified by the user. After each iteration, the predicted PSF is rendered at each star location and the mean squared error (Equation \ref{eq:MSE}) is computed using the training stars. The worst $10\%$ of training stars are filtered out. 

\begin{equation}
    \text{Mean Squared Error} = \frac{1}{N_\mathrm{pix}} \sum _{(i,j)}^{N_\mathrm{pix}} (v_{i,j} - p_{i,j})^2
    \label{eq:MSE}
\end{equation}

Alternatively, we provide a mode that removes $N \times \sigma$ outliers from the interpolation. Polynomial interpolation for high-degree polynomials can be the most expensive part of the whole PSF fitting procedure. For this reason, \texttt{ShOpt} is strict about filtering outliers in polynomial interpretation. Training stars are excluded based on the value of $s$ that was found during the analytic-profile interpolation step, which eliminates the need for additional iterations to clean the training data. Additionally, we make the conscious decision not to continue filtering our training stars until we reach a given $n \sigma$ confidence.  The JWST PSFs contain lots of background noise between the wings of the stars we are trying to model. Consequently, an excessive number of iterations may be required to reach the desired level of confidence. 

\section{Data pre-processing and outlier rejection} \label{sec:shDH}
In \texttt{ShOpt}'s outlier rejection process, three distinct phases are employed: (1) SNR-based filtering on input stars; (2) a Gaussian fit size filter on the input stars; and (3) a separate Gaussian fit size filter on pixel grid models. Following these phases, the process advances to the iterative refinement for polynomial fitting, as detailed in Section \ref{sec:FOV}.

Before doing anything else, \texttt{ShOpt} filters out bad stars on the basis of SNR; the default SNR metric is the SExtractor \texttt{SNR\_WIN} parameter. By default, \texttt{ShOpt} filters out the noisiest 33\% of entries, a percentage determined through experimentation. Users can adjust this threshold to fit their specific data sets. 

The remaining vignettes are then fit with a multivariate Gaussian, as described in Section \ref{sec:psfHat}. We filter again based on the object size $s$. A very small or very large $s$ probably corresponds to objects that are not point sources. By default, vignettes with $s < 0.075$ or $s > 1$ are filtered out. This ensures further computation time is not wasted on bad data. We apply the same size filtering after fitting Gaussians to the pixel grid models. This prevents poor pixel grid models from being used in our polynomial interpolation, as mentioned in Section \ref{sec:FOV}.

\texttt{ShOpt} also offers several methods for efficiently cleaning useful data without discarding it. Given that SExtractor sets the flux of interloping sources to a sentinel value of $-10^{32}$, \texttt{ShOpt} sets pixels less than $-1000$ to \texttt{NaN}. Before doing any analytic or pixel grid fits, we also smooth the image according to the kernel introduced in Equation \ref{eq:lk} to avoid any hot pixels in the vignettes. Finally, we recommend that users select \texttt{true} in the yaml configuration for the setting \texttt{sum\_pixel\_grid\_and\_inputs\_to\_unity}. This enforces intensity to sum to unity in each of our model vignettes and their pixel grid fits.

\section{Runtime Analysis} \label{sec:runtime}
In this section, we compare the algorithmic complexity and convergence properties of \texttt{ShOpt}, PIFF and PSFEx, examining the analytic profile fits, pixel grid fits, and polynomial interpolation steps separately. Runtime analysis of the three PSF fitters allow us to argue for the algorithmic choices implemented in \texttt{ShOpt} without needing to factor in programming language or computing power. The results will serve as predictions for the speed tests of Section \ref{subsec:speedTests}.

\subsection{Analytic Profile Fit Runtime} \label{subsec:analyticprofileruntime}
Let $n$ denote the number of pixels on one side of a square vignette, so that there are $n^2$ total pixels. In our optimization scheme, we compute the loss between every pixel in the vignette and the analytic profile prediction an average of $I$ times. Our nominal runtime is thus $\mathcal{O}(n^2 \times I)$. 

PIFF solves for $\left[s, g_1, g_2\right]$ using iterative nonlinear least squares minimization and takes advantage of SciPy's \citep{virtanen2020scipy} optimization package \citep{Jarvis_2020}. By default, SciPy's least squares minimizer uses the trust region reflective (TRF) algorithm, which relies on first order information via the Jacobian. On the other hand, \texttt{ShOpt} uses LBFGS, which uses super-linear approximations to calculate the direction of improvement. Since each iteration of LBFGS should take us closer to the solution than conjugate gradient descent, we should expect LBFGS to take fewer steps on average to compute. The LBFGS algorithm and the reparameterization step introduced in Section \ref{subsec:analytic} allows us to argue that $I_{ShOpt} < I_{PIFF}$. While we refrain from claiming that \texttt{ShOpt} will always converge to the correct $\left[s, g_1, g_2\right]$ faster than PIFF, we have designed \texttt{ShOpt} with the hopes that using a more memory expensive technique to compute its update steps while keeping the solution within a constraint gives us better convergence. 

\subsection{Pixel Grid Fit Runtime}
\subsubsection{PCA Runtime}
Principal component analysis relies on computing the singular value decomposition of the covariance matrix of a given data set. Singular value decomposition is $\mathcal{O}(n^3)$ for an $n \times n$ matrix, but the fit is typically much cheaper to compute for an approximation using the first $k$ principal components. While this does not lower the big-O complexity to the $\chi^2$ minimization pixel-grid fitting implemented in PIFF and PSFEx \citep{2011ASPC, Jarvis_2020}, we do not encounter any noticeable speed bottle necks at this step.\footnote{A $\chi^2$ minimization pixel-grid fit will be incorporated in a future \texttt{ShOpt} release.} We will explore this more in in \ref{subsec:speedTests}.

\subsubsection{Autoencoder Runtime}
As in \ref{subsec:analyticprofileruntime}, we may observe that the complexity is $\mathcal{O}(n^2 \times I)$ because the loss is computed an average of $I$ times over $n^2$ pixels. The number of parameters between layers as a function of neurons $m$ in layer $i$ in our network is given by 
\begin{equation}
N_{params} = \left(m_{i-1} \times m_{i}\right) + m_i
    \label{eq:paramtercount}        
\end{equation}
where the first term corresponds to the number of weights and the second term corresponds to the biases. In our network the number of nodes for the input and output is set by a flattened version of the input image. Therefore, the number of parameters to learn grows as $\left[ \left(n^2 \times 128\right) + 128 \right] + \left[ \left(128 \times n^2\right) + n^2 \right] $. On the other hand, PIFF and PSFEx employ a form of $\chi^2$ minimization, wherein each pixel in the image is a learnable parameter; for an $(n, n)$ image there are only $n^2$ learnable parameters and $n^2$ pixels computed in the loss function. Even though the loss function in \texttt{ShOpt} is also computed over $n^2$ pixels, there are much more than $n^2$ learnable parameters, so we expect the autoencoder to take more iterations to converge on average than PIFF and PSFEx. For this reason, \texttt{autoencoder} is not the default mode for pixel grid fits in \texttt{shopt}. It should be reserved for cases that demand precision and where the images are small enough to be learned efficiently so the transfer learning effect is more pronounced. 

\subsection{Polynomial Interpolation Runtime}
Almost all PSF fitting software makes use of polynomial interpolation to model the variation of the PSF across the camera's field of view. Any performance gains in this area are primarily derived from the efficiency of the polynomial fitting process to the data. In PSFEx, these fits are implemented using PCA, which \cite{2011ASPC} argues requires the lowest number of basis vectors to approximate an image if the data is sufficiently well-behaved. 

In \texttt{ShOpt}, we use the design matrix and the known least squares solution to the matrix equation $Ax=b$ to fit each pixel with an independent polynomial. This process is inherently parallelizable because the polynomial found for one pixel is independent of the polynomial found for any other pixel. \texttt{ShOpt} will automatically use as many threads are made available to it. The idea to break these operations into computation blocks is inspired by Tiny Machine Learning problems described in \cite{sabot2023mema}  and the CAKE algorithm outlined in \cite{9910104}. Currently, the number of threads is not configurable because it has to be specified before the program is run. On UNIX, we ran \texttt{export JULIA\_NUM\_THREADS=auto} before running the program, on Windows you can similarly run \texttt{set JULIA\_NUM\_THREADS=auto}.
\begin{figure}[!htb]
    \centering
    \includegraphics[width=0.4\textwidth]{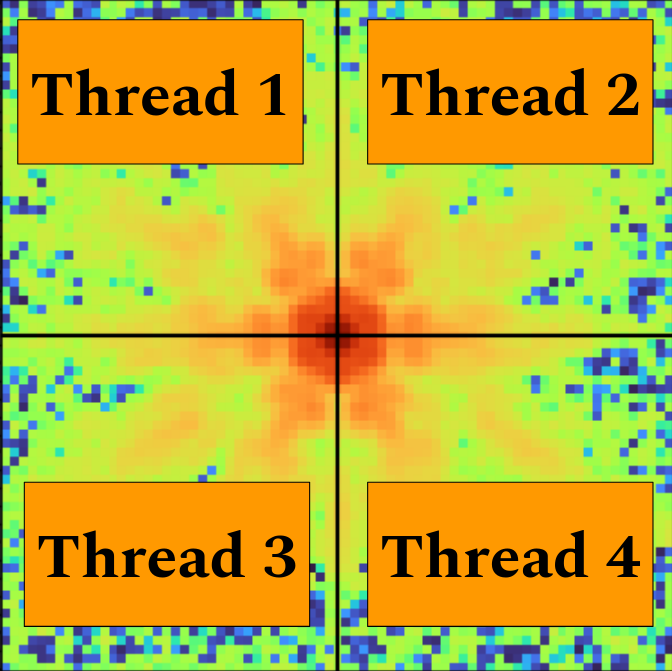}
    \caption{Pictoral representation of multi-threading in \texttt{ShOpt}. In this example, all of the pixels in the upper left of the pixel grid renderings are interpolated across the field of view in thread $1$, all of the pixels in the upper right of the pixel grid renderings are interpolated across the field of view in thread $2$, and so on. }
    \label{fig:mt}
\end{figure}
\subsection{Order of Operations} \label{sec:ooo}
We are also deliberate about the order of operations in \texttt{ShOpt}. The initial analytic profile fitting serves as a data cleaning method, which allows us to not waste compute time fitting pixel-grid models to outlier stars. A second round of analytic profile fits to the resulting PSF models further refines the data set, ensuring that the (computationally expensive) polynomial interpolation is applied only to the highest quality PSF models. By contrast, PIFF adopts an integrated approach where analytic and pixel grid fits are interwoven with iterations of polynomial interpolation. While this method is thorough, our approach is designed to prioritize computational expediency.

\section{Benchmarking, Data, and Analysis} \label{sec:datacleaning}
Our PSF analysis utilizes NIRCam imaging from the COSMOS-Web survey \citep{casey2023cosmosweb}, which we have chosen for its expansive coverage (more than three times the area of all other JWST surveys combined) in four bandpasses as well as the fact that many of its science cases require careful PSF characterization. This section presents an overview of the three COSMOS-Web data sets employed in our PSF benchmarking analysis---simulated single exposures, simulated observation mosaics, and real observation mosaics---followed by a description of the benchmarking methods we apply to each.

We employ simulated observations for PSF code unit testing because they provide a controlled environment with fully known input parameters. Unlike real observations, simulations allow for us to control for observational imperfections such as saturation, noise, or defective pixels; this allows for the establishment of a ground truth against which the accuracy of PSF model fits can be measured directly. Although real observations feature the true PSF and are integral to our study, their inherent uncertainties and the absence of a definitive PSF ground truth make them less suitable for initial code validation. Instead, the real data serves as a stress-test for the PSF fitting codes, ensuring they remain effective when confronted with the vagaries of real observations.

\begin{figure}[!htb]
    \centering
    \includegraphics[width=0.8\textwidth]{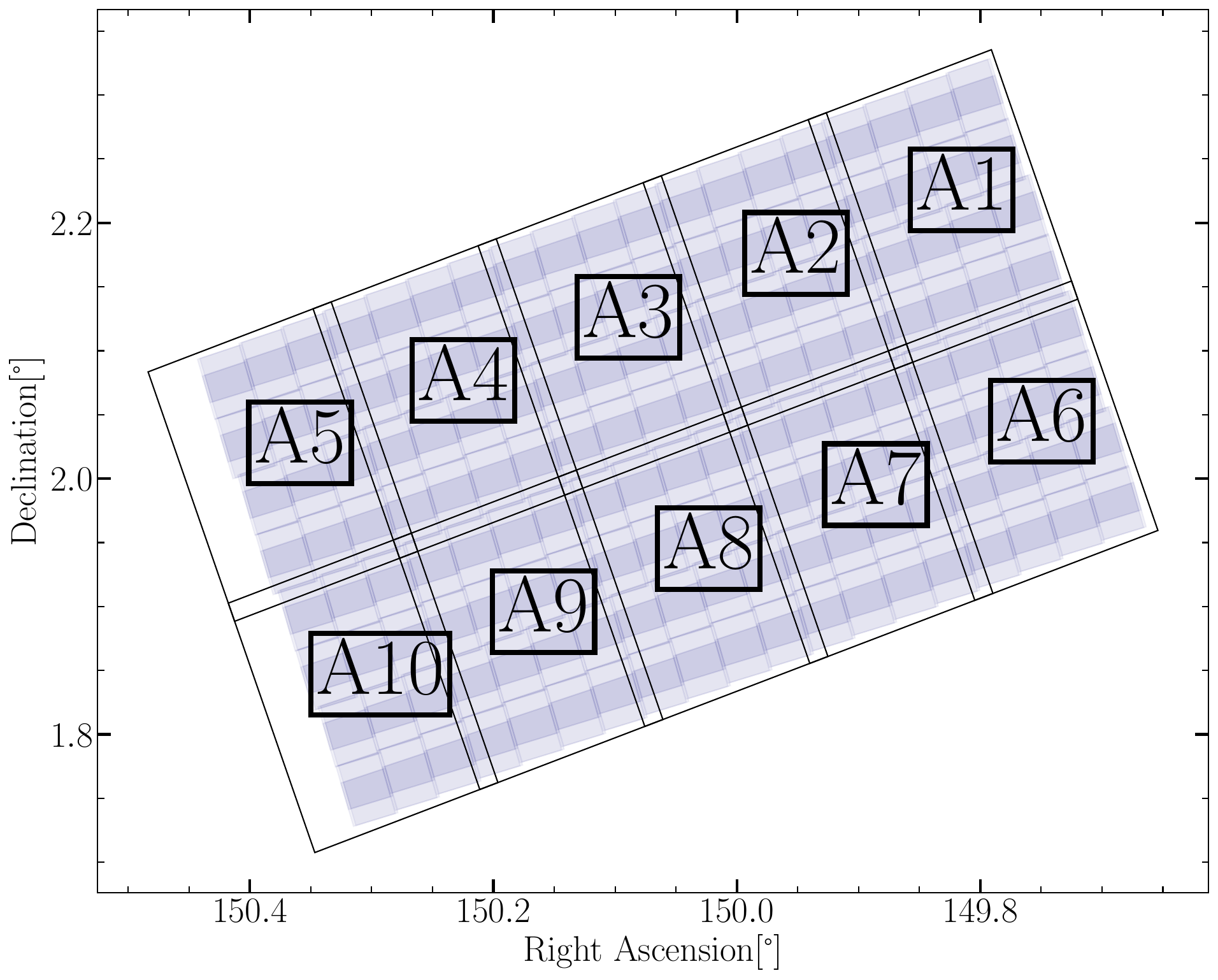}
    \caption{April data tiling scheme.}
    \label{fig:AprTiles}
\end{figure}

\begin{enumerate}
    \item \textbf{Simulated Data:} These simulations are based on the COSMOS 2020 data \citep{weaver2022}, and contain all the known galaxies and stars in this field. Galaxy fluxes in JWST filters were calculated from \texttt{Stardust} SEDs if they existed \citep{kokorev2021}; otherwise, fluxes were found by interpolating across existing photometry in other filters. Star fluxes were similarly modeled by interpolating available photometry. Galaxies with counterparts in the Zurich Structure \& Morphology catalog \citep{scarlata2007, sargent2007} were assigned the according morphological parameters (Sérsic indices, ellipticity, position angle, size), or from observed distributions if they did not.  
    
    COSMOS-Web-like NIRCam images were generated using the Multi-Instrument Ramp Generator \citep[MIRAGE;][]{Mirage}, which includes PSF modeling with WebbPSF \citep{perrin2014}, sky background, detector noise, dark current, and Poisson noise.  After the images were generated, the JWST calibration pipeline \citep{jwstcalibrationpipeline2023} was used to reduce the raw NIRCam data and create mosaics, with some modifications, like 1/f noise and subtraction of low-level background \cite[e.g.,][]{finkelstein2022b,bagley2022}. We employ two sets of simulated images:
    \begin{enumerate}
        \item \textbf{Simulated Single Exposures:} Stars are generated by convolving a MIRAGE WebbPSF model with an idealized point sources, so these images allow us to measure success directly by comparing the learned PSF to the input PSF. We employ images from stage 2 of the JWST calibration pipeline, so there are no effects from dithering or distortion (i.e., \texttt{tweakreg}). The main challenge in using this data is low star density, particularly for the short wavelength channel. With a training/validation split of $90\%$, this led to an average of $1$ validation star for each subarray in the F115W filter, $1$ validation star for the F150W  filter, $2.2$ stars for the F277W filter, and $2.3$ stars for the F444W filter. To gather enough data for meaningful summary statistics, we combine exposures from 156 different visits.

        \item \textbf{Simulated Mosaics:} Image mosaics are \texttt{i2d}-format data cubes built from single exposures that have passed through stage 3 of the JWST science calibration pipeline.  

        As such, the mosaics do reflect normal dithering and distortion effects. The simulated mosaics used in our study cover a contiguous area of 76 arcmin$^2$ at the full COSMOS-Web depth ($\sim$ 27th magnitude) and provide a reasonable 0.5 star per square arcminute.  
    \end{enumerate}
For an in-depth discussion of the simulation process, refer to Drakos et al. (in prep).
      
    \item \textbf{Real Mosaics:} The real data used in our analysis was taken in April 2023 and includes visits $77 - 152$, covering $0.28$ deg$^2$ in the bottom right of our allocated area of the COSMOS field \citep{casey2023cosmosweb}, JWST program ID 1727\footnote{Available from MAST at STScI, \url{http://mast.stsci.edu}}. Three of the planned visits were skipped, so the data includes a total of $72$ visits. As with the simulated mosaics, we use the \texttt{i2d}-format data cubes produced by stage 3 of the JWST science calibration pipeline. We restrict ourselves to an approximately $0.11$ deg$^2$ area of the full field of view, corresponding to tiles $\{A1, A5, A6, A10\}$ in Figure \ref{fig:AprTiles}; we chose these particular tiles to test the relative performance of the PSF fitters where we expect astrometric distortions to be the most severe. 
\end{enumerate}

\begin{figure}[!htb]
    \centering
    \includegraphics[width=0.8\linewidth]{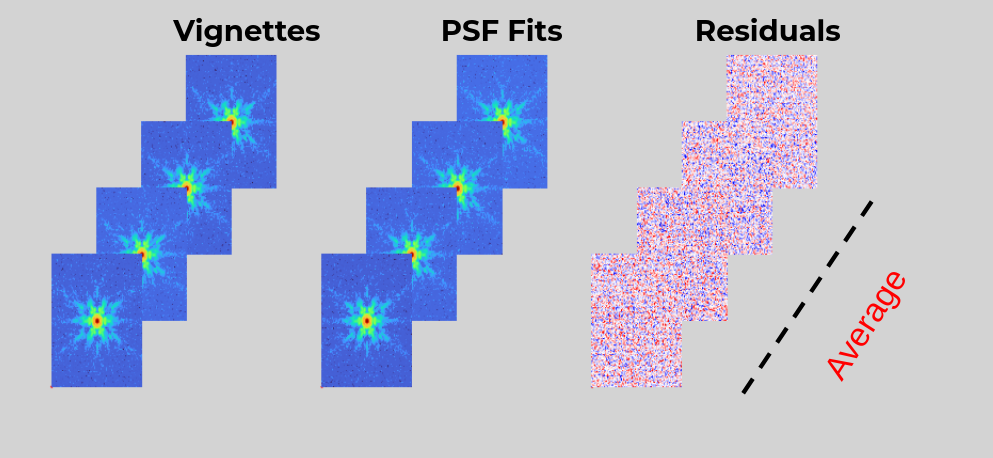}
    \caption{Schema illustrating the computation of summary statistics for PSF model assessment. Individual star-PSF model pairs are compared to produce residual images, which are then averaged into a mean residual image for the ensemble.}
    \label{fig:SummaryStatistics}
\end{figure}

Our benchmarking procedure has four steps: 
\begin{enumerate}
    \item Run Source Extractor (SExtractor) on our images to generate star catalogs \citep{1996A&AS}. For the simulated data, stars are identified by matching the SExtractor catalogs the input point source catalogs. For real data, star catalogs are created using cuts on the stellar locus, an example of which is shown in Figure \ref{fig:sizemap}. The config file for SExtractor is given in the Appendix. Note that we run source extractor on the individual tiles and aggregate our results over all of them.

\begin{figure}[!htb]
    \centering
    \includegraphics[width=0.9\linewidth]{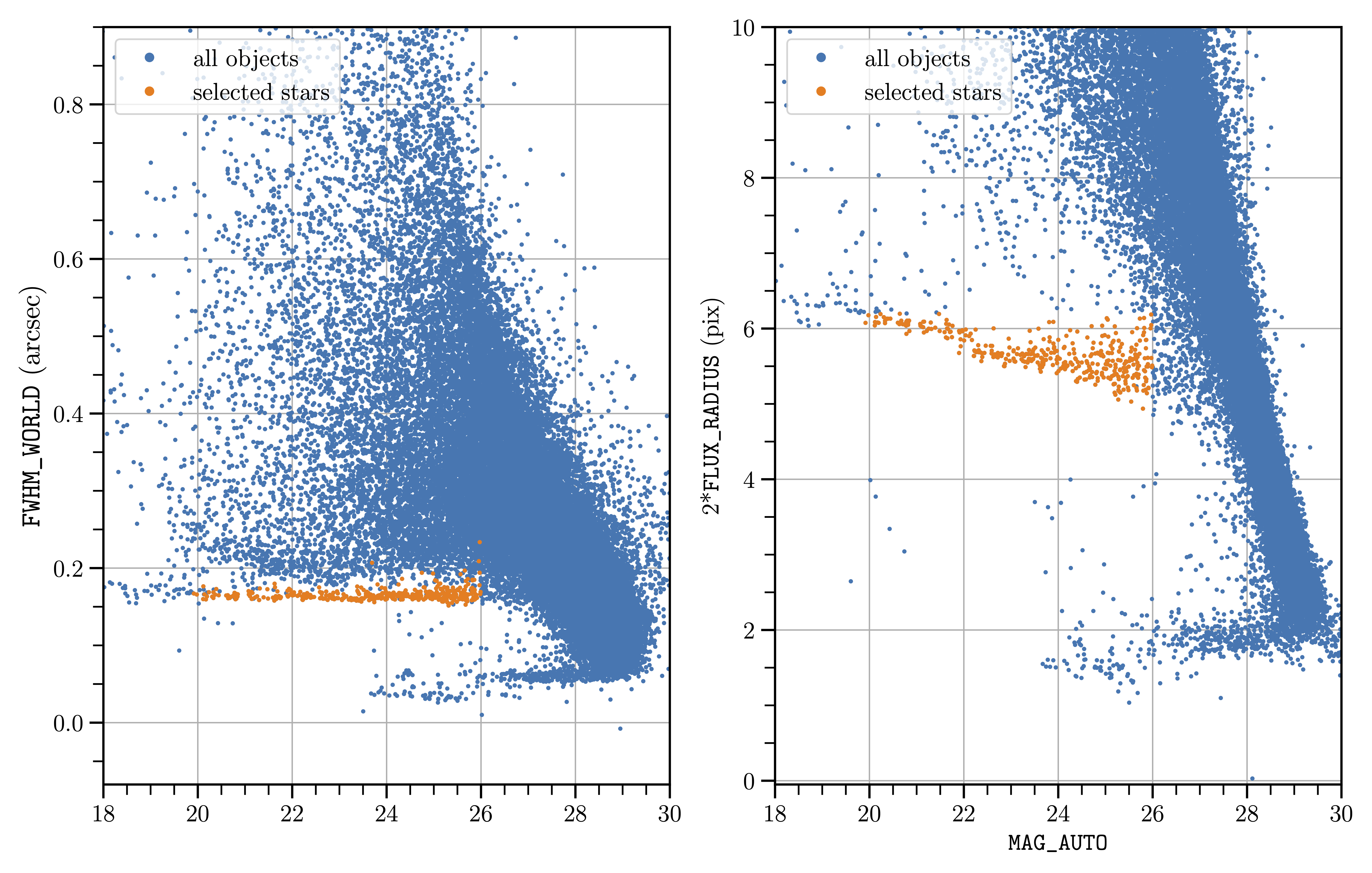}
    \caption{Size-magnitude diagram for objects detected in the A6 mosaic in F444W A6. Blue points represent all sources; orange points show stars selected by stellar locus parameter cuts.}
    \label{fig:sizemap}
\end{figure}
    
    \item The resulting star catalogs are segregated into training (90\%) and validation (10\%) catalogs. We are careful to filter out saturated stars in our catalogs by searching for sentinel values of $0$ in the \texttt{ERR} extension (for i2d-format mosaics) or ${1, 2}$ in the \texttt{DQ} extension (for single exposures). Vignettes that contain any pixels set to a sentinel value are removed from the catalogs.
    
    \item We use the training catalogs to get empirical PSF models for each fitter. We run \texttt{ShOpt} \citep{berman2023shoptjl}, PIFF \citep{Jarvis_2020}, and PSFEx \citep{2011ASPC}. 
    
    \item We then use the reserved validation star catalogs to calculate the summary statistics of Equations \ref{eq:Reducedchisq}--\ref{eq:MAE}. Figure \ref{fig:SummaryStatistics} illustrates this process: we calculate the residuals between the star vignettes (or MIRAGE cutouts for simulations) and renderings of the PSF models, then take the mean. To obtain single residual scores, we compute the mean and standard error of all pixels in residual images. 

\end{enumerate}

\noindent
The code for creating star catalogs, calculating statistics, and creating the associated PSF diagnostic 
figures can be found on GitHub\footnote{\href{https://github.com/mcclearyj/cweb_psf}{https://github.com/mcclearyj/cweb\_psf}}.

\section{Results} \label{sec:results}

In this section, we report the outcomes of the PSF benchmarking analysis detailed in Section \ref{sec:datacleaning}. The comprehensive PSF model fidelity analysis is presented in Section \ref{subsec:fidelityresults}, and the assessment of the computational efficiency of the different PSF fitters is presented in Section \ref{subsec:speedTests}.

We note that \texttt{ShOpt} was run in \texttt{smoothing} mode throughout. We found that \texttt{PCA} mode and \texttt{Autoencoder} mode added additional complexity to the PSF fitting process that resulted in poorer fits. Appendix \ref{configs} contains representative configuration files supplied to each PSF fitter.

\subsection{Non-parametric Model Fidelity}\label{subsec:fidelityresults}

As described above, we evaluate the relative performance of PSF models produced by PIFF, PSFEx, and \texttt{ShOpt} using mean and median reduced $\chi^2$ residuals, the MRE, and the MAE (cf. Equations \ref{eq:Reducedchisq}--\ref{eq:MAE}). These statistics are computed using the reserved validation stars, which number from 35 for the simulated F277W and mosaics to 156 for the F150W real mosaics (Table \ref{tab:num_val_stars}). Statistics and diagnostic plots are based on PSF vignette sizes of $(75, 75)$ pixels, which encloses the majority of the relevant star and PSF light profiles without including excessive sky background or a large number of interloping objects. 

\begin{table}
\centering
\begin{tabular}{|l|l|c|}
\hline
\textbf{Data Type} & \textbf{Filter}                                 & \textbf{Validation Stars} \\ \hline
\multirow{4}{*}{Simulated single exposures} & F115W             & 3373                    \\ \cline{2-3} 
                                                  & F150W             & 7186                    \\ \cline{2-3} 
                                                  & F277W           & 2573                   \\ \cline{2-3} 
                                                  & F444W             & 740                   \\ \hline
\multirow{4}{*}{Simulated mosaics}         & F115W         & 37                   \\ \cline{2-3} 
                                                  & F150W            & 79                   \\ \cline{2-3} 
                                                  & F277W          & 35                    \\ \cline{2-3} 
                                                  & F444W          & 35                   \\ \hline
\multirow{4}{*}{Real mosaics}          & F115W     & 155                   \\ \cline{2-3} 
                                                  & F150W      & 156                   \\ \cline{2-3} 
                                                  & F277W     & 148                   \\ \cline{2-3} 
                                                  & F444W     & 136                   \\ \hline
\end{tabular}
\caption{Number of reserved validation stars for the three imaging data sets under consideration in each of the four COSMOS-Web NIRCam bandpasses.}
\label{tab:num_val_stars}
\end{table}

\subsubsection{Simulated single exposures}

\begin{figure}[htpb]
    \centering
    \includegraphics[width=\linewidth]{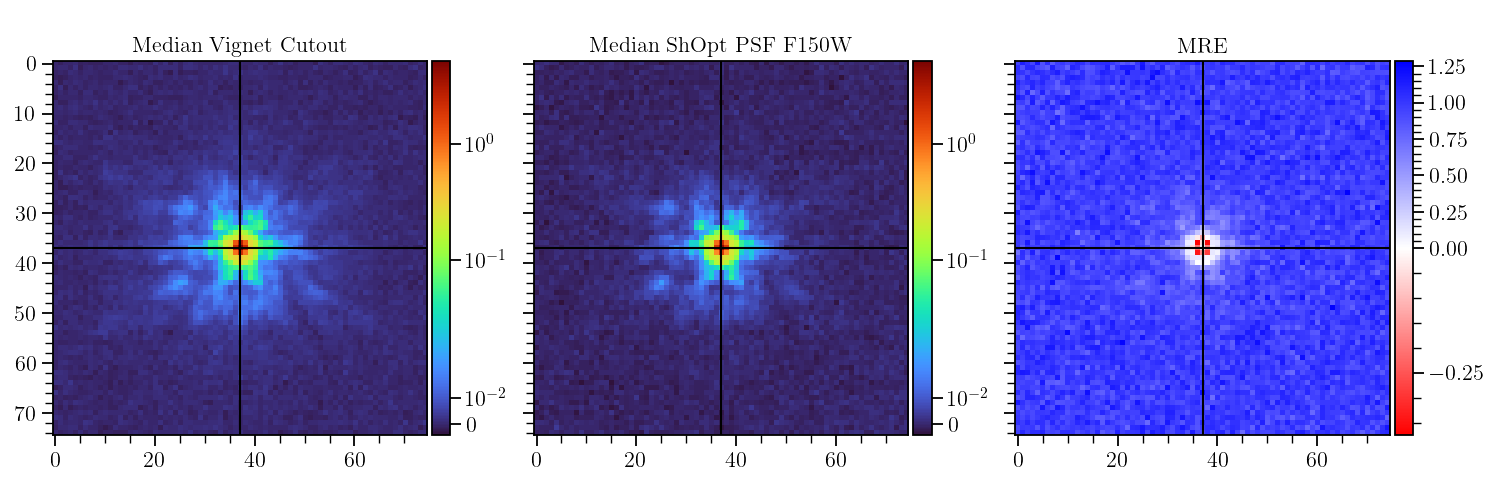}\label{fig:f444wShoptMirage}\\
    %(b) F444W Input Mirage PSF vs PSFEx model comparison
    \includegraphics[width=\linewidth]{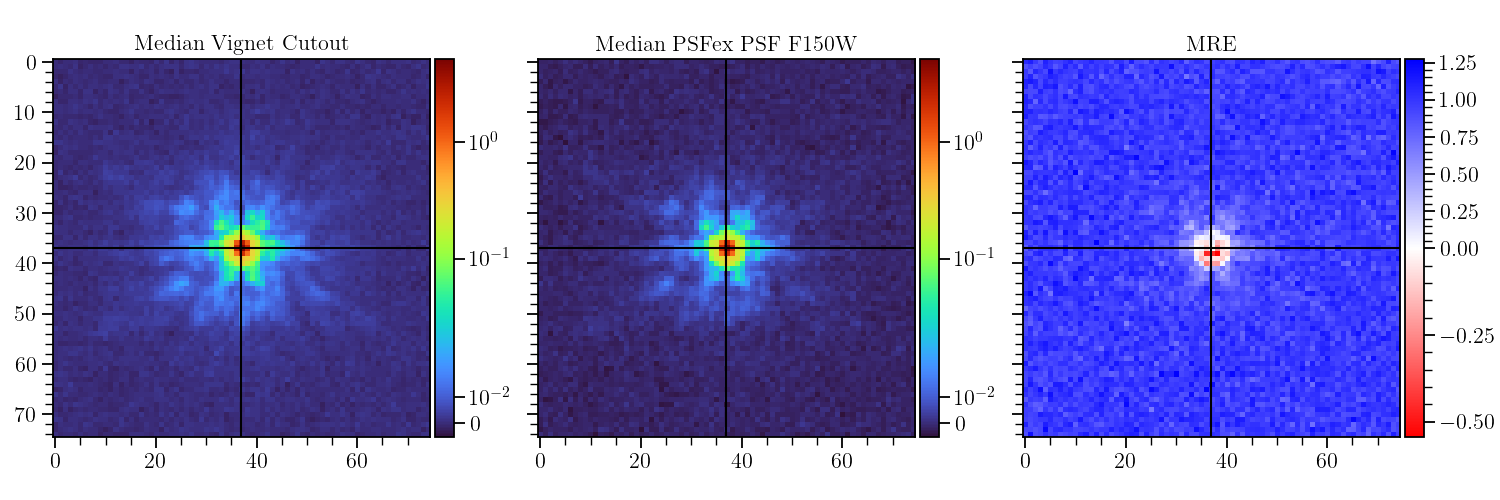}
\caption{Mean relative error between MIRAGE input point source images and PSF models for the F277W filter. Left panels show the median MIRAGE cutout; center panels show the median PSF cutout. Right panels show the average relative error between the MIRAGE and PSF cutouts. Top two rows are \texttt{ShOpt}, bottom two rows are PSFEx. Color bars for left and center panels show pixel intensity values in units of MJy/sr. Color bars in the right panels show the (dimensionless) relative error. The MREs defined in Equation \ref{eq:MRE} are displayed in the titles of the mean residual images.}\label{fig:F444MiragePSFEx}
\end{figure}

\begin{figure}
    \centering
    \includegraphics[width=0.65\linewidth]{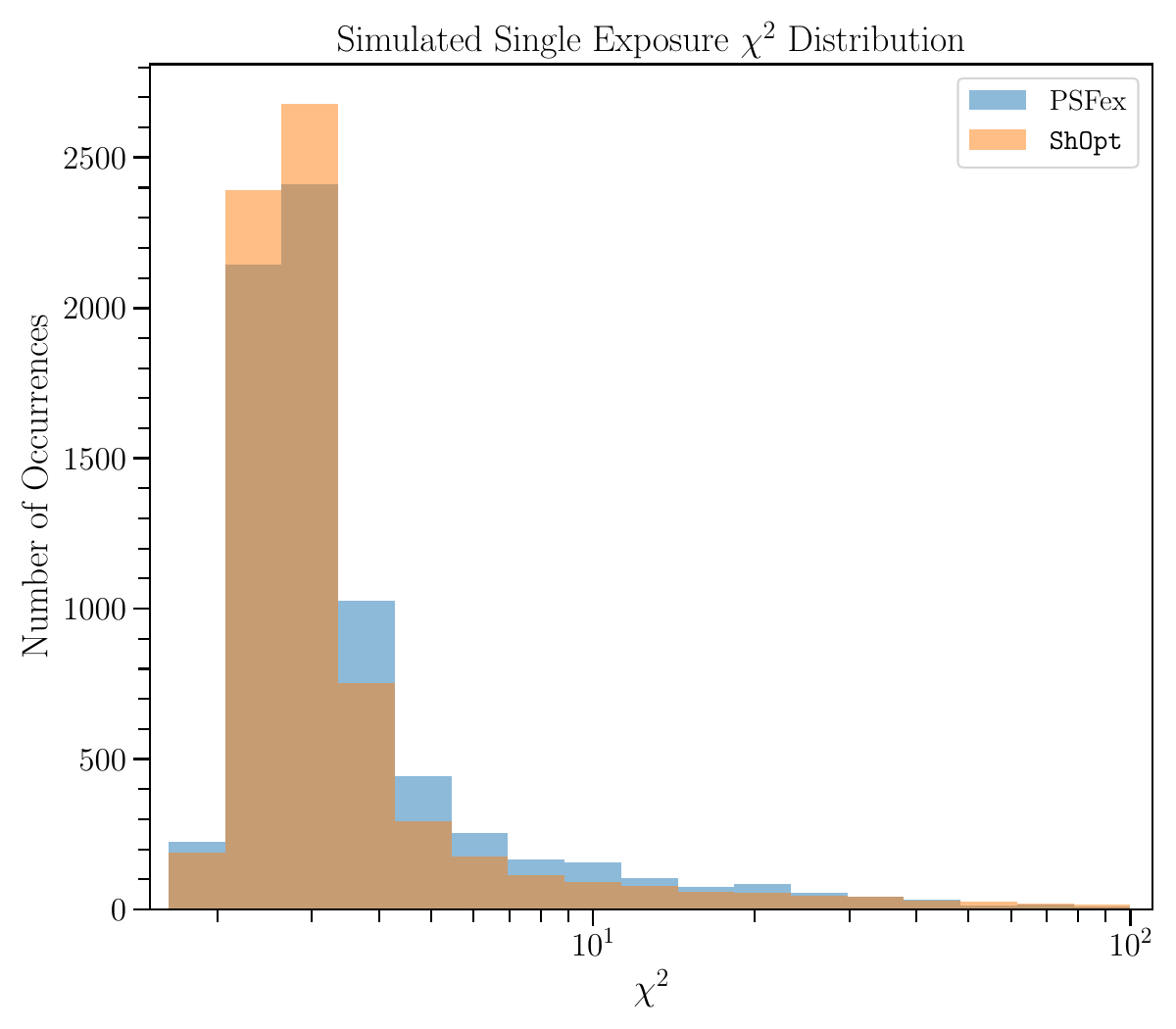}
    \caption{Distribution of $\chi^2_\nu$ for simulated single exposure images. Not shown are $42$ PSFEx $\chi^2_\nu$ values greater than $100$ and $100$ \texttt{ShOpt} $\chi^2_\nu$ value greater than $100$.}
    \label{fig:sse_chisq_distribution}
\end{figure}

While the simulated single exposures yield ample training data in aggregate, training data on individual detectors from individual visits is sparse. Despite this sparsity of training data, Figures \ref{fig:F444MiragePSFEx}, \ref{fig:f115wShoptMirage}, and \ref{fig:f277wPsfexMirage} show that both PSFEx and \texttt{ShOpt} produce reasonable models of the PSF. While both PSF fitters seem to slightly underfit the center of the PSF, they are otherwise able to model the finer details of the PSF.
The distribution of reduced $\chi^2$ shown in Figure \ref{fig:sse_chisq_distribution} illustrates that \texttt{ShOpt} produces a fit that is just as strong, if not stronger than PSFex.

\subsubsection{Simulated mosaics}

\begin{figure}[htpb]
    \centering
        \includegraphics[width=\linewidth]{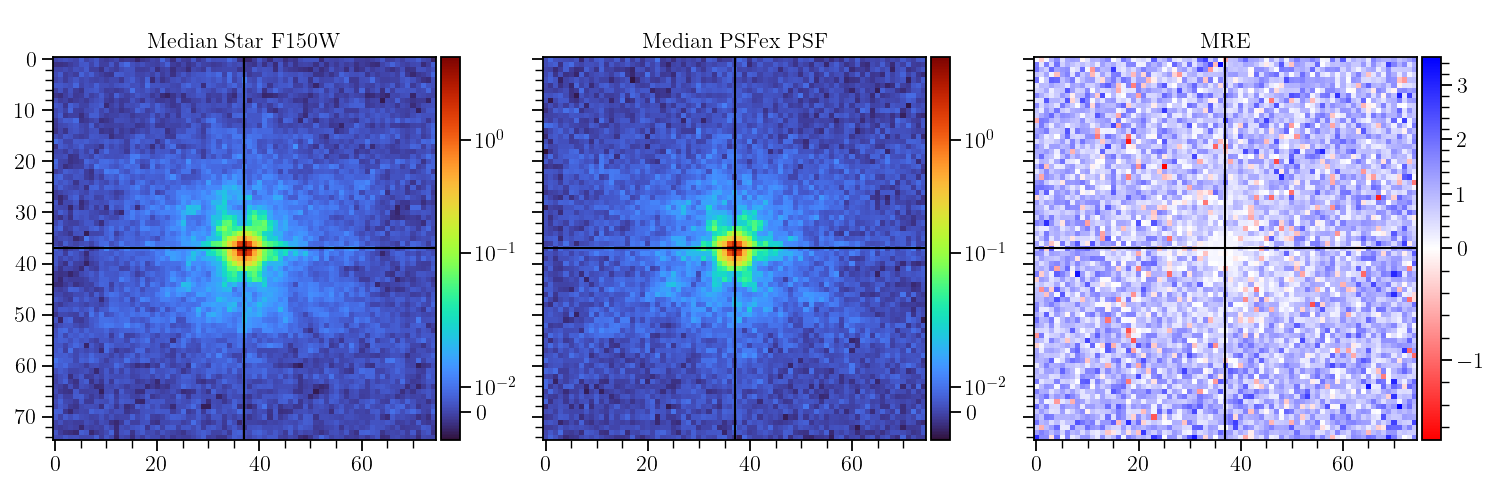}\\

    \includegraphics[width=\linewidth]{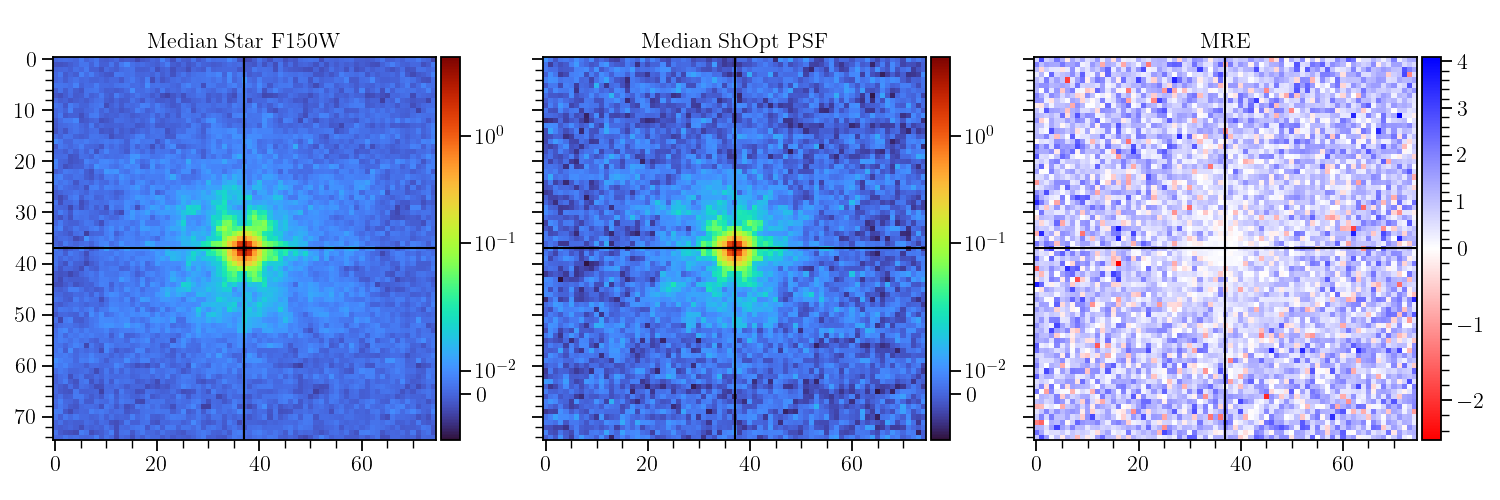}\\
    
         \includegraphics[width=\linewidth]{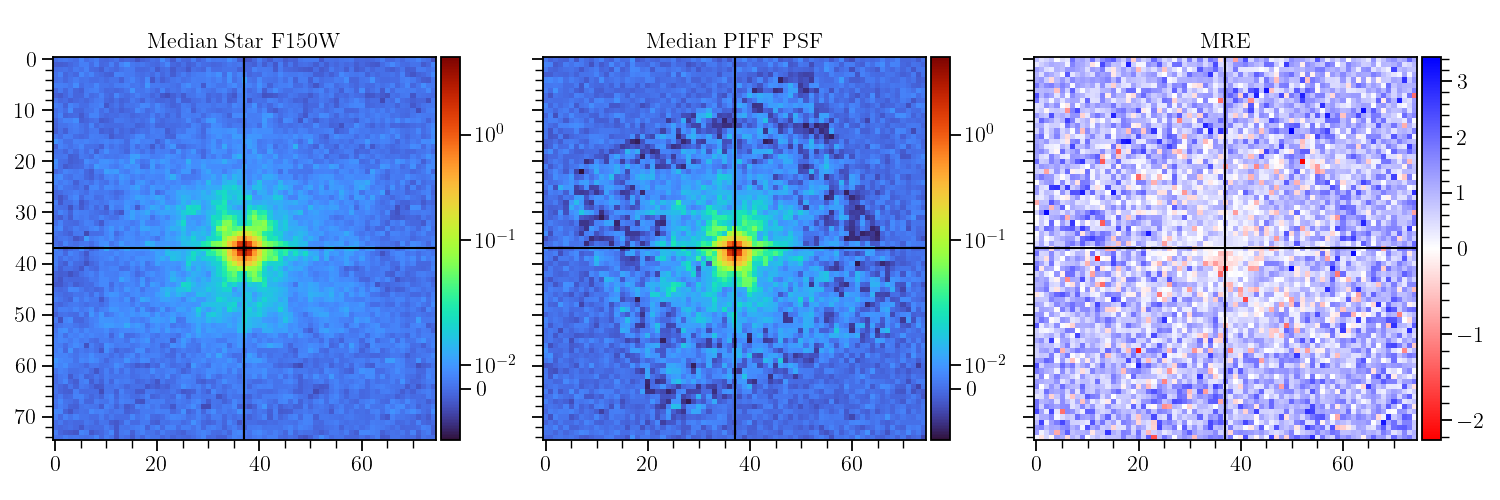}
        
    \caption{Evaluation of mean relative error between stars and PSF models for simulated mosaics in the F150W bandpass. Left panels show the median of the vignettes. Center panels show the median PSF cutout. Right panels show the average relative error between the vignette cutouts and the PSF cutouts. Top is PSFEx, middle is \texttt{ShOpt}, bottom is PIFF. }
    \label{fig:f150wCOSMOSSIMS}
\end{figure}

The simulated data mosaics have a higher density of training stars, as evinced by significantly better fits for PSFEx and \texttt{ShOpt} than the simulated single exposures. We also supply PIFF fits for these images. Figures \ref{fig:f150wCOSMOSSIMS} and \ref{fig:f115wCOSMOSSIMS}--\ref{fig:f444wCOSMOSSIMS} do not show the same over-concentration visible in Figures \ref{fig:F444MiragePSFEx}, \ref{fig:f115wShoptMirage}, and \ref{fig:f277wPsfexMirage}.

\begin{table}[hbtp]
\centering
\begin{tabular}{cccccc}
\toprule
Filter & PSF Fitter & MAE & MRE & $\overline{\chi^2_\nu}$ & Median $\chi^2_\nu$ \\
\midrule
\multirow{3}{*}{F115W} 
                       & ShOpt & 3.41$_{-1.79}^{+5.17}$ & 0.80$\pm$1.18 &  8.89$_{-1.16}^{+25.85}$ & 1.78 \\
                       & PIFF  & 3.11$_{-1.72}^{+4.78}$ & 0.87$\pm$1.20 & 12.76 $_{-1.26}^{+49.40}$ & 1.82 \\
                       & PSFEx & 2.67$_{-1.54}^{+3.90}$ & 0.81$\pm$0.94 & 3.81$_{-1.26}^{+10.90}$ & 1.88 \\
\midrule
\multirow{3}{*}{F150W} 
                       & ShOpt & 3.22$_{-2.08}^{+4.39}$ & 0.88$\pm$0.79 & 5.82$_{-0.93}^{+3.73}$ & 1.42 \\
                       & PIFF  & 2.87$_{-2.01}^{+3.83}$ & 0.88$\pm$0.72 & 5.93$_{-0.94}^{+4.90}$ & 1.39 \\
                       & PSFEx & 2.70$_{-1.90}^{+3.54}$ & 0.87$\pm$0.64 & 5.40$_{-0.92}^{+2.07}$ & 1.30 \\
\midrule
\multirow{3}{*}{F277W} 
                       & ShOpt & 3.60$_{-1.83}^{+5.49}$ & 0.77$\pm$1.26 & 15.21$_{-1.25}^{+20.29}$ & 1.98 \\
                       & PIFF  & 2.75$_{-1.68}^{+4.01}$ & 0.85$\pm$1.06 & 36.97$_{-1.38}^{+48.55}$ & 2.09 \\
                       & PSFEx & 2.67$_{-1.58}^{+3.92}$ & 0.83$\pm$0.96 & 16.99$_{-1.12}^{+37.48}$ & 1.88 \\
\midrule
\multirow{3}{*}{F444W} 
                       & ShOpt & 3.08$_{-1.86}^{+4.43}$ & 0.80$\pm$1.07 & 6.11$_{-1.03}^{+6.14}$ & 1.70 \\
                       & PIFF  & 2.84$_{-1.76}^{+4.10}$ & 0.89$\pm$1.05 & 13.49$_{-1.12}^{+15.76}$ & 1.71 \\
                       & PSFEx & 2.76$_{-1.70}^{+3.99}$ & 0.86$\pm$1.00 &  2.01$_{-1.03}^{+3.31}$ & 1.46 \\
\bottomrule
\end{tabular}
\caption{Simulated mosaic summary statistics. The MAE and $\overline{\chi^2_\nu}$ statistics are reported with $10\%$ and $90\%$ errors.}
\label{tab:extended_example_reformatted}
\end{table}

\begin{figure}[htpb]
    \centering
    \includegraphics[width=0.7\linewidth]{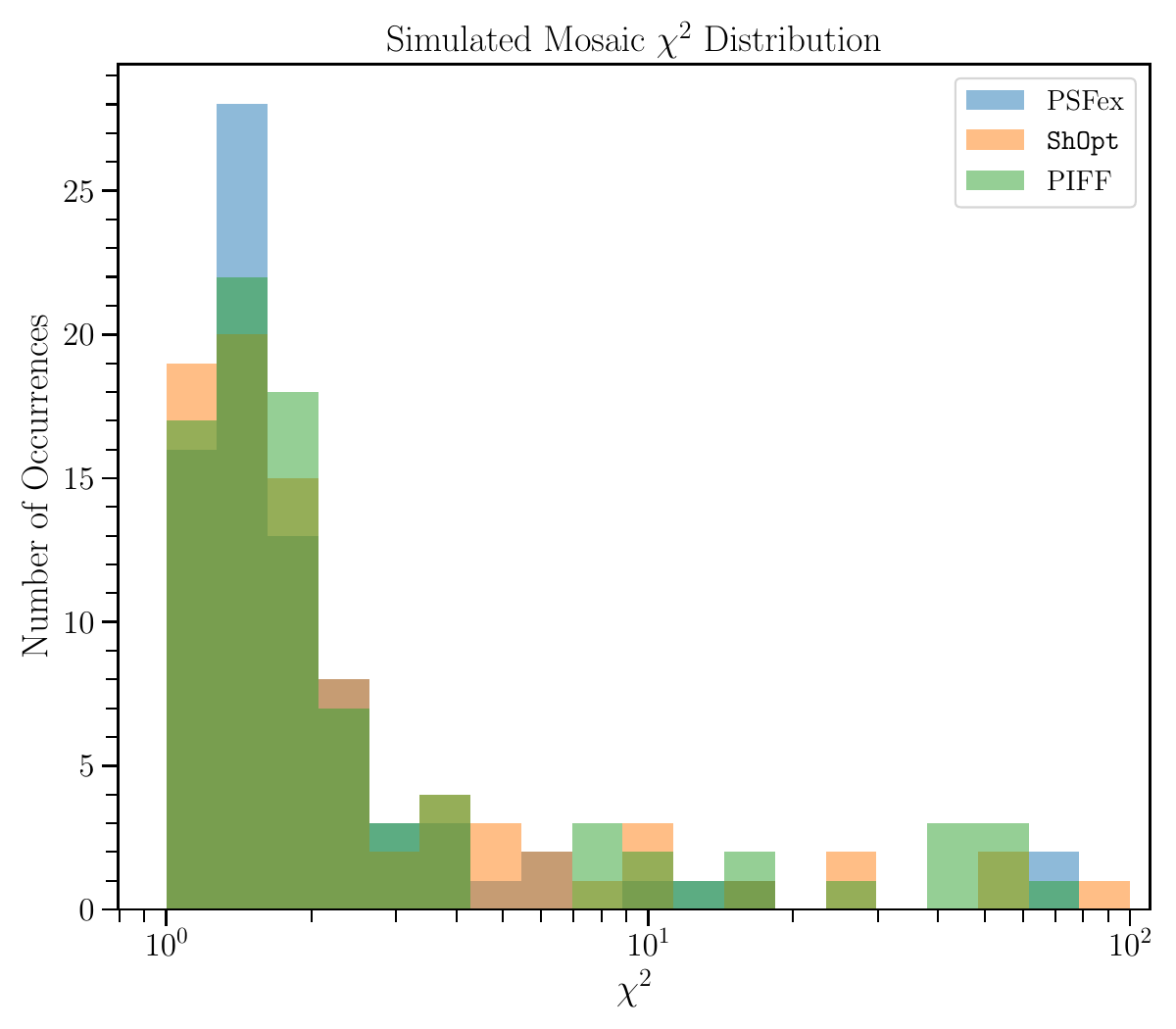}
    \caption{Distribution of $\chi^2_\nu$ for simulated data mosaics. PSFEx values are shown in blue, \texttt{ShOpt} values are shown in orange, and PIFF values are shown in green. For clarity, the plot excludes two PSFEx $\chi^2_\nu$ values greater than $100$, one \texttt{ShOpt} $\chi^2_\nu$ value greater than $100$, and two PIFF $\chi^2$ values greater than $100$.}
    \label{fig:mosaic_chisq_distribution}
\end{figure} 

Figure \ref{fig:mosaic_chisq_distribution} suggests minimal statistical difference in performance among the PSF fitters, with the same heavy tailed distribution of $\chi^2_\nu$ for each. 

Table \ref{tab:extended_example_reformatted} shows MRE and MAE values are consistent with zero for all \texttt{ShOpt} and PSFEx models, suggesting minimal bias. While the PIFF model residuals are also consistent with zero, the $90\%$ bounds tend to be larger than \texttt{ShOpt} and PSFEx, indicating an increased incidence of catastrophic fits. We also point out that the $10\%$ errors tend to be close the median error among each of the PSF fitters.

In general, the presence of outlier fits obscures a clear ranking based on MAE, MRE, and $\overline{\chi^2}$ alone. The median $\chi^2$ and its distribution, as shown in Figure \ref{fig:mosaic_chisq_distribution} and Table \ref{tab:extended_example_reformatted}, seem to be more indicative of performance, demonstrating no significant difference in the reliability of the PSF fitters for the simulated mosaics. 

\subsubsection{Real mosaics}

\begin{figure}[hbtp]
    \centering
        \includegraphics[width=\linewidth]{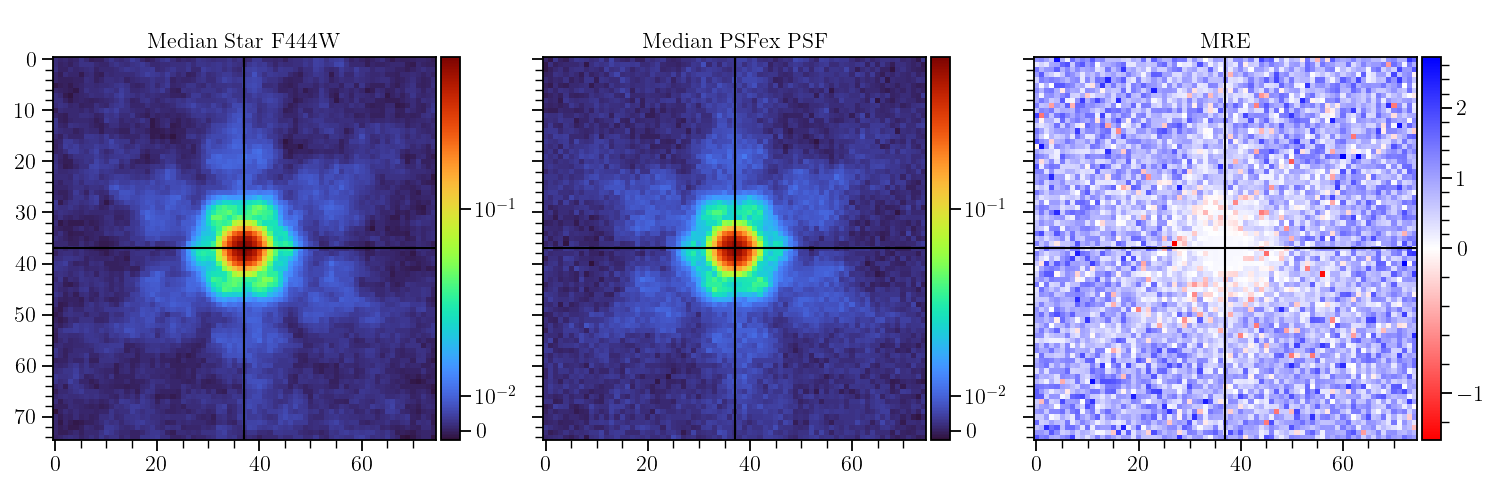}
        \includegraphics[width=\linewidth]{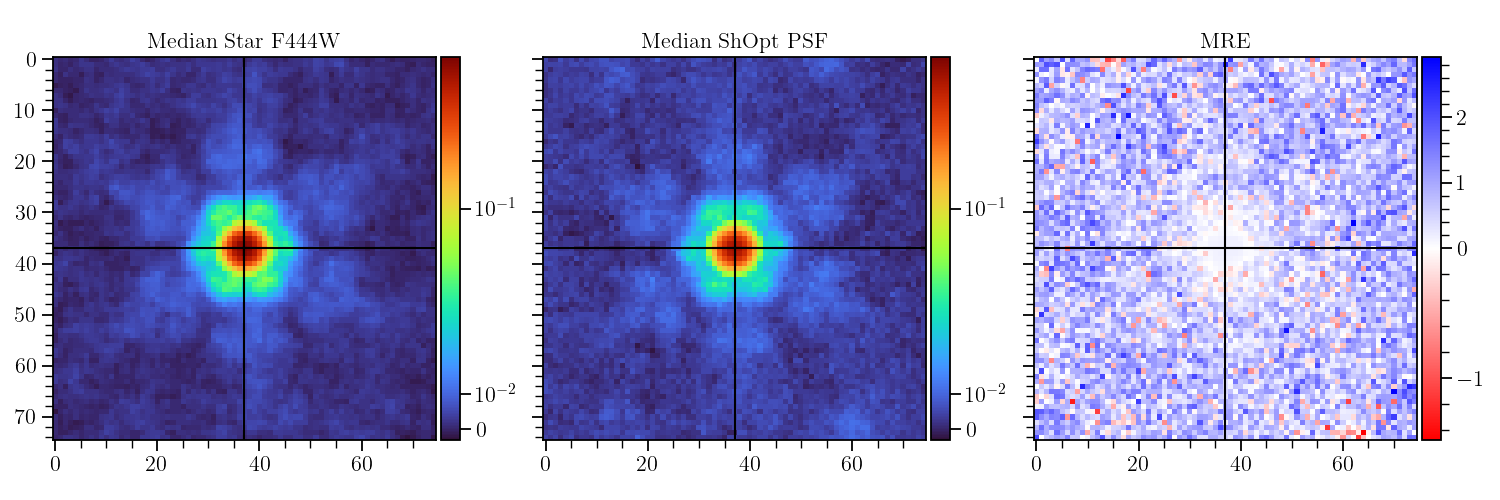}
        \centering
    \caption{Evaluation of mean relative error between stars and PSF models for real data mosaics in the F444W bandpass. Left panels show the median star vignette; center panels show the median PSF model; right panels show the mean residuals of individual star-PSF vignettes. Top rows is PSFEx, bottom row is \texttt{ShOpt}. }
    \label{fig:f444wAprMos}
\end{figure}

We find similar results for the real data mosaics as for the simulated mosaics exposures, namely that each PSF fitter produces similarly high quality models. Figures \ref{fig:f444wAprMos} and \ref{fig:f115wAprMos}-\ref{fig:f277wAprMos} suggest that in the main, both \texttt{ShOpt} and PSFEx are able to model the finer details of the PSF at all bandpasses analyzed.

\begin{figure}[hbtp]
    \centering
    \includegraphics[width=0.7\linewidth]{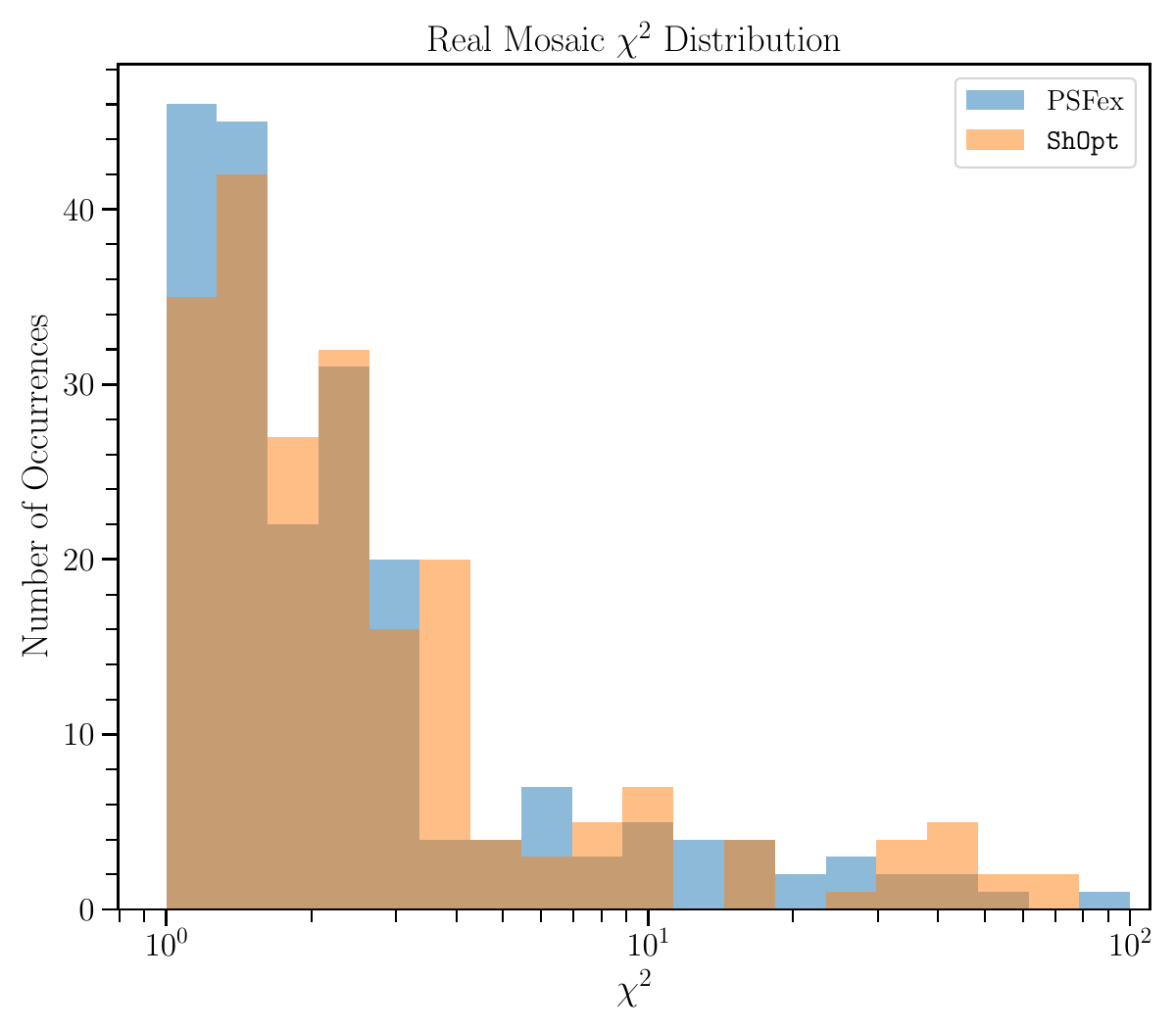}
    \caption{Distribution of $\chi^2_\nu$ for real data mosaics. Not shown are two PSFEx $\chi^2_\nu$ values greater than $100$ and four ShOpt $\chi^2_\nu$ values greater than $100$.}
    \label{fig:apr_mosaic_chisq_distribution}
\end{figure}

The high model fidelity for each fitter is further supported by the values of median and mean reduced $\chi^2$ in Table \ref{tab:Apr_Real_table}, as well as the distributions of reduced $\chi^2$ in Figure and \ref{fig:apr_mosaic_chisq_distribution}. We do not run PIFF on the real mosaics due to timing costs; the real data mosaics cover tens of thousands of pixels, and PIFF fits did not reliably converge. 

\begin{table} [hbtp]
\centering
\begin{tabular}{cccccc}
\toprule
Filter & PSF Fitter & MAE & MRE & $\overline{\chi^2}_\nu$ & Median $\chi^2_\nu$ \\
\midrule
\multirow{2}{*}{F115W} 
                       & ShOpt & 2.96$_{-2.25}^{+3.65}$ & 0.88$\pm$0.50 & 3.12$_{-0.99}^{+3.96}$ & 1.49 \\
                       & PSFEx & 2.76$_{-2.15}^{+3.38}$ & 0.90$\pm$0.49 & 4.49$_{-0.96}^{+5.30}$ & 1.45 \\
\midrule
\multirow{2}{*}{F150W} 
                       & ShOpt & 2.96$_{-2.24}^{+3.69}$ & 0.86$\pm$0.51 & 2.05$_{-1.02}^{+3.00}$ & 1.43 \\
                       & PSFEx & 2.76$_{-2.15}^{+3.42}$ & 0.87$\pm$0.48 & 1.65$_{-0.93}^{+2.09}$ & 1.27 \\
\midrule
\multirow{2}{*}{F277W} 
                       & ShOpt & 2.96$_{-1.87}^{+3.92}$ & 0.63$\pm$0.55 & 29.28$_{-2.59}^{+92.44}$ & 4.15 \\
                       & PSFEx & 2.56$_{-1.81}^{+3.27}$ & 0.75$\pm$0.50 & 13.55$_{-2.46}^{+34.46}$ & 3.60 \\
\midrule
\multirow{2}{*}{F444W} 
                       & ShOpt & 2.89$_{-1.91}^{+3.83}$ & 0.70$\pm$0.60 & 10.28$_{-1.67}^{+32.84}$ & 2.86 \\
                       & PSFEx & 2.58$_{-1.84}^{+3.30}$ & 0.78$\pm$0.52 & 7.34$_{-1.71}^{+15.22}$ & 2.49 \\
\bottomrule
\end{tabular}
\caption{Real mosaic summary statistics.}
\label{tab:Apr_Real_table}
\end{table}

\clearpage
\subsection{Size and Shape Analysis}
Although the NIRCam PSFs are obviously not elliptical Gaussians, adaptive second moments are a common way of evaluating the quality of PSF fits \citep{hirata2003shear, mandelbaum2005systematic}, and so in this section we explore residuals in the second moment fits of stars and PSF models. Specifically, we use the the GalSim software's \citep{rowe2015galsim} \texttt{FindAdaptiveMom} function to measure the size ($\sigma_{\rm HSM}$) and shape ($(g1, g2)$) of the best-fit elliptical Gaussians of all validation stars and PSF models, then compute the average of the differences between the two. As demonstrated by Table \ref{tab:HSM_apr}, \texttt{ShOpt} is able to produce fits that are just as good if not better than PSFEx and PIFF across the different data sets and wavelengths. The only area where the \texttt{ShOpt} models struggled was with the shape adaptive moments on some of the simulated single exposures. Otherwise, \texttt{ShOpt} produced just as good if not better size and shape statistics compared to the other PSF fitters.
\begin{table}[!htbp] 
\centering
\begin{tabular}{@{}llllll@{}}
\toprule
& & & \multicolumn{3}{c}{MSE}       \\ \cmidrule(lr){4-6}
Data & Fitter & Wavelength & $\sigma_{\rm HSM} \times 10^{-2}$   & $g_1 \times 10^{-2}$  & $g_2 \times 10^{-2}$   \\ \midrule
Simulated Single Exposure & PSFEx & F115W & $0.134 \pm 0.138$ & $ 0.06 \pm 0.01$ & $0.17 \pm 0.05$ \\
Simulated Single Exposure & ShOpt & F115W & $0.248 \pm 0.032$ & $0.37 \pm 0.01$ & $0.23 \pm 0.01$ \\ \hline 
Simulated Single Exposure & PSFEx & F150W & $0.46 \pm 0.12$ & $0.17 \pm 0.01$ & $0.24 \pm 0.01$\\
Simulated Single Exposure & ShOpt & F150W & $0.56 \pm 0.13$ & $0.35 \pm 0.01$ & $0.23 \pm 0.01$\\ \hline 
Simulated Single Exposure & PSFEx & F277W & $2.48 \pm 4.30$ & $0.12 \pm 0.02$ & $0.10 \pm 0.02$\\
Simulated Single Exposure &  ShOpt & F277W & $15.3 \pm 10.90$ & $0.61 \pm 0.03$ & $1.68 \pm 0.06$ \\ \hline 
Simulated Single Exposure &  PSFEx & F444W & $1.75 \pm 0.67$ & $0.16 \pm 0.02$ & $0.29 \pm 0.04$\\
Simulated Single Exposure &  ShOpt & F444W & $3.46 \pm 0.89$ & $0.67 \pm 0.05$ & $1.12 \pm 0.09$\\ \hline
Simulated Mosaics & PSFEx & F115W & $17.70 \pm 9.65$ & $1.63 \pm 0.63$ & $0.86 \pm 0.28$ \\  
Simulated Mosaics & ShOpt & F115W & $18.44 \pm 10.60$ & $1.57 \pm 0.62$ & $0.83 \pm 0.26$ \\  
Simulated Mosaics & PIFF & F115W & $18.80 \pm 10.90$ & $1.52 \pm 0.60$ & $0.93 \pm 0.31$ \\ \hline 
Simulated Mosaics & PSFEx & F150W & $7.96 \pm 5.58$ & $0.72 \pm 0.51$ & $0.07 \pm 0.02$ \\  
Simulated Mosaics & ShOpt & F150W & $7.35 \pm 5.00$ & $0.70 \pm 0.50$ & $0.07 \pm 0.02$ \\  
Simulated Mosaics & PIFF & F150W & $7.53 \pm 5.35$ & $0.76 \pm 0.53$ & $0.07 \pm 0.02$ \\ \hline 
Simulated Mosaics & PSFEx & F277W & $2.29 \pm 0.58$ & $0.42 \pm 0.13$ & $0.17 \pm 0.05$ \\  
Simulated Mosaics & ShOpt & F277W & $2.83 \pm 0.83$ & $0.44 \pm 0.13$ & $0.16 \pm 0.04$ \\  
Simulated Mosaics & PIFF & F277W & $5.19 \pm 0.81$ & $0.43 \pm 0.17$ & $0.13 \pm 0.04$ \\ \hline 
Simulated Mosaics & PSFEx & F444W & $1.08 \pm 0.33$ & $0.17 \pm 0.05$ & $0.30 \pm 0.17$ \\  
Simulated Mosaics & ShOpt & F444W & $0.98 \pm 0.27$ & $0.16 \pm 0.05$ & $0.32 \pm 0.18$ \\  
Simulated Mosaics & PIFF & F444W & $2.28 \pm 0.31$ & $0.30 \pm 0.08$ & $0.30 \pm 0.17$ \\ \hline 
April Mosaics & PSFEx & F115W & $0.67 \pm 0.10$ & $0.33 \pm 0.05$ & $0.19 \pm 0.05$ \\
April Mosaics & ShOpt & F115W & $0.78 \pm 0.10$ & $0.31 \pm 0.05$ &  $0.20 \pm 0.05$\\ \hline 
April Mosaics & PSFEx & F150W & $0.38 \pm 0.06$ & $0.12 \pm 0.02$& $0.04 \pm 0.01$\\
April Mosaics & ShOpt & F150W & $0.56 \pm 0.06$ & $0.11 \pm 0.02$& $0.03 \pm 0.00$\\ \hline 
April Mosaics & PSFEx & F277W & $1.19 \pm 0.23$ & $0.05 \pm 0.01$ & $0.04 \pm 0.01$\\
April Mosaics &  ShOpt & F277W & $0.90 \pm 0.13$ & $0.04 \pm 0.01$ & $0.03 \pm 0.01$\\ \hline 
April Mosaics &  PSFEx & F444W & $2.02 \pm 1.49$ & $0.02 \pm 0.01$ & $0.01 \pm 0.00$\\
April Mosaics &  ShOpt & F444W & $ 2.00 \pm 1.63$ & $0.03 \pm 0.01 $& $0.01 \pm 0.00$\\
\bottomrule
\end{tabular}
\caption{The residuals in the adaptive second moments of stars and PSF models across all data sets used in this work.} 
\label{tab:HSM_apr}
\end{table}
\clearpage
\subsection{Program Speed and Scalability} \label{subsec:speedTests}
We investigate how the different PSF fitters handle the simulated mosaic data as the number of pixels in the star vignettes and the degree of the polynomial used to model spatial variations increase. Each PSF fitter is run multiple times to get an average time in seconds for pixel basis fits. For \texttt{ShOpt}, we set the size of the side length $n$ of the analytic fit stamp to be $30$. That is, we use $30 \times 30$ pixels to fit the multivariate Gaussian. This is separate from the number of pixels we use for the pixel basis. We do not use any GPU compute power to accelerate the PSF fitters even though \texttt{ShOpt autoencoder} mode can be GPU-accelerated. All of these tests are run on the same hardware using Northeastern's Discovery Cluster on the zen2 CPUs \citep{northeastern_university_research_computing_2024}. These tests aim to test our calculations in Section \ref{sec:runtime}.  

\subsubsection{Variation with vignette size}
\begin{figure}
    \centering
    \includegraphics[width=1\linewidth]{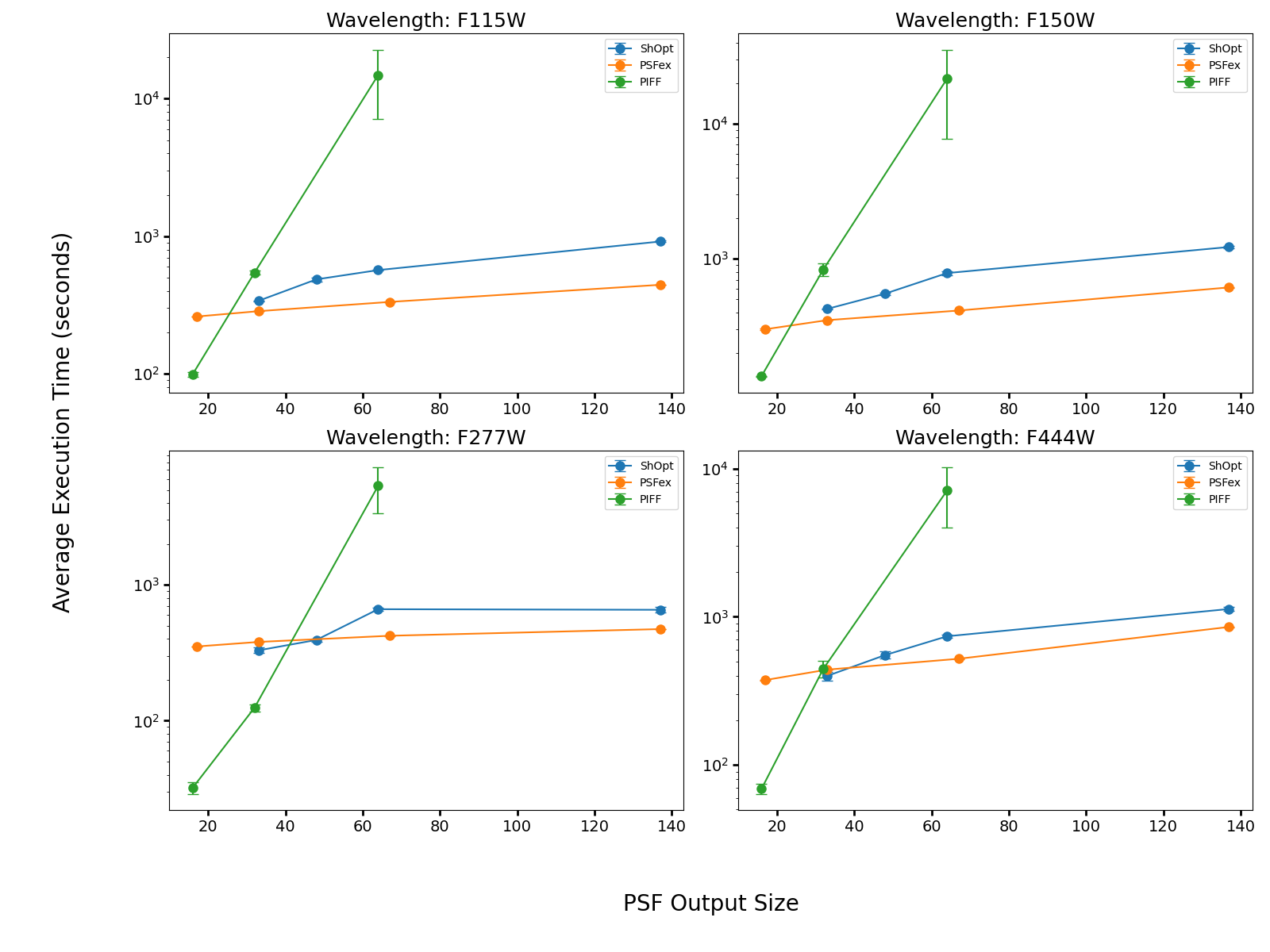}
    \caption{Average execution time as a function of the output size of the PSF sidelength plotted on a log scale.}
    \label{fig:psfOut}
\end{figure}

While the runtime for both \texttt{ShOpt} and PSFEx scales modestly as the output PSF size increases, the PIFF runtime does not. For larger PSF sizes, PIFF's average \texttt{BasisInterp} execution time is an order of magnitude longer than \texttt{ShOpt} and PSFEx. While \texttt{ShOpt} and PSFEx required approximately $5$ to $10$ minutes to process PSF sizes of $(137, 137)$ pixels for each of the four wavelengths, PIFF consistently took between $1$ and $3$ hours for PSF sizes of $(64, 64)$ (Figure \ref{fig:psfOut}). In the worst case, PIFF can take as much as an entire day to finish. While most PSF fitters had consistent completion times, PIFF exhibits greater variability for $(64,64)$-pixel PSF output sizes, particularly with shorter wavelengths. It is worth noting that PIFF was developed using DECam imaging, for which PSF sizes of $(17, 17)$ are sufficient. For those vignette sizes, PIFF is actually faster than the other PSF fitters, see Figure \ref{fig:psfOut}. It is also worth noting that PIFF contains alternatives to the \texttt{BasisInterp} algorithm benchmarked here and that these algorithms may be faster.
 
\begin{figure}
    \centering
    \includegraphics[width=0.9\linewidth]{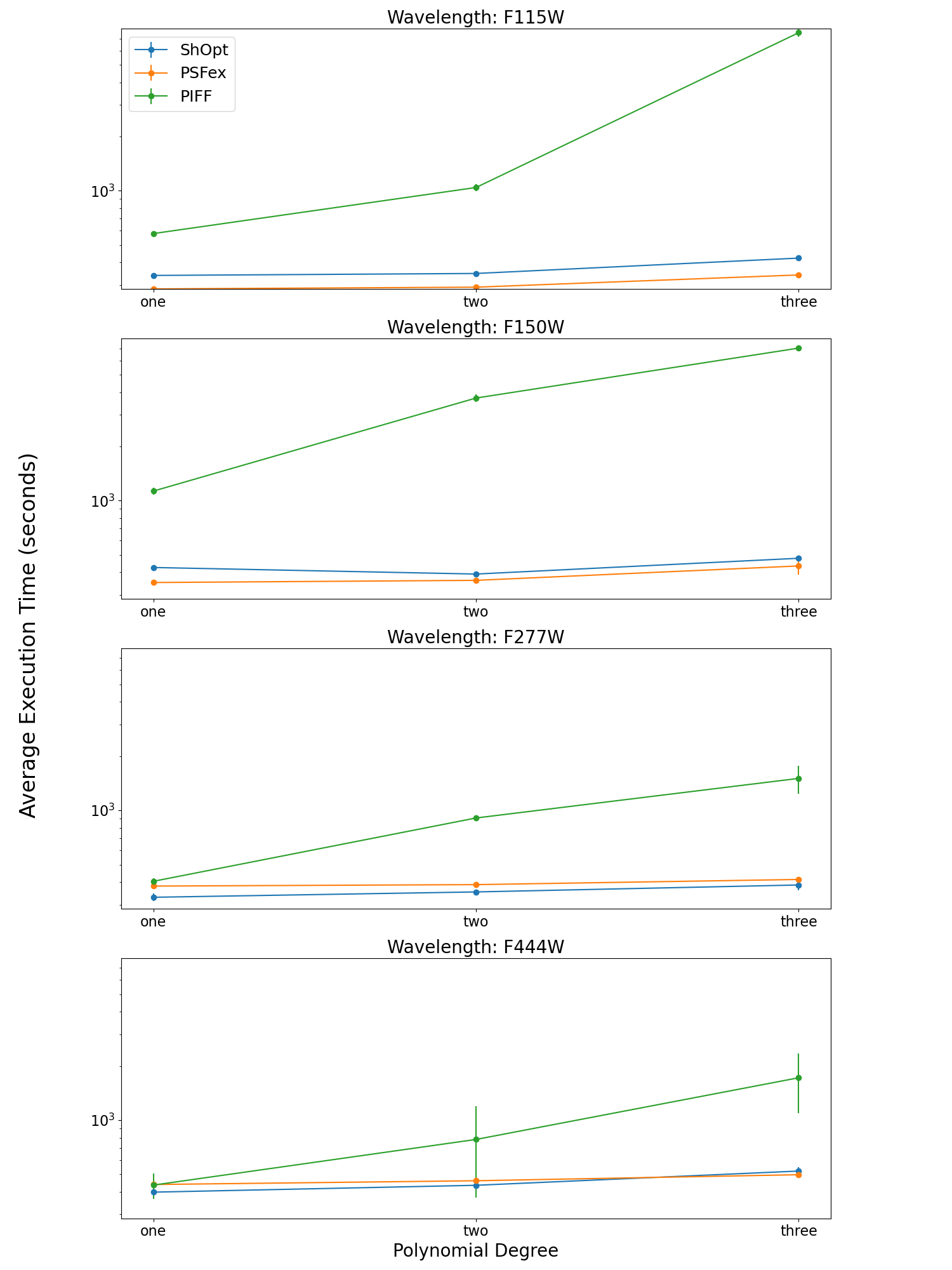}
    \caption{Average execution time as a function of the degree of the interpolating polynomial on $(33, 33)$ PSF models plotted on a log scale.}
    \label{fig:polynomial_degree}
\end{figure} 

\subsubsection{Polynomial Degree}

As in the case of vignette size, while both \texttt{ShOpt} and PSFEx scale as the the degree of the spatial interpolation polynomial increases, PIFF does not. For a degree-$1$ interpolation, all three PSF fitters have similar performance for $(33, 33)$ PSF sizes. However, as we increase the degree to $2$ and $3$, PIFF takes an order of magnitude longer on average to execution, see Figure \ref{fig:polynomial_degree}. For larger degree sizes, PIFF often did not converge to a PSF model with $4\sigma$ confidence after its maximum of $30$ iterations. \texttt{ShOpt} and PSFEx maintain a consistent runtime of about $5$ minutes, whereas PIFF experiences a dramatic increase in runtime, going up to hours for short wavelengths with higher degrees of the interpolating polynomial, and by around thirty minutes for longer wavelengths.

\section{Discussion and Conclusions} \label{sec:conclusions}

We have presented \texttt{ShOpt}, a new empirical PSF characterization tool, and with it, a methodology for benchmarking the accuracy and computational efficiency of PSF fitting software. 

In our assessment of PSF model quality, we produced a series of mean residual images, including normalized mean error, mean absolute error, and chi-squared error.  
We further condense this information into aggregate statistics that quantify the pixel-level discrepancies between the modeled and actual PSFs across the ensemble of stars being modeled. We supplement these statistics with second moment HSM fit statistics because of their wide use in the literature.

Though convenient, single-number figures of merit can mask discrepancies between a PSF model and the true PSF. Accordingly, we recommend a holistic approach for evaluating the quality of a particular PSF model, using both aggregate statistics and mean residual images. Among the various single-number figures of merit we examined, the average and median chi-squared values were the most illustrative of mismatched PSF models. Additionally, the distributions of these statistics are usually indicative of the performance of the PSF modeling.

In Section \ref{sec:runtime}, we predicted that \texttt{ShOpt}'s algorithmic design would allow it to scale as the PSF model size and the degree of the interpolating polynomial increased; this assertion was borne out in our speed testing. Our analysis does not distinguish between the contributions of architectural choices and the Julia language itself to \texttt{ShOpt}'s fast execution time. Regardless, we find that \texttt{ShOpt} delivers speed performance on par with PSFEx, with only marginal differences in processing times for PSF models $(137,137)$ pixels in size (large enough to enclose most of the NIRCam PSF). \texttt{ShOpt}'s competitive speed comes from a combination of multi-threading, imposed geometric constraints, the implementation of the LBFGS algorithm, and a thorough data cleaning pipeline. Conversely, PIFF---optimized for the larger pixel scale of the Dark Energy Camera---does not execute in a reasonable timeframe even for PSF sizes that are $(64,64)$ pixels in size.

In its current state, \texttt{ShOpt} ~produces an excellent model of the NIRCam PSF for all three data sets, and is able to produce these models extremely fast.

Several enhancements are planned for \texttt{ShOpt} that may improve the quality of its PSF models. For example, \texttt{ShOpt} builds models strictly in the native pixel scale of the image; implementing super- or sub-sampling of the image pixel scale would align \texttt{ShOpt} with the standards of other PSF software. A more advanced PCA method for PSF reconstruction, proposed in \cite{lin2023hybpsf}, is slated for future integration in our \texttt{PCA} mode. There is also a method of non-negative PCA outlined in \citep{https://doi.org/10.1002/(SICI)1099-128X(199709/10)11:5<393::AID-CEM483>3.0.CO;2-L} that may be useful to incorporate. These upgrades, among others, will be featured in future \texttt{ShOpt} releases.

\texttt{ShOpt} incorporates several innovative techniques that could benefit other PSF modeling efforts. These include leveraging manifold properties when fitting analytic profiles; using low-dimensional reconstructions for pixel-basis fits; and pixel-by-pixel parallelization during fitting of spatial variation. It has demonstrated the ability to bridge the speed of PSFEx with the advances of PIFF, making it a viable PSF fitter for next-generation astronomical observatories.

\begin{acknowledgments}
This work was supported by a Northeastern University Undergraduate Research and Fellowships PEAK Experiences Award. E.B. was also supported by a Northeastern University Physics Department Co-op Research Fellowship and by a Dean's College of Science Undergraduate Research Fellowship. This paper underwent internal review from the COSMOS-Web collaboration, and J.M. and E.B. thank everyone on this team for their insights. Additionally, E.B. thanks Professor David Rosen for valuable insights during the early stages of this work. Support for the COSMOS-Web survey was provided by NASA through grant JWST-GO-01727 and HST-AR-15802 awarded by the Space Telescope Science Institute, which is operated by the Association of Universities for Research in Astronomy, Inc., under NASA contract NAS 5-26555. This work was made possible by utilizing the CANDIDE cluster at the Institut d’Astrophysique de Paris, which was funded through grants from the PNCG, CNES, DIM-ACAV, and the Cosmic Dawn Center and maintained by Stephane Rouberol. Further support was provided by Research Computing at Northeastern University.

\end{acknowledgments}

\appendix \label{sec:appendix}

\section{Additional PSF Diagnostic Figures}

\begin{figure}[ht]
    \centering
        \includegraphics[width=\linewidth]{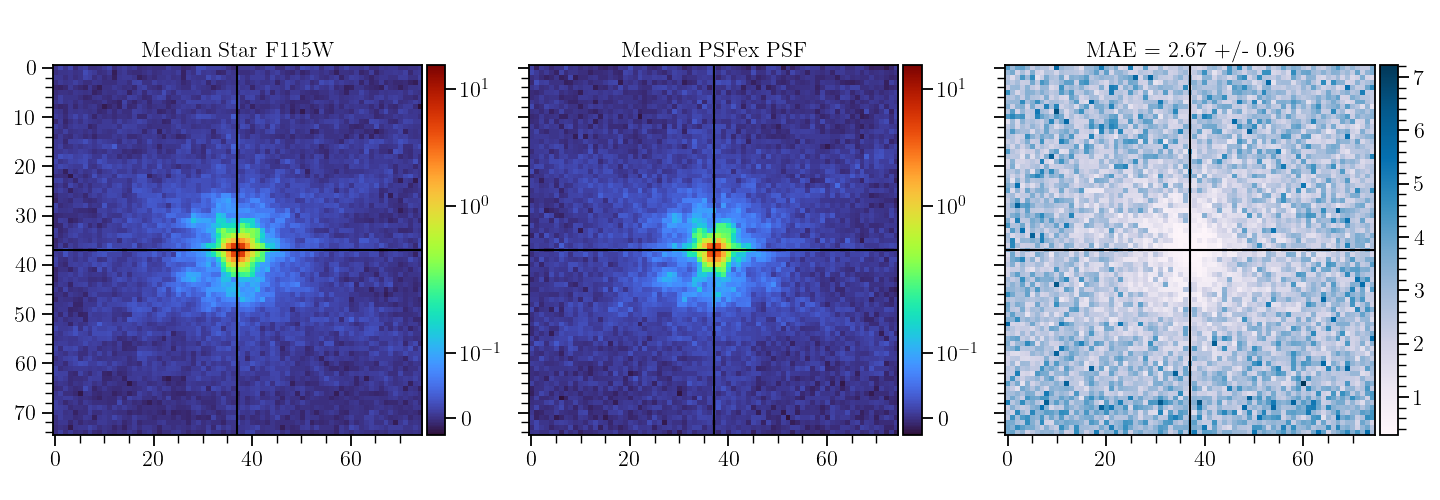}\\
        \includegraphics[width=\linewidth]{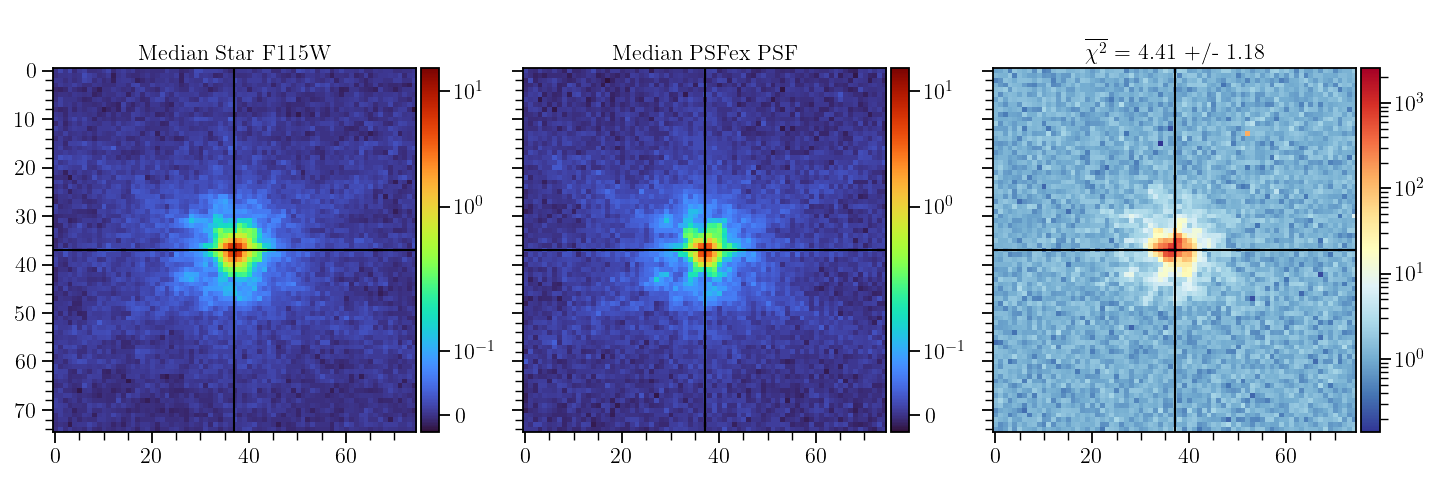}
    \caption{Examples of mean average error (top panel) and $\chi^2$ residual figures (bottom panel), shown here for the simulated mosaics in F115W.}\label{fig:mae_chi2_plots}
\end{figure}

\begin{figure}[hb]
    \centering
    \includegraphics[width=\textwidth]{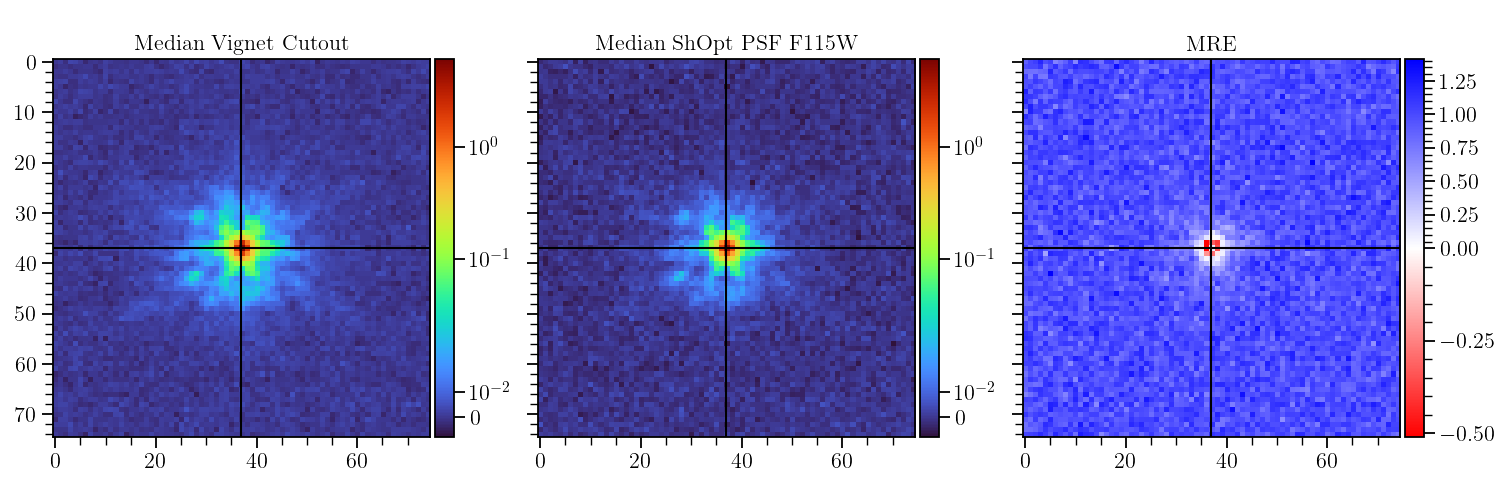}\\
    \includegraphics[width=\textwidth]{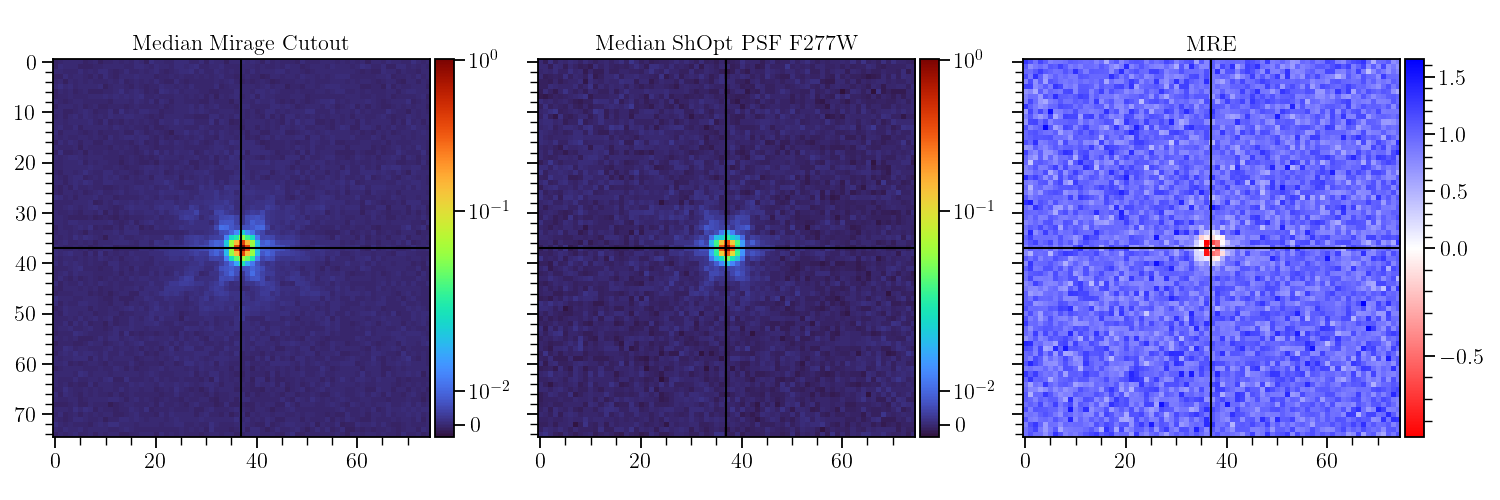}\label{fig:f150wShoptMirage}\\
    \includegraphics[width=\textwidth]{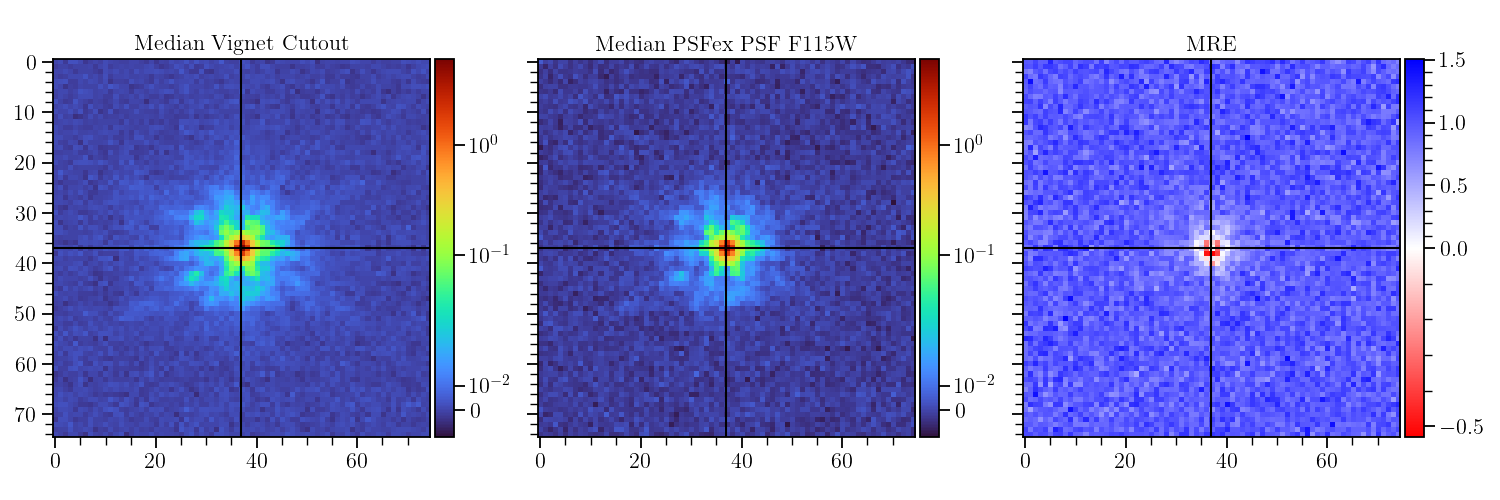} \\
    \includegraphics[width=\textwidth]{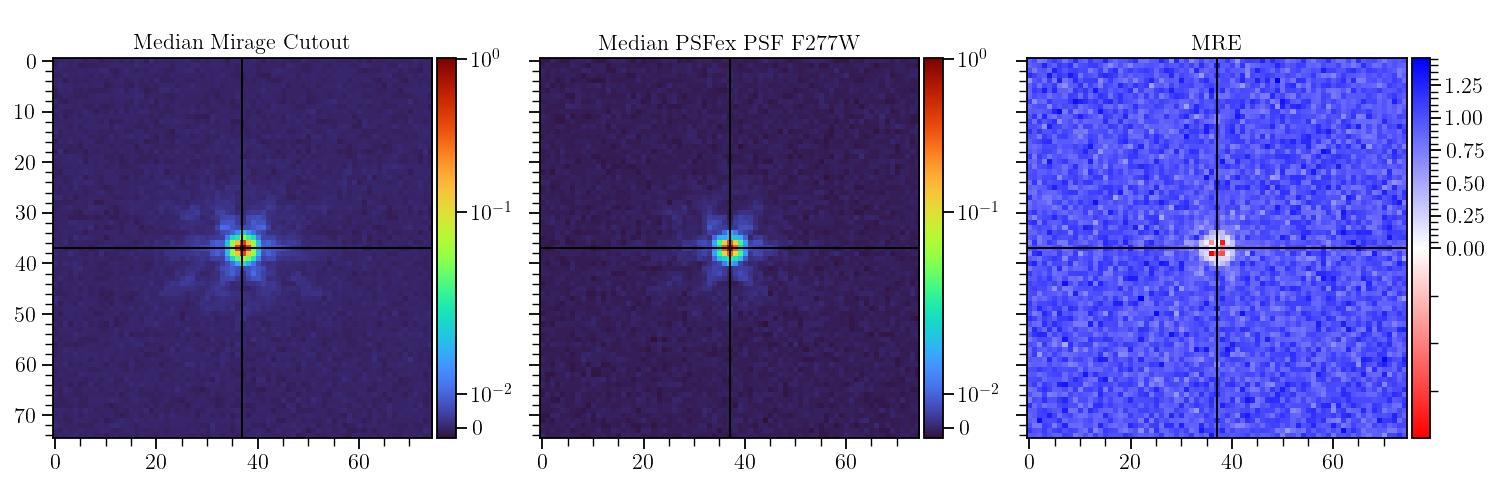}\label{fig:f150wPsfexMirage}
\caption{Mean relative error between MIRAGE input point source images and PSF models for the F115W and F150W bandpasses. Panels and color bars are the same as Figure \ref{fig:F444MiragePSFEx}.}
\label{fig:f115wShoptMirage}
\end{figure}

\begin{figure}[htpb]
    \centering
    
    \includegraphics[width=\textwidth]{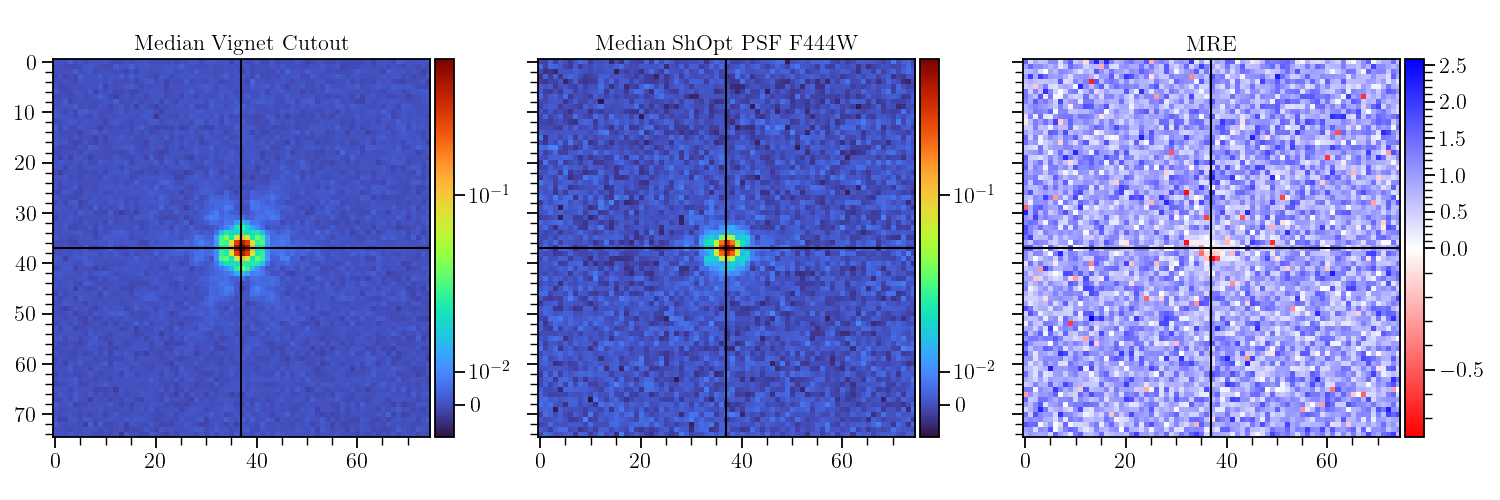}\label{fig:f277wShoptMirage}\\
    \includegraphics[width=\linewidth]{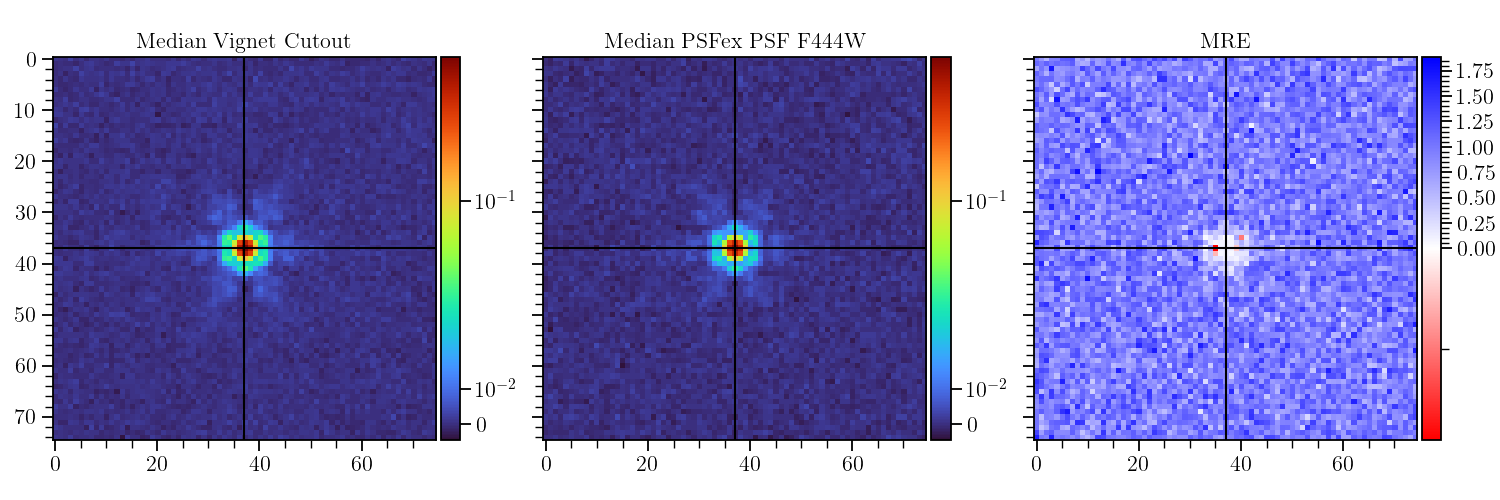}
\caption{Mean relative error between MIRAGE input point source images and PSF models for the F444W filter. Panels and color bars are the same as Figure \ref{fig:F444MiragePSFEx}.}\label{fig:f277wPsfexMirage}
\end{figure}

\begin{figure}[htpb]
    \centering
    \includegraphics[width=\linewidth]{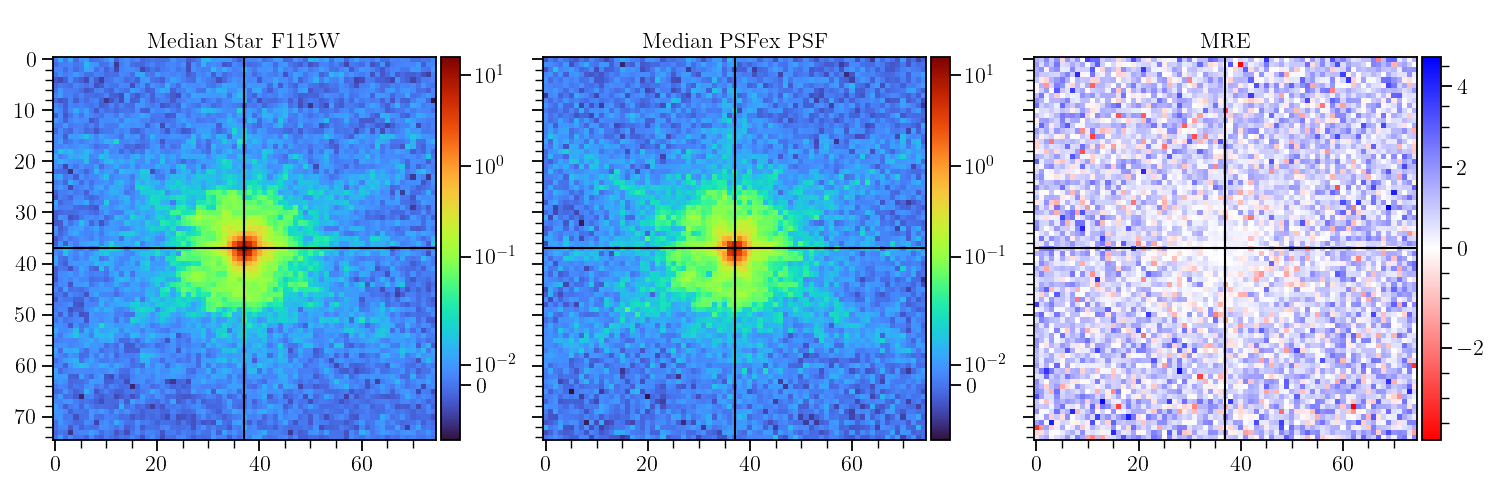}\\
      
        \includegraphics[width=\linewidth]{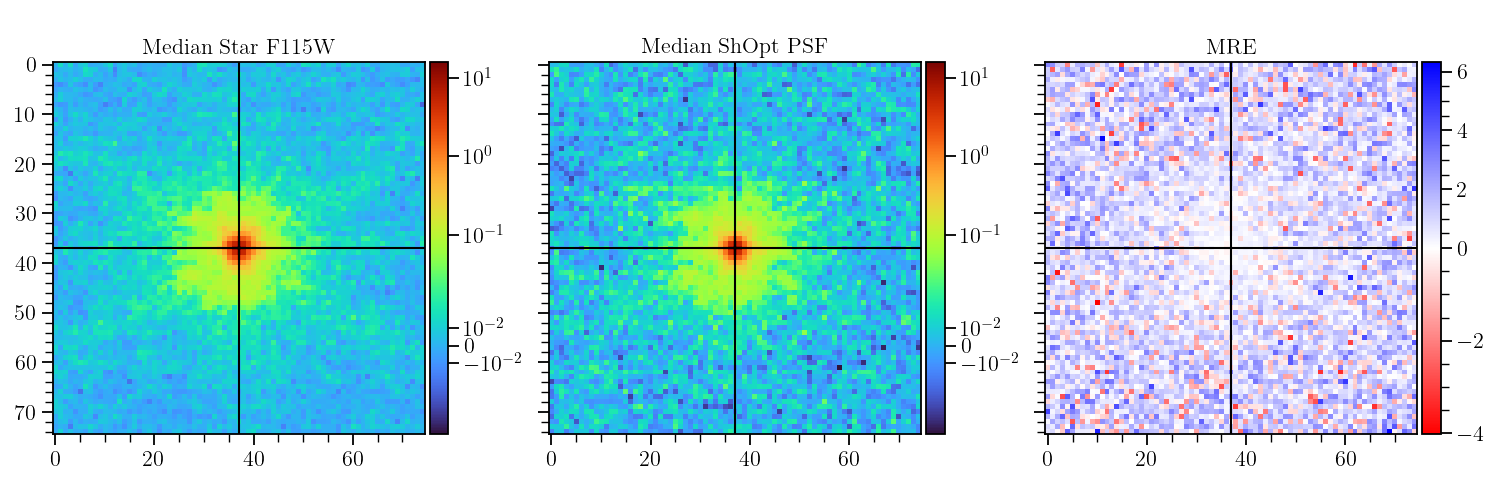}\\

        \includegraphics[width=\linewidth]{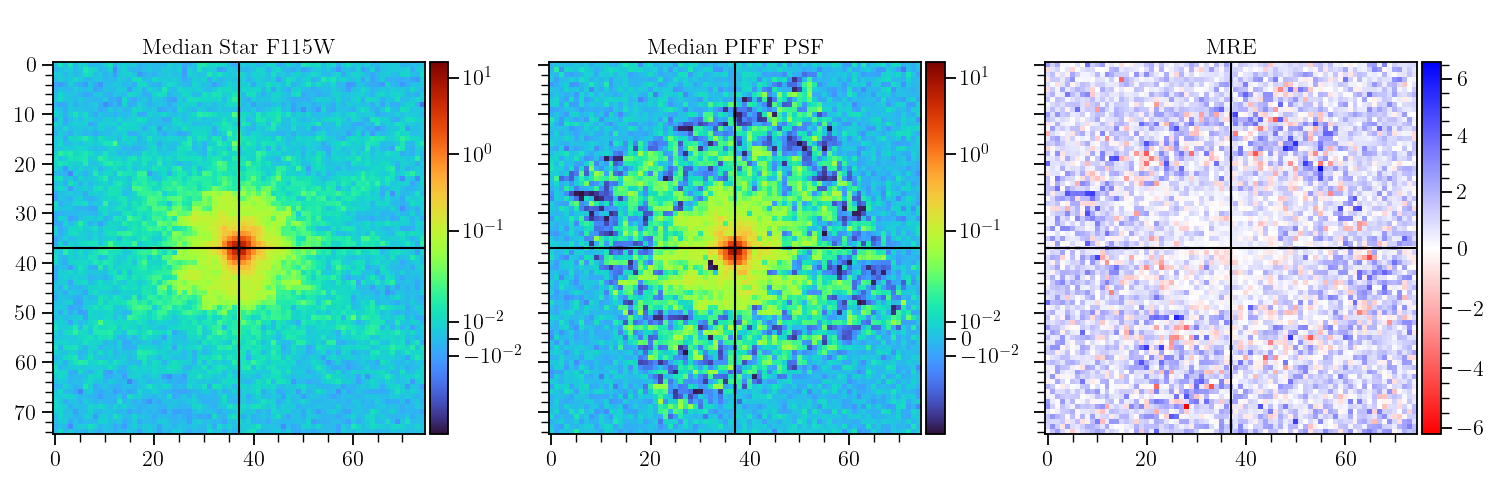}

    \caption{Mean relative error between stars and PSF models for simulated mosaics in the F115W bandpass. Left panels show the median of the vignettes. Panels and color bars are the same as in Figure \ref{fig:f150wCOSMOSSIMS}.}
    \label{fig:f115wCOSMOSSIMS}
\end{figure}

\begin{figure}[htpb]
    \centering
        \includegraphics[width=\linewidth]{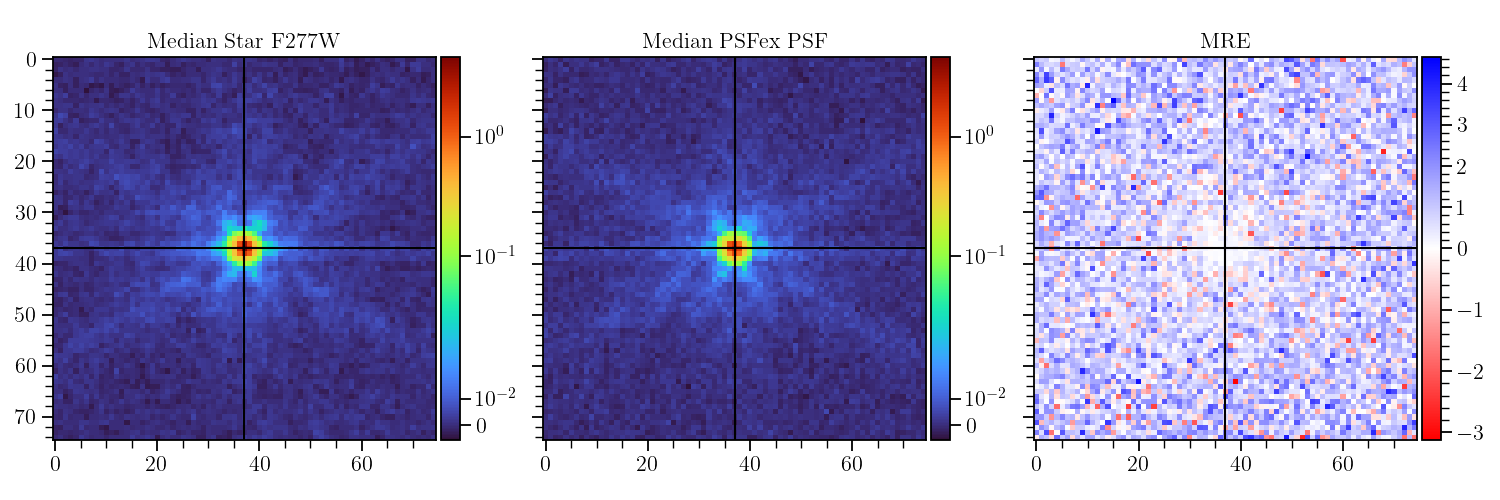}\\
        \includegraphics[width=\linewidth]{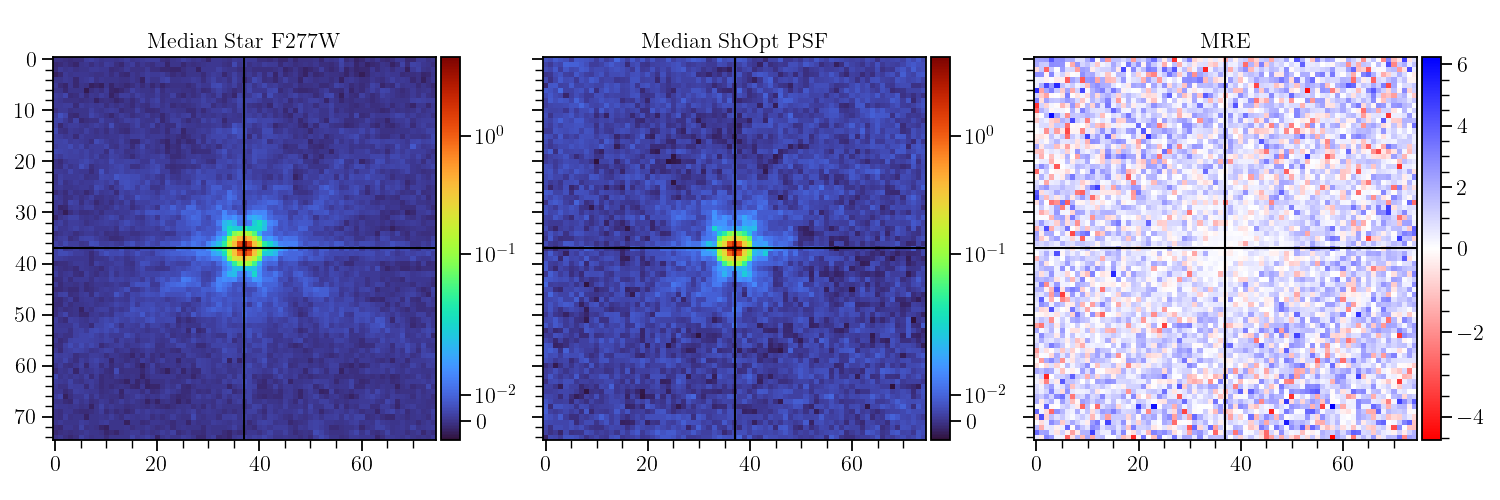}\\

        \includegraphics[width=\linewidth]{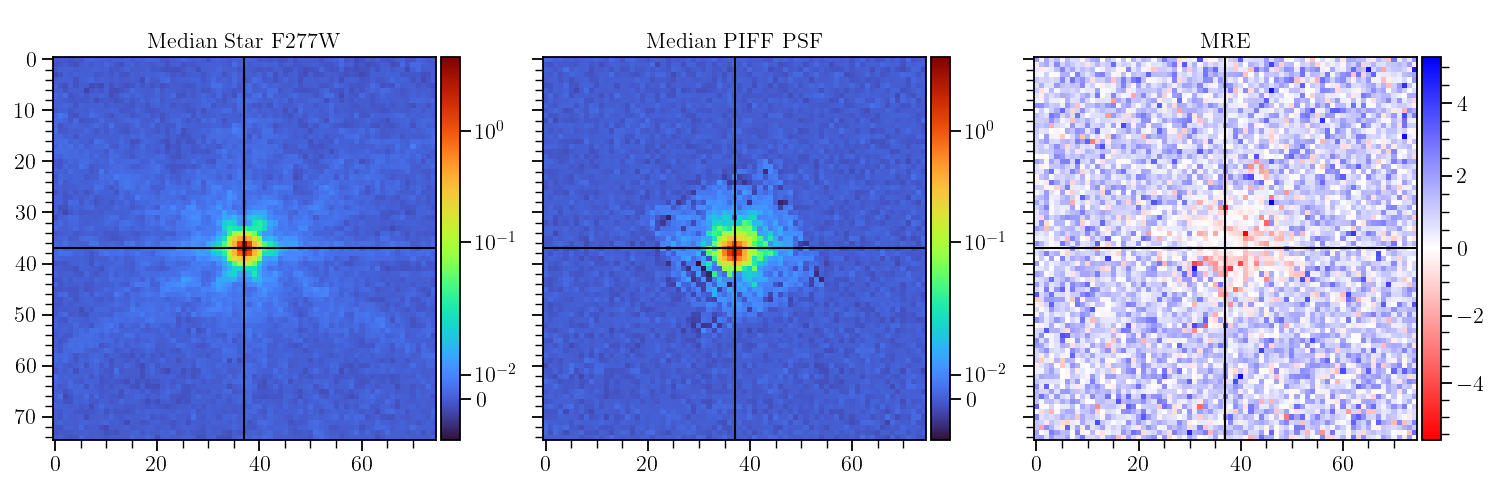}
    \caption{Evaluation of mean relative error between stars and PSF models for simulated mosaics in the F277W bandpass. Left panels show the median of the vignettes. Panels and color bars are the same as in Figure \ref{fig:f150wCOSMOSSIMS}.}
    \label{fig:f277wCOSMOSSIMS}
\end{figure}

\begin{figure}[htpb]
    \centering
        \includegraphics[width=\linewidth]{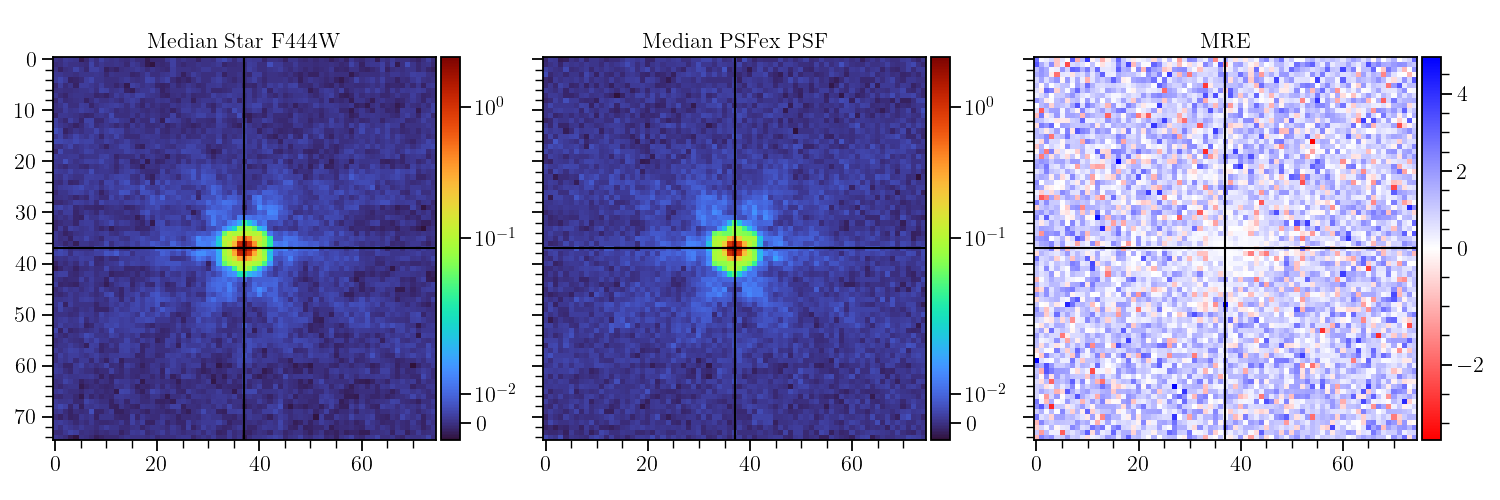}

           \includegraphics[width=\linewidth]{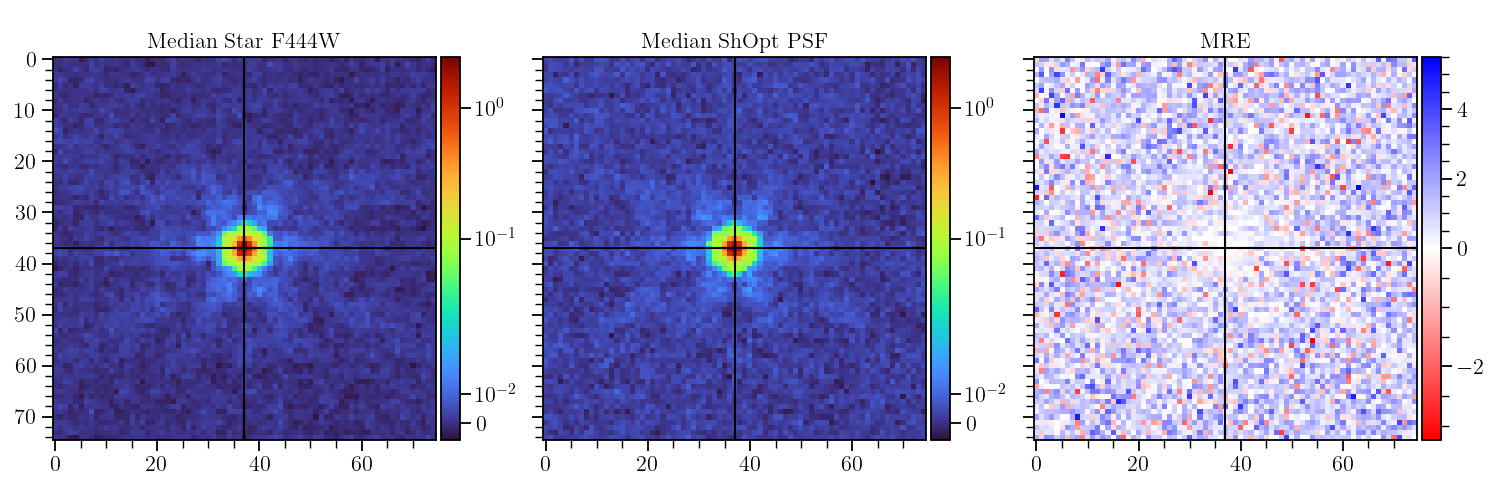}
   
          \includegraphics[width=\linewidth]{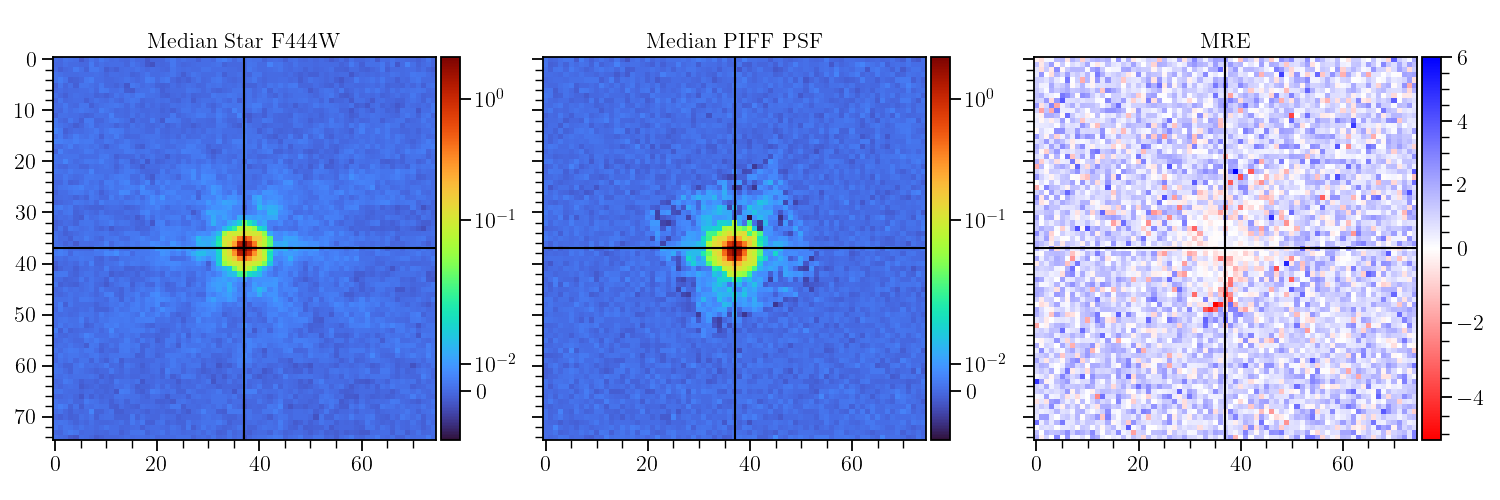}
  
    \caption{Evaluation of mean relative error between stars and PSF models for simulated mosaics in the F444W bandpass. Left panels show the median of the vignettes. Panels and color bars are the same as in Figure \ref{fig:f150wCOSMOSSIMS}.}
    \label{fig:f444wCOSMOSSIMS}
\end{figure}

\begin{figure}[hbtp]
    \centering
         \includegraphics[width=\linewidth]{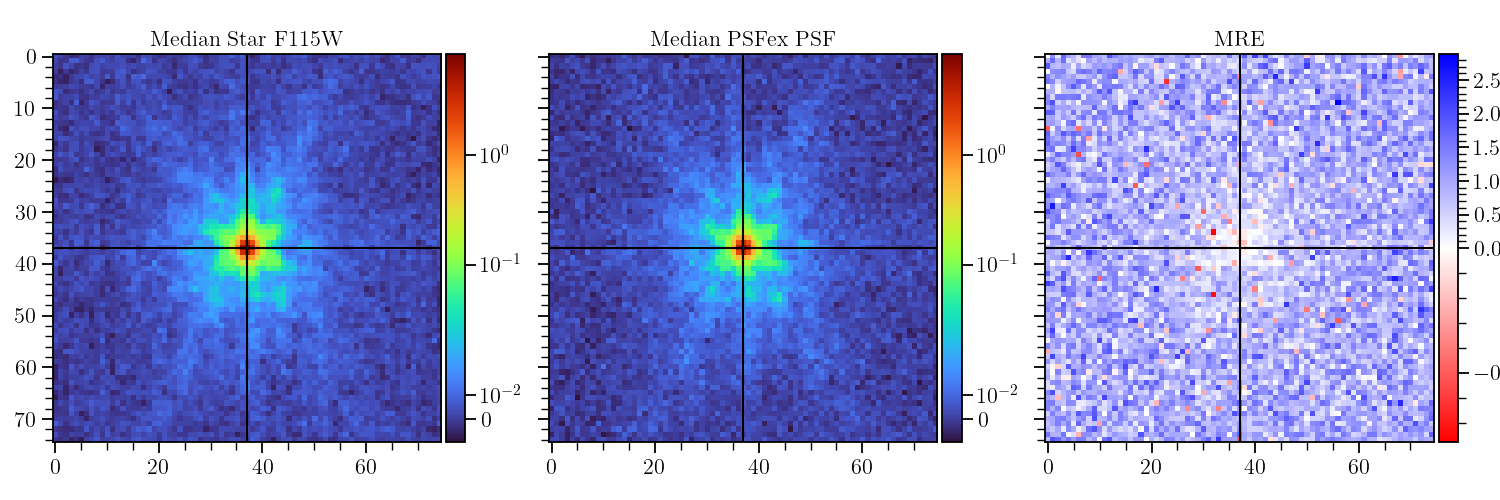}\\
        \includegraphics[width=\linewidth]{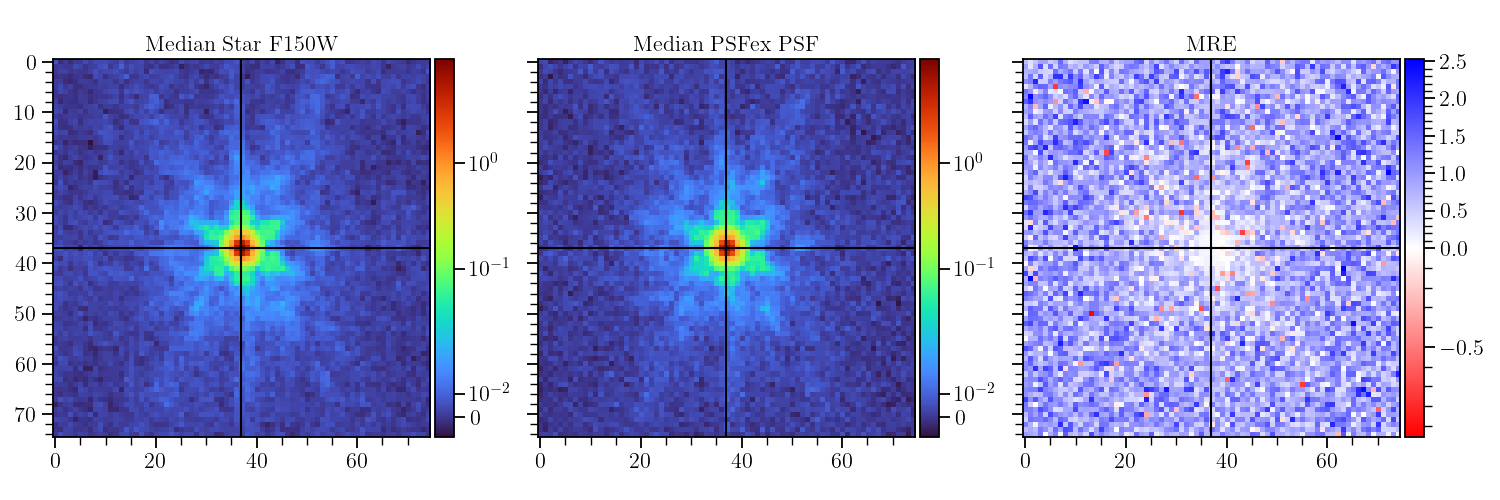}\\
        \includegraphics[width=\linewidth]{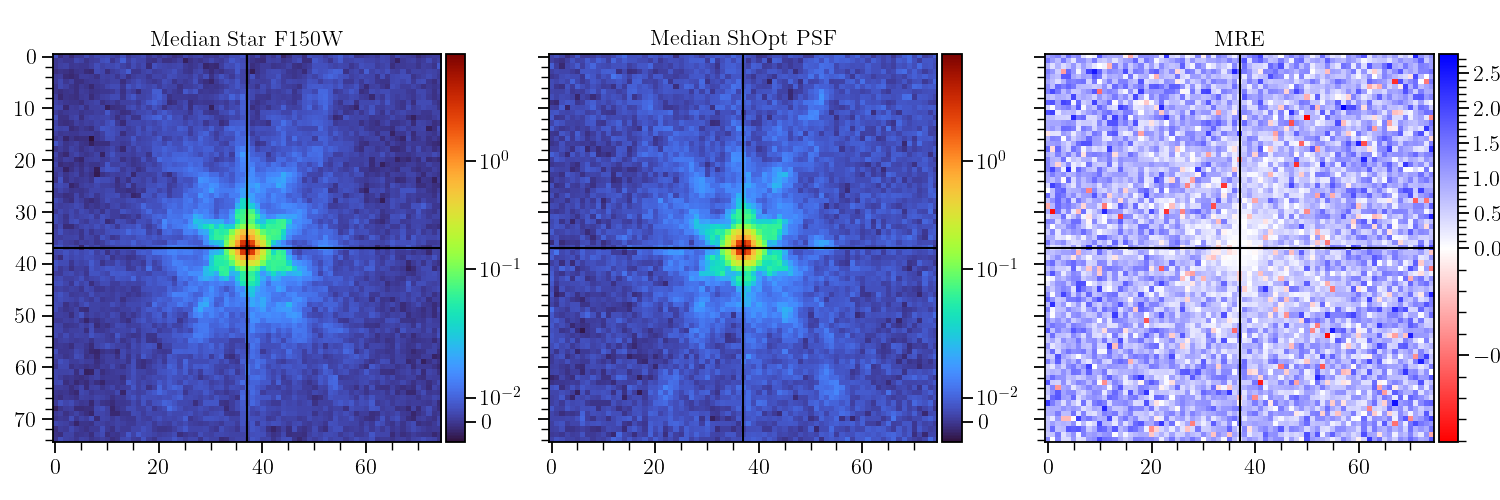}\\
         \includegraphics[width=\linewidth]{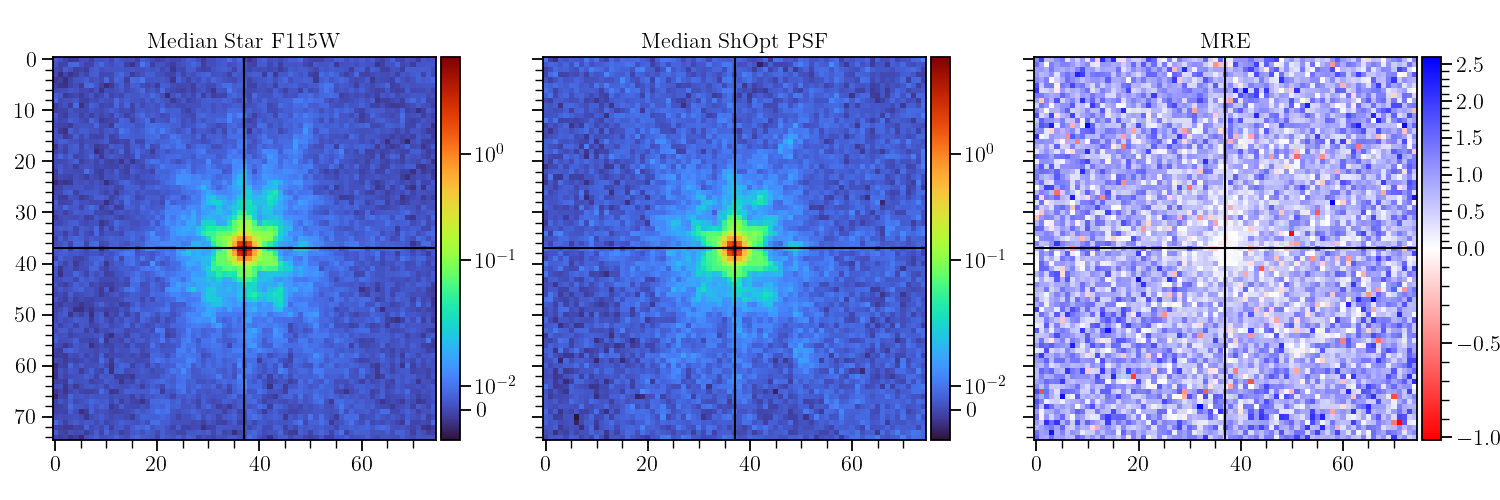}

    \caption{Mean relative error between stars and PSF models for real data mosaics in the F115W and F150W bandpasses. Color bars are the same as Figure \ref{fig:f444wAprMos}.}
    \label{fig:f115wAprMos}
\end{figure}

\begin{figure}
    \centering
        \includegraphics[width=\linewidth]{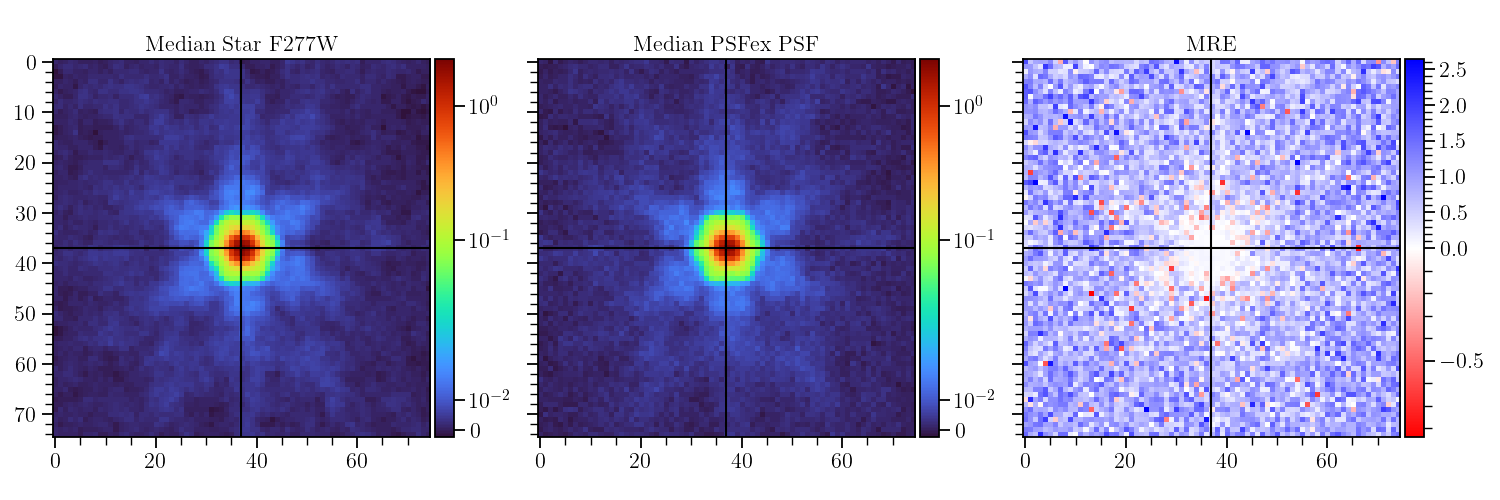}
        \includegraphics[width=\linewidth]{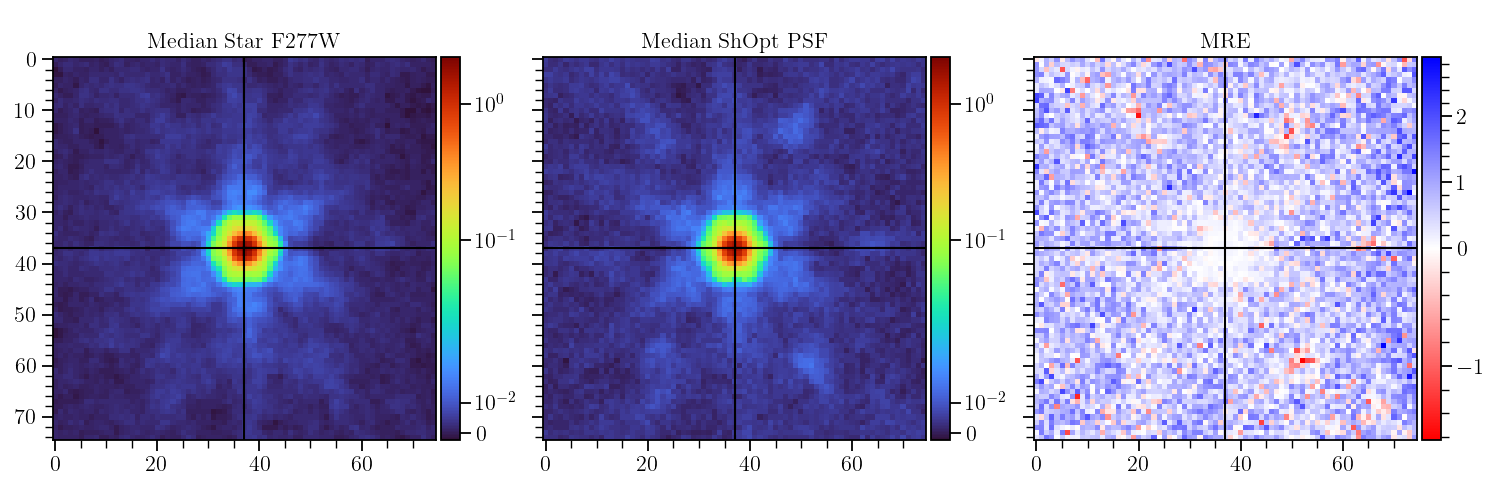}
    \caption{Evaluation of mean relative error between stars and PSF models for real data mosaics in the F277W bandpass. Panels and color bars are the same as in Figure \ref{fig:f444wAprMos}.}
    \label{fig:f277wAprMos}
\end{figure}

\clearpage

\section{Running ShOpt}
\small
\begin{python}
export JULIA_NUM_THREADS=auto # On Windows set JULIA_NUM_THREADS=auto 
julia shopt.jl [configdir] [outdir] [catalog.fits]  
\end{python}
\normalsize
See our \texttt{TutorialNotebook.ipynb} or our \texttt{README.md} for more detailed setup instructions on our GitHub.
\section{Testing Configs}\label{configs}
\subsection{Source Extractor config}
\footnotesize
\begin{verbatim}
# Default configuration file for SExtractor 2.25.0
#
#-------------------------------- Catalog ------------------------------------

CATALOG_NAME     catalog.fits  # name of the output catalog
CATALOG_TYPE     FITS_LDAC     
PARAMETERS_NAME  sextractor.param  # name of the file containing catalog contents

#------------------------------- Extraction ----------------------------------

DETECT_TYPE      CCD            # CCD (linear) or PHOTO (with gamma correction)
DETECT_MINAREA   8              # min. # of pixels above threshold
DETECT_MAXAREA   0              # max. # of pixels above threshold (0=unlimited)
THRESH_TYPE      RELATIVE       # threshold type: RELATIVE (in sigmas)
                                # or ABSOLUTE (in ADUs)
DETECT_THRESH    2             # <sigmas> or <threshold>,<ZP> in mag.arcsec-2
ANALYSIS_THRESH  2            # <sigmas> or <threshold>,<ZP> in mag.arcsec-2

FILTER           Y              # apply filter for detection (Y or N)?
FILTER_NAME  gauss_2.5_5x5.conv # name of the file containing the filter
FILTER_THRESH                   # Threshold[s] for retina filtering

DEBLEND_NTHRESH  32             # Number of deblending sub-thresholds
DEBLEND_MINCONT  0.005          # Minimum contrast parameter for deblending
CLEAN            Y              # Clean spurious detections? (Y or N)?
CLEAN_PARAM      1.0            # Cleaning efficiency

MASK_TYPE        CORRECT        # type of detection MASKing: can be one of
                                # NONE, BLANK or CORRECT

#-------------------------------- WEIGHTing ----------------------------------

WEIGHT_TYPE    MAP_WEIGHT       # type of WEIGHTing: NONE, BACKGROUND,
                                # MAP_RMS, MAP_VAR or MAP_WEIGHT
RESCALE_WEIGHTS  Y              # Rescale input weights/variances (Y/N)?
WEIGHT_IMAGE     weight.fits    # weight-map filename
WEIGHT_GAIN      Y              # modulate gain (E/ADU) with weights? (Y/N)
WEIGHT_THRESH    0              # weight threshold[s] for bad pixels

#-------------------------------- FLAGging -----------------------------------

FLAG_IMAGE       flag.fits      # filename for an input FLAG-image
FLAG_TYPE        OR             # flag pixel combination: OR, AND, MIN, MAX
                                # or MOST

#----------------------- Differential Geometry Map ---------------------------

DGEO_TYPE        NONE           # Differential geometry map type: NONE or PIXEL
DGEO_IMAGE       dgeo.fits      # Filename for input differential geometry image

#------------------------------ Photometry -----------------------------------

PHOT_APERTURES   31            # MAG_APER aperture diameter(s) in pixels
PHOT_AUTOPARAMS  2, 2.5         # MAG_AUTO parameters: <Kron_fact>,<min_radius>
PHOT_PETROPARAMS 2.0, 3.5       # MAG_PETRO parameters: <Petrosian_fact>,
                                # <min_radius>
PHOT_AUTOAPERS   0.0,0.0        # <estimation>,<measurement> minimum apertures
                                # for MAG_AUTO and MAG_PETRO
PHOT_FLUXFRAC    0.5            # flux fraction[s] used for FLUX_RADIUS

SATUR_LEVEL      37000.0        # level (in ADUs) at which arises saturation
SATUR_KEY        SATURATE       # keyword for saturation level (in ADUs)

MAG_ZEROPOINT    28.086519392           # magnitude zero-point
MAG_GAMMA        4.0            # gamma of emulsion (for photographic scans)
GAIN             0.0            # detector gain in e-/ADU
GAIN_KEY         GAIN           # keyword for detector gain in e-/ADU
PIXEL_SCALE      0              # size of pixel in arcsec (0=use FITS WCS info)

#------------------------- Star/Galaxy Separation ----------------------------

SEEING_FWHM      0.07           # stellar FWHM in arcsec
STARNNW_NAME     default.nnw    # Neural-Network_Weight table filename

#------------------------------ Background -----------------------------------

BACK_TYPE        AUTO           # AUTO or MANUAL
BACK_VALUE       0.0            # Default background value in MANUAL mode
BACK_SIZE        128            # Background mesh: <size> or <width>,<height>
BACK_FILTERSIZE  3              # Background filter: <size> or <width>,<height>

BACKPHOTO_TYPE   LOCAL          # can be GLOBAL or LOCAL
BACKPHOTO_THICK  24             # thickness of the background LOCAL annulus
BACK_FILTTHRESH  0.0            # Threshold above which the background-
                                # map filter operates

#------------------------------ Check Image ----------------------------------

CHECKIMAGE_TYPE  -BACKGROUND, APERTURES     # Check-image type(s)
CHECKIMAGE_NAME  im.sub.fits, im.aper.fits  # Filename for the check-image(s)
\end{verbatim}
\normalsize
\subsection{ShOpt config}
\footnotesize
\begin{verbatim}
saveYaml: true #save this file with each run

# Options: auotoencoder, PCA, smoothing. 
# Make sure mode is a string with double quotes
mode: "smoothing"

# If PCA mode is enabled, how many moments do you 
# want to use for your pixel grid fit
PCAterms: 50 

# The size of the smoothing kernel
lanczos: 5 

# For Autoencoder mode, when enabled
NNparams: 
  # max number of training epochs for each pixel grid fit
  epochs: 100 
  # The stopping gradient of the loss function for the pixel grid fit 
  minGradientPixel: 1e-5 

# For fitting analytic profile
AnalyticFitParams: 
  # Stopping gradient for LBFGS on vignettes
  minGradientAnalyticModel: 1e-6 
  # Stopping gradient for LBFGS on pixel grid models
  minGradientAnalyticLearned: 1e-6 
  # The subset of pixels you wish to fit the analytic profile to
  analyticFitStampSize: 64  

dataProcessing:
  # Filter this % of stars based off of signal to noise
  SnRPercentile: 0.33 
  # Filter stars with analytic profile fit size s exceeding this value 
  sUpperBound: 1 
  # Filter stars with analytic profile fit size s below this value 
  sLowerBound: 0.075 

# What plots do you want?
plots: 
  unicodePlots: true
  normalPlots:
    parametersHistogram: true
    parametersScatterplot: true
  cairomakiePlots:
    streamplots: false
  pythonPlots: false

# Degree of polynomial for spatial interpolation of PSF model 
polynomialDegree: 1 
# Size of pixel grid PSF model
stampSize: 130 

# How many stars are you using the train versus to validate the PSF fit
training_ratio: 0.9 
# Sum flux to unity true or false
sum_pixel_grid_and_inputs_to_unity: false 

# stopping gradient for LFBGS used for polynomial interpolation
polynomial_interpolation_stopping_gradient: 1e-12 

# Name to prefix summary.shopt
summary_name: ''
# Do you want to save storage by only storing essential information, how to reconstruct the PSF and analytic models
truncate_summary_file: true 

CommentsOnRun: "** This is where you can leave comments or notes to self on the run! **"
\end{verbatim} 
\normalsize
\subsection{PIFF config}
\footnotesize
\begin{verbatim}
# modules and input.wcs fields, in which case, the code will use the (less
# accurate) WCS that ships with the image in the fits file.

input:

    # Input file directory
    dir: "./"

    # Input filename(s) and HDU extensions 
    image_file_name: "mosaic_nircam_f277w_COSMOS-Web_i2d.fits"
    image_hdu: 1 
    weight_hdu: 4

    # Input catalog and HDU extension 
    cat_file_name: "mosaic_nircam_f277w_COSMOS-Web_i2d_starcat.fits"
    cat_hdu: 2

    # What columns in the catalog have things we need?
    x_col: XWIN_IMAGE
    y_col: YWIN_IMAGE
    ra_col: ALPHAWIN_J2000
    dec_col: DELTAWIN_J2000

    # The telescope pointing
    ra: 149.9303551903936
    dec: 2.380272767453749

    # Leave blank if you don't know what it is!
    # gain: 1

    # How large should the postage stamp cutouts of the stars be?
    stamp_size: 100

    # Use all cores for reading the input files
    nproc: -1

select:

    # For bright stars, weight them equivalent to snr=100 stars, not higher.
    max_snr: 100

    # Remove stars with snr < 10
    min_snr:  10

    # Reserve 15% of the stars for diagnostics
    reserve_frac: 0.15

    # Reject size outliers
    hsm_size_reject: True

psf:

    # This type of PSF will use a separate model/interp solution for each chip.
    # But all the solutions will be given in a single output file.
    type: SingleChip

    # Also use all cores when finding psf
    nproc: -1

    outliers:

        type: Chisq

        # The threshold is given in terms of nsigma equivalent
        nsigma: 4

        # Only remove at most 3% of the stars per iteration.
        max_remove: 0.03

    model:
        # This model uses a grid of pixels to model the surface brightness distribution.
        type: PixelGrid
        scale: 0.03      # NIRCam ative pixel scale
        size: 75      

    interp:

        # This interpolator does some of the model solving when interpolating
        # to handle degenerate information from masking 
        # and the fact that the pixels are smaller than native.
        type: BasisPolynomial
        order: 1
\end{verbatim} 
\normalsize
\subsection{PSFEx Config}
\footnotesize
\begin{verbatim}
# Default configuration file for PSFEx 3.17.1
# EB 2016-06-28
#

#-------------------------------- PSF model ----------------------------------

BASIS_TYPE      PIXEL           # NONE, PIXEL, GAUSS-LAGUERRE or FILE
#BASIS_NUMBER   30              # Basis number or parameter
PSF_SAMPLING    0               # Sampling step in pixel units (0.0 = auto)
PSF_SIZE        261             # Image size of the PSF model
PSF_RECENTER    Y
#------------------------- Point source measurements -------------------------

CENTER_KEYS     XWIN_IMAGE,YWIN_IMAGE # Catalogue parameters for source pre-centering
PHOTFLUX_KEY    FLUX_APER(1)    # Catalogue parameter for photometric norm.
PHOTFLUXERR_KEY FLUXERR_APER(1) # Catalogue parameter for photometric error

#----------------------------- PSF variability -------------------------------

PSFVAR_KEYS     XWIN_IMAGE,YWIN_IMAGE # Catalogue or FITS (preceded by :) params
PSFVAR_GROUPS   1,1             # Group tag for each context key
PSFVAR_DEGREES  1               # Polynom degree for each group

#----------------------------- Sample selection ------------------------------

SAMPLE_AUTOSELECT  Y           # Automatically select the FWHM (Y/N) ?
SAMPLEVAR_TYPE    NONE         # File-to-file PSF variability: NONE or SEEING
SAMPLE_FWHMRANGE   1,20        # Allowed FWHM range (2.7,3.2)
SAMPLE_VARIABILITY 0.3         # Allowed FWHM variability (1.0 = 100%)
SAMPLE_MINSN       50         # Minimum S/N for a source to be used
SAMPLE_MAXELLIP    0.3         # Maximum (A-B)/(A+B) for a source to be used

#----------------------------- Output catalogs -------------------------------

OUTCAT_TYPE        FITS_LDAC      # NONE, ASCII_HEAD, ASCII, FITS_LDAC
OUTCAT_NAME        psfex_out.cat  # Output catalog filename


#------------------------------- Check-plots ----------------------------------

CHECKPLOT_DEV       PDF         # NULL, XWIN, TK, PS, PSC, XFIG, PNG,
                                # JPEG, AQT, PDF or SVG
CHECKPLOT_RES       0           # Check-plot resolution (0 = default)
CHECKPLOT_ANTIALIAS Y           # Anti-aliasing using convert (Y/N) ?
CHECKPLOT_TYPE     NONE 
CHECKPLOT_NAME      

#------------------------------ Check-Images ---------------------------------

CHECKIMAGE_TYPE CHI,SAMPLES,RESIDUALS,SNAPSHOTS, -SYMMETRICAL
                                # or MOFFAT,-MOFFAT,-SYMMETRICAL
CHECKIMAGE_NAME chi.fits,samp.fits,resi.fits,snap.fits, minus_symm.fits
                                # Check-image filenames
CHECKIMAGE_CUBE Y
\end{verbatim}
\normalsize
\subsection{Shell Script}
This is given to inform how much memory we requested from Discovery for purposes of speed testing.
\footnotesize
\begin{verbatim}
#!/bin/bash
#SBATCH --nodes=1
#SBATCH --mem=16G
#SBATCH --cpus-per-task=4
#SBATCH --time=24:00:00
#SBATCH --partition=short
#SBATCH --job-name=150_A6
#SBATCH --ntasks=1
#SBATCH --constraint=zen2            # Requesting specific CPU architecture
pwd
source /work/mccleary_group/berman.ed/minicondaInstall/bin/activate
python get_galaxy_cutouts.py -config configs/box_cutter.yaml file #file is a placeholder for a fits file
\end{verbatim}
\normalsize
\section{Additional ShOpt Checkplots and Outputs}
\subsection{Diagnostic Material}

\begin{figure}[!htb]
    \centering

    \begin{minipage}[b]{0.75\textwidth}
        \includegraphics[width=1\linewidth]{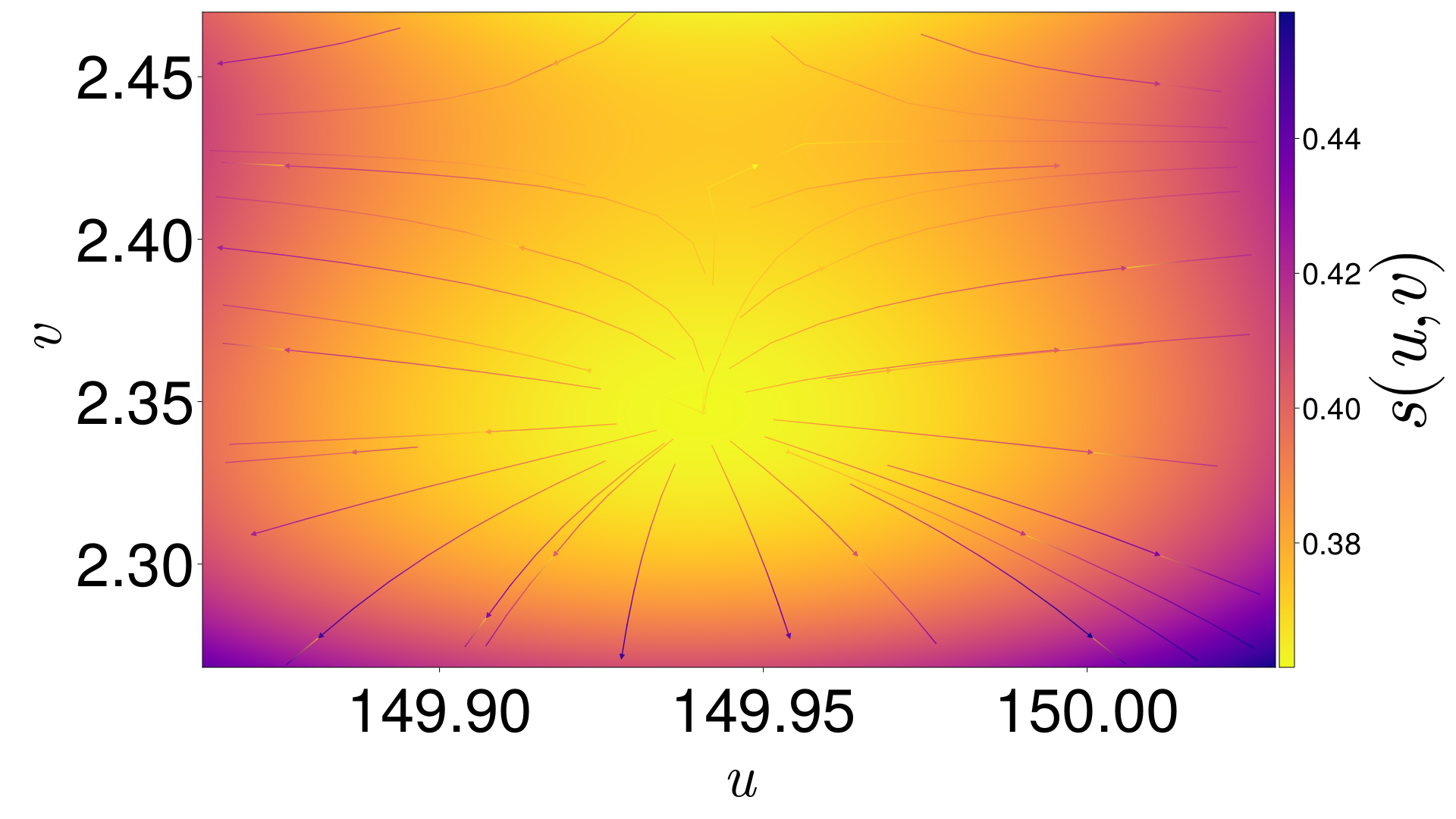}
        \centering
        (a) $s(u,v)$
        \label{fig:subfig1}
    \end{minipage}
    
    \vspace{\baselineskip}

    \begin{minipage}[b]{0.75\textwidth}
        \includegraphics[width=\textwidth]{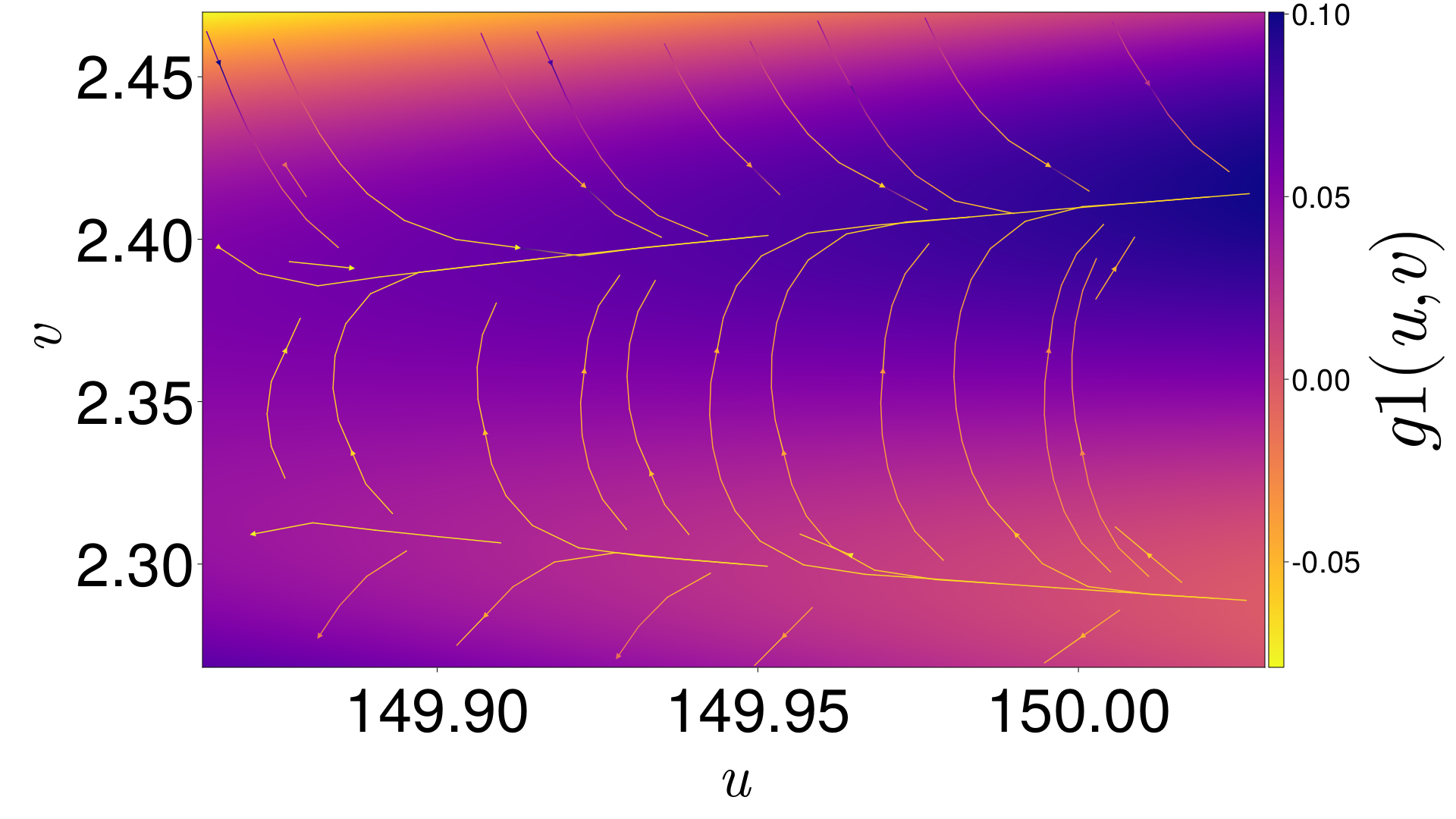}
        \centering
        (b) $g_1(u,v)$
        \label{fig:subfig2}
    \end{minipage}

    \vspace{\baselineskip}

    \begin{minipage}[b]{0.75\textwidth}
        \includegraphics[width=\textwidth]{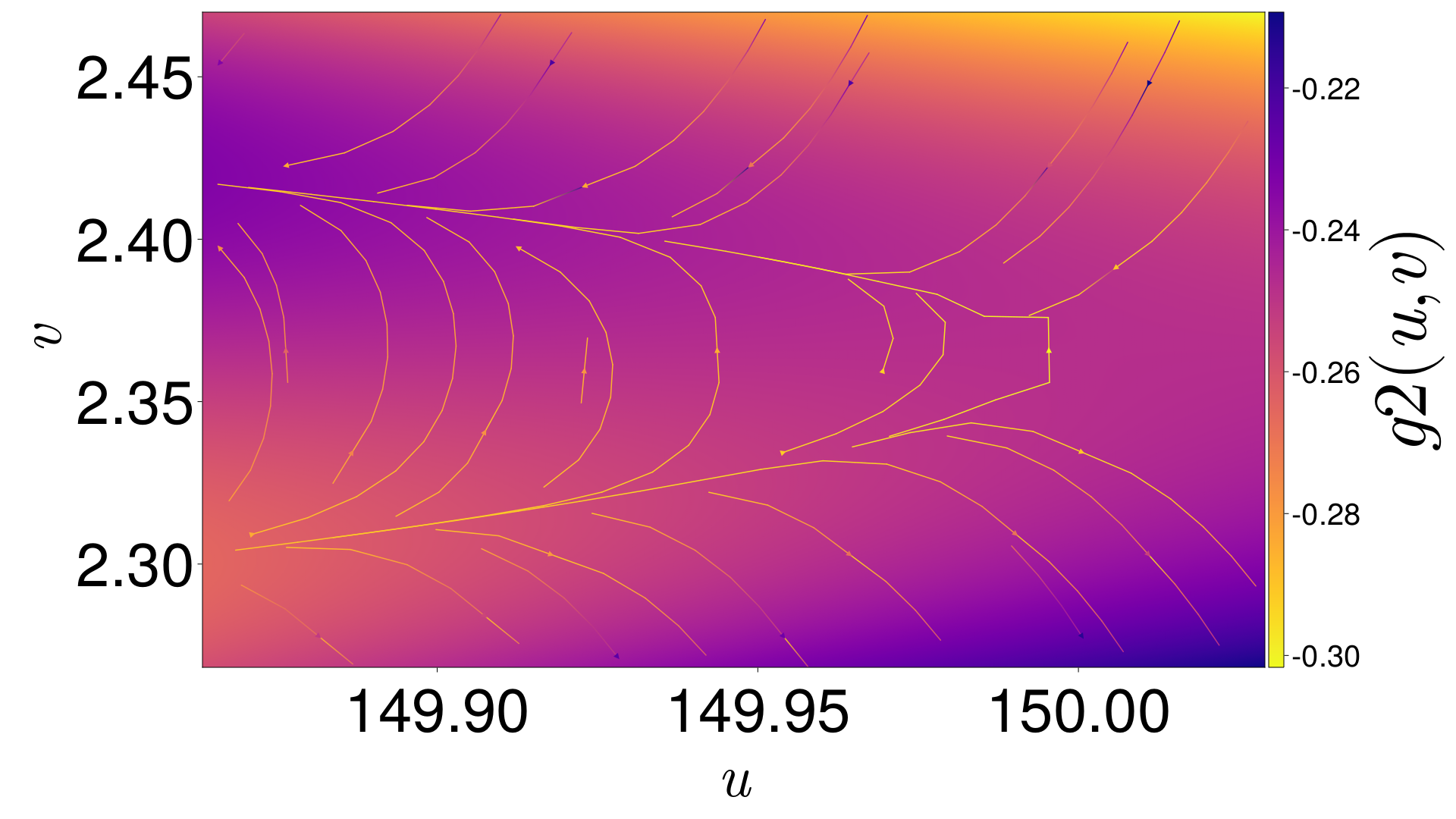}
        \centering
        (c) $g_2(u,v)$
        \label{fig:subfig3}
    \end{minipage}

    \caption{Stream plots demonstrating how variables $\left[s, g_1, g_2\right]$ vary across the field of view (in astrometric coordinates $(u,v)$) for the F115W simulated mosaic image. Recall that $s$ corresponds to size and $g_1, g_2$ correspond to shear.}
    \label{fig:streamplots}

\end{figure}

\texttt{ShOpt} provides the following streamplots (Figure \ref{fig:streamplots}) to give the user an inclination toward how the PSF is changing across the field of view. We also have a Julia script, \texttt{reader.jl} that reads in the \texttt{summary.shopt} file and provides easy PSF reconstruction. If you want to do your analysis in Python, we also have Python code available for reading in \texttt{summary.shopt} files here \href{https://github.com/EdwardBerman/sigma-Eta-Shopt-Reader}{https://github.com/EdwardBerman/sigma-Eta-Shopt-Reader}.  

\subsection{summary.shopt}
\texttt{summary.shopt} contains $6$ relevant extensions. The first extension is named polynomial matrix, and it contains a $3-$dimension matrix. $2-$dimensions correspond to the dimensions of the input vignettes and the third dimension corresponds to the coefficients of the polynomial at that pixel. The second extension contains all data relevant to learned parameters $\left[s, g_1, g_2 \right]$ as well as the $(u,v)$ coordinates at each star. We also measure the mean relative error between stars and their pixel grid fits before the polynomial interpolation step. Note that only stars that make it through all filters are contained. The third, and fourth extensions contain $3-$dimensional arrays corresponding to the input vignettes and the pixel grid fits of the vignettes. The fifth extension provides flags that tell you the indices of stars that were filtered out of the final interpolation step. The sixth extension tells you how to find $\left[s, g_1, g_2 \right]$ at an arbitrary $(u,v)$. There is also a mode that only outputs the first, second, and sixth extensions for reasons related to storage concern. This is enabled by default.

\subsection{Command Line Outputs}
As an extra convenience, we give users the option to display some of the diagnostic material to the terminal using \texttt{UnicodePlots.jl}. This may be useful for less scrupulous more exploratory runs of our software or for users looking for a quick sanity check that everything ran correctly without having to navigate to an output directory and open all of the saved checkplots.

We also print out to the terminal some key information about what is happening as the program runs, including what the program is doing, how many and which stars are failing or being filtered, progress on fitting, and how long particular portions of the code took to run.

\clearpage

\section{Petal Diagrams}

We speculate that petal diagrams may be able to approximate the spiky natures of JWST PSFS. Consider $r = A \cos(k\theta + \gamma)$, shown below in Figure \ref{fig:petals} for different $\left[A, k\right]$ values where $\gamma = 0$. In practice, $\left[A, k, \gamma \right]$ could be learnable parameters. We could then choose some $f(r) \propto \frac{1}{r}$ such that the gray fades from black to white. We would define $f(r)$ piecewise such that it is $0$ outside of the petal and decreases radially with $r$ inside the petal. The upshot of this approach is that we can just look at the learned $k$ and immediately know if our PSF captures the correct number of wings. Alternatively, \cite{berge2019exponential} introduced exponential shapelets with an orthogonal separation of $r$ and $\theta$, that may also be useful.

\begin{figure}[!htb]
    \centering
    \includegraphics[width=0.75\linewidth]{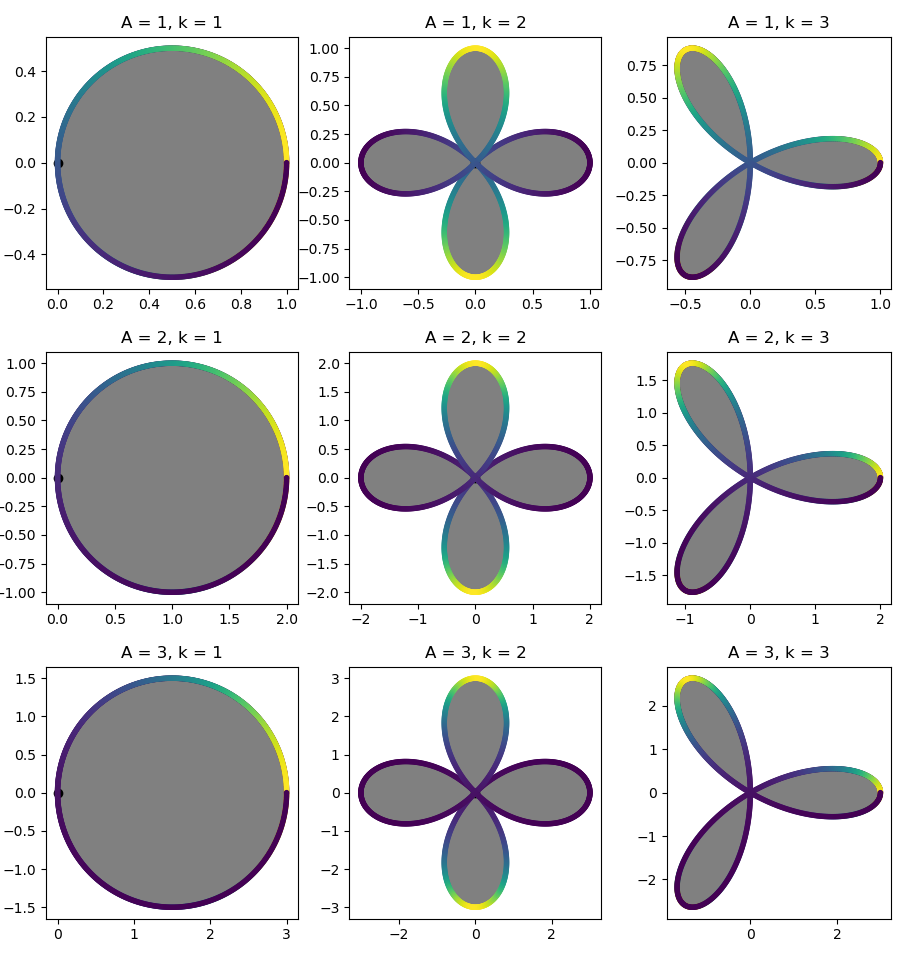}
    \caption{Petal diagrams approximating different PSFs. Different k values correspond to PSF with different numbers of spikes. The A value determines how long the spikes are. Here, all of the pixels inside the petals (gray) are set to a constant value, and everything outside is $0$.}
    \label{fig:petals}
\end{figure}

\clearpage

\bibliography{main}{}
\bibliographystyle{aasjournal}

\end{document}